\documentclass[11pt,a4paper]{elsarticle}
\usepackage[colorlinks]{hyperref}
\usepackage{multirow,pifont,lscape}

\usepackage[dvipsnames]{xcolor}

\AtBeginDocument{%
\usepackage{color}
\definecolor{mgray}{cmyk}{0,0,0,.8}
\hypersetup{
    bookmarks=true,         
    unicode=false,          
    pdftoolbar=true,        
    pdfmenubar=true,        
    pdffitwindow=true,      
    pdfnewwindow=true,      
    colorlinks=true,       
    linkcolor=mgray,          
    citecolor=blue,        
    filecolor=magenta,      
    urlcolor=blue          
}
}

\usepackage[utf8]{inputenc}
\usepackage{eso-pic}
\usepackage{amsmath,amsfonts,amsthm,graphicx,ulem,tocvsec2}
\usepackage{titlesec}
\usepackage{sectsty}

\allsectionsfont{\normalfont\sffamily\bfseries}

\journal{\hspace{-10em}\colorbox{white}{\phantom{Preprint for submission to Nuclear Physics B}}}

\makeatletter
\def\paragraph{\secdef{\els@aparagraph}{\els@bparagraph}}
\def\els@aparagraph[#1]#2{\elsparagraph[#1]{#2.}}
\def\els@bparagraph#1{\elsparagraph*{#1.}}

\usepackage{wrapfig}

\usepackage{multirow}

\usepackage{mathrsfs}
\usepackage{empheq}

\usepackage[font=footnotesize,labelfont=bf]{caption}

\newcommand\secref[1]{section~\ref{#1}}

\newcommand{\bal}{\begin{align}}
\newcommand{\eal}{\end{align}}
\newcommand{\ba}{\begin{align}}
\newcommand{\ea}{\end{align}}
\newcommand{\beq}{\begin{equation}}
\newcommand{\eeq}{\end{equation}}
\newcommand\beqa{\begin{eqnarray}}
\newcommand\eeqa{\end{eqnarray}}
\newcommand\bea{\begin{array}}
\newcommand\eea{\end{array}}
\newcommand\comment[1]{{}}

\newcommand{\algu}{\mathfrak{u}}
\newcommand{\su}{\mathfrak{su}}
\renewcommand{\sl}{\mathfrak{sl}}
\newcommand{\psu}{\mathfrak{psu}}
\newcommand{\pu}{\mathfrak{pu}}
\newcommand{\gl}{\mathfrak{gl}}

    \newcommand{\COMMENT}[1]{}
    
    \newcommand{\neqa}{\nonumber\end{eqnarray}}
    
\def\a{{\alpha}}

\def\[{\left[}
\def\]{\right]}

\def\l{\lambda}

\def\a{\alpha}
\def\b{\beta}

\def\<{\langle}
\def\>{\rangle}

\def\i2{\frac{i}{2}}

\def\<{\langle}
\def\>{\rangle}

\def\i2{\frac{i}{2}}

\DeclareMathOperator{\Tr}{Tr}

\def\1h{\hat 1}
\def\2h{{\hat 2}}
\def\3h{{\hat 3}}
\def\4h{{\hat 4}}

\def\be{\begin{eqnarray}}
\def\ee{\end{eqnarray}}
\def\no{\nonumber}

    \def\CO{{\cal O}}

    \def\<{\left\langle\,}
    \def\>{\, \right\rangle}
    \def\[{\left[}
    \def\]{\right]}

   \def\gl{{\mathfrak{gl}}}
   \def\su{{\mathfrak{su}}}
   \def\sl{{\mathfrak{sl}}}

\def\i{{\mathsf{i}}}

\renewcommand{\Im}{{\rm Im}\,}

\newcommand{\emp}{\emptyset}
\def\mi{\text{-}}

\colorlet{comp}{blue}
\newcommand{\suici}{{\color{comp}\ding{108}}}
\newcommand{\suii}{{\color{red}\ding{108}}}

\newcommand{\ZZ}{\mathcal{Z}}
\newcommand{\ZZb}{\bar{\mathcal{Z}}}
\newcommand{\XX}{\mathcal{X}}
\newcommand{\XXb}{\bar{\mathcal{X}}}
\newcommand{\YY}{\mathcal{Y}}
\newcommand{\YYb}{\bar{\mathcal{Y}}}
\newcommand{\DD}{\mathcal{D}}
\newcommand{\FF}{\mathcal{F}}
\newcommand{\FFb}{\bar{\mathcal{F}}}

\newcommand{\dQ}{\mathbb{Q}}

\newcommand{\aaa}{\mathbf{a}}
\newcommand{\bbb}{\mathbf{b}}
\newcommand{\fff}{\mathbf{f}}

\newcommand{\xZ}{{\color{NavyBlue}\mathcal{Z}}}
\newcommand{\xX}{{\color{ForestGreen}\mathcal{X}}}
\newcommand{\xY}{{\color{Violet}\mathcal{Y}}}

\newcommand{\xPsi}{{\color{Maroon}\Psi}}

\newcommand{\xD}{{\color{BurntOrange}\mathcal{D}}}

\usepackage{amssymb}
\usepackage{subfig}
\usepackage{overpic,rotating}
\usepackage{multicol}

\usepackage{varioref}

\begin{document}

\begin{frontmatter}

\title{\sffamily\LARGE{\bf The full spectrum of  AdS$_5$/CFT$_4$ I:} \\ Representation theory and one-loop Q-system}

\author[1,2]{Christian Marboe}
\ead{marboec@tcd.ie}
\author[1,3]{Dmytro Volin}
\address[1]{School of Mathematics, Trinity College Dublin, College Green, Dublin 2, Ireland \vspace{2mm}
}
\address[2]{Institut f\"{u}r Mathematik \& Institut f\"{u}r Physik, Humboldt-Universit\"{a}t zu Berlin,\\ Zum Gro\ss en Windkanal 6, 12489 Berlin, Germany \vspace{2mm}
}
\address[3]{Bogolyubov Institute for Theoretical Physics, 14-b, Metrolohichna str.Kiev, 03680, Ukraine\\[8em]}
\ead{volind@tcd.ie}

\begin{abstract}
With the formulation of the quantum spectral curve for the  AdS$_5$/CFT$_4$ integrable system, it became potentially possible to compute its full spectrum with high efficiency. This is the first paper in a series devoted to the explicit design of such computations, with no restrictions to particular subsectors being imposed.

We revisit the representation theoretical classification of possible states in the spectrum and map the symmetry multiplets to solutions of the quantum spectral curve at zero coupling. To this end it is practical to introduce a generalisation of Young diagrams to the case of non-compact representations and define algebraic Q-systems directly on these diagrams. Furthermore, we propose an algorithm to explicitly solve such Q-systems that circumvents the traditional usage of  Bethe equations and simplifies the computation effort.

For example, our algorithm quickly obtains explicit analytic results 
for all 495 multiplets that accommodate single-trace operators in  $\mathcal{N}=4$ SYM with classical conformal dimension up to $\frac{13}{2}$. We plan to use these results as the seed for solving the quantum spectral curve perturbatively to high loop orders in the next paper of the series.


\end{abstract}

%
%


%
%

\end{frontmatter}

\AddToShipoutPictureBG*{%
  \AtPageUpperLeft{%
    \hspace{0.9\paperwidth}%
    \raisebox{-5\baselineskip}{%
      \makebox[0pt][r]{\texttt{ TCDMATH 17-02}}}
}}%

\AddToShipoutPictureBG*{%
  \AtPageUpperLeft{%
    \hspace{0.7\paperwidth}%
    \raisebox{-25.5\baselineskip}{%
      \makebox[0pt][r]{

      \begin{picture}(324,100)
            \setlength{\unitlength}{0.22mm}
\linethickness{0.5mm}

\color{gray}
\put(0,70){\line(0,1){30}}
\put(10,30){\line(0,1){70}}
\put(20,0){\line(0,1){100}}
\put(30,0){\line(0,1){70}}
\put(40,0){\line(0,1){30}}
\put(0,90){\line(1,0){20}}
\put(0,80){\line(1,0){20}}
\put(0,70){\line(1,0){30}}
\put(10,60){\line(1,0){20}}
\put(10,50){\line(1,0){20}}
\put(10,40){\line(1,0){20}}
\put(10,30){\line(1,0){30}}
\put(20,20){\line(1,0){20}}
\put(20,10){\line(1,0){20}}
\color{black}
\put(10,30){\line(0,1){40}}
\put(20,30){\line(0,1){40}}
\put(30,30){\line(0,1){40}}
\put(10,70){\line(1,0){20}}
\put(10,60){\line(1,0){20}}
\put(10,50){\line(1,0){20}}
\put(10,40){\line(1,0){20}}
\put(10,30){\line(1,0){20}}

\color{gray}
\put(50,70){\line(0,1){30}}
\put(60,50){\line(0,1){50}}
\put(70,30){\line(0,1){70}}
\put(80,0){\line(0,1){100}}
\put(90,0){\line(0,1){70}}
\put(100,0){\line(0,1){50}}
\put(110,0){\line(0,1){30}}
\put(50,90){\line(1,0){30}}
\put(50,80){\line(1,0){30}}
\put(50,70){\line(1,0){40}}
\put(60,60){\line(1,0){30}}
\put(60,50){\line(1,0){40}}
\put(70,40){\line(1,0){30}}
\put(70,30){\line(1,0){40}}
\put(80,20){\line(1,0){30}}
\put(80,10){\line(1,0){30}}
\color{black}
\put(60,50){\line(0,1){20}}
\put(70,30){\line(0,1){40}}
\put(80,30){\line(0,1){40}}
\put(90,30){\line(0,1){40}}
\put(100,30){\line(0,1){20}}
\put(60,70){\line(1,0){30}}
\put(60,60){\line(1,0){30}}
\put(60,50){\line(1,0){40}}
\put(70,40){\line(1,0){30}}
\put(70,30){\line(1,0){30}}


\color{gray}
\put(120,70){\line(0,1){30}}
\put(130,50){\line(0,1){50}}
\put(140,50){\line(0,1){50}}
\put(150,30){\line(0,1){70}}
\put(160,0){\line(0,1){100}}
\put(170,0){\line(0,1){70}}
\put(180,0){\line(0,1){50}}
\put(190,0){\line(0,1){50}}
\put(200,0){\line(0,1){30}}
\put(120,90){\line(1,0){40}}
\put(120,80){\line(1,0){40}}
\put(120,70){\line(1,0){50}}
\put(130,60){\line(1,0){40}}
\put(130,50){\line(1,0){60}}
\put(150,40){\line(1,0){40}}
\put(150,30){\line(1,0){50}}
\put(160,20){\line(1,0){40}}
\put(160,10){\line(1,0){40}}
\color{black}
\put(130,50){\line(0,1){20}}
\put(140,50){\line(0,1){20}}
\put(150,30){\line(0,1){40}}
\put(160,30){\line(0,1){40}}
\put(170,30){\line(0,1){40}}
\put(180,30){\line(0,1){20}}
\put(190,30){\line(0,1){20}}
\put(130,70){\line(1,0){40}}
\put(130,60){\line(1,0){40}}
\put(130,50){\line(1,0){60}}
\put(150,40){\line(1,0){40}}
\put(150,30){\line(1,0){40}}

\color{gray}
\put(210,80){\line(0,1){20}}
\put(220,50){\line(0,1){50}}
\put(230,20){\line(0,1){80}}
\put(240,0){\line(0,1){100}}
\put(250,0){\line(0,1){80}}
\put(260,0){\line(0,1){50}}
\put(270,0){\line(0,1){20}}
\put(210,90){\line(1,0){30}}
\put(210,80){\line(1,0){40}}
\put(220,70){\line(1,0){30}}
\put(220,60){\line(1,0){30}}
\put(220,50){\line(1,0){40}}
\put(230,40){\line(1,0){30}}
\put(230,30){\line(1,0){30}}
\put(230,20){\line(1,0){40}}
\put(240,10){\line(1,0){30}}
\color{black}
\put(220,50){\line(0,1){20}}
\put(230,20){\line(0,1){50}}
\put(240,20){\line(0,1){60}}
\put(250,30){\line(0,1){50}}
\put(260,30){\line(0,1){20}}
\put(240,80){\line(1,0){10}}
\put(220,70){\line(1,0){30}}
\put(220,60){\line(1,0){30}}
\put(220,50){\line(1,0){40}}
\put(230,40){\line(1,0){30}}
\put(230,30){\line(1,0){30}}
\put(230,20){\line(1,0){10}}

\color{gray}
\put(280,90){\line(0,1){10}}
\put(290,10){\line(0,1){90}}
\put(300,0){\line(0,1){100}}
\put(310,0){\line(0,1){90}}
\put(320,0){\line(0,1){10}}
\put(280,90){\line(1,0){30}}
\put(290,80){\line(1,0){20}}
\put(290,70){\line(1,0){20}}
\put(290,60){\line(1,0){20}}
\put(290,50){\line(1,0){20}}
\put(290,40){\line(1,0){20}}
\put(290,30){\line(1,0){20}}
\put(290,20){\line(1,0){20}}
\put(290,10){\line(1,0){30}}
\color{black}
\put(290,10){\line(0,1){60}}
\put(300,10){\line(0,1){80}}
\put(310,30){\line(0,1){60}}
\put(300,90){\line(1,0){10}}
\put(300,80){\line(1,0){10}}
\put(290,70){\line(1,0){20}}
\put(290,60){\line(1,0){20}}
\put(290,50){\line(1,0){20}}
\put(290,40){\line(1,0){20}}
\put(290,30){\line(1,0){20}}
\put(290,20){\line(1,0){10}}
\put(290,10){\line(1,0){10}}

\color{gray}
\put(330,70){\line(0,1){30}}
\put(340,60){\line(0,1){40}}
\put(350,40){\line(0,1){60}}
\put(360,30){\line(0,1){70}}
\put(370,0){\line(0,1){100}}
\put(380,0){\line(0,1){70}}
\put(390,0){\line(0,1){60}}
\put(400,0){\line(0,1){40}}
\put(410,0){\line(0,1){30}}
\put(330,90){\line(1,0){40}}
\put(330,80){\line(1,0){40}}
\put(330,70){\line(1,0){50}}
\put(340,60){\line(1,0){50}}
\put(350,50){\line(1,0){40}}
\put(350,40){\line(1,0){50}}
\put(360,30){\line(1,0){50}}
\put(370,20){\line(1,0){40}}
\put(370,10){\line(1,0){40}}
\color{black}
\put(340,60){\line(0,1){10}}
\put(350,40){\line(0,1){30}}
\put(360,30){\line(0,1){40}}
\put(370,30){\line(0,1){40}}
\put(380,30){\line(0,1){40}}
\put(390,30){\line(0,1){30}}
\put(400,30){\line(0,1){10}}
\put(340,70){\line(1,0){40}}
\put(340,60){\line(1,0){50}}
\put(350,50){\line(1,0){40}}
\put(350,40){\line(1,0){50}}
\put(360,30){\line(1,0){40}}

\color{gray}
\put(420,90){\line(0,1){10}}
\put(430,30){\line(0,1){70}}
\put(440,30){\line(0,1){70}}
\put(450,0){\line(0,1){100}}
\put(460,0){\line(0,1){90}}
\put(470,0){\line(0,1){30}}
\put(480,0){\line(0,1){30}}
\put(420,90){\line(1,0){40}}
\put(430,80){\line(1,0){30}}
\put(430,70){\line(1,0){30}}
\put(430,60){\line(1,0){30}}
\put(430,50){\line(1,0){30}}
\put(430,40){\line(1,0){30}}
\put(430,30){\line(1,0){50}}
\put(450,20){\line(1,0){30}}
\put(450,10){\line(1,0){30}}
\color{black}
\put(450,90){\line(1,0){10}}
\put(450,80){\line(1,0){10}}
\put(430,70){\line(1,0){30}}
\put(430,60){\line(1,0){30}}
\put(430,50){\line(1,0){30}}
\put(430,40){\line(1,0){30}}
\put(430,30){\line(1,0){30}}
\put(430,30){\line(0,1){40}}
\put(440,30){\line(0,1){40}}
\put(450,30){\line(0,1){60}}
\put(460,30){\line(0,1){60}}

\color{gray}
\put(490,70){\line(0,1){30}}
\put(500,70){\line(0,1){30}}
\put(510,10){\line(0,1){90}}
\put(520,0){\line(0,1){100}}
\put(530,0){\line(0,1){70}}
\put(540,0){\line(0,1){70}}
\put(550,0){\line(0,1){10}}
\put(490,90){\line(1,0){30}}
\put(490,80){\line(1,0){30}}
\put(490,70){\line(1,0){50}}
\put(510,60){\line(1,0){30}}
\put(510,50){\line(1,0){30}}
\put(510,40){\line(1,0){30}}
\put(510,30){\line(1,0){30}}
\put(510,20){\line(1,0){30}}
\put(510,10){\line(1,0){40}}
\color{black}
\put(510,10){\line(0,1){60}}
\put(520,10){\line(0,1){60}}
\put(530,30){\line(0,1){40}}
\put(540,30){\line(0,1){40}}
\put(510,70){\line(1,0){30}}
\put(510,60){\line(1,0){30}}
\put(510,50){\line(1,0){30}}
\put(510,40){\line(1,0){30}}
\put(510,30){\line(1,0){30}}
\put(510,20){\line(1,0){10}}
\put(510,10){\line(1,0){10}}

\color{gray}
\put(560,70){\line(0,1){30}}
\put(570,70){\line(0,1){30}}
\put(580,30){\line(0,1){70}}
\put(590,30){\line(0,1){70}}
\put(600,0){\line(0,1){100}}
\put(610,0){\line(0,1){70}}
\put(620,0){\line(0,1){70}}
\put(630,0){\line(0,1){30}}
\put(640,0){\line(0,1){30}}
\put(560,90){\line(1,0){40}}
\put(560,80){\line(1,0){40}}
\put(560,70){\line(1,0){60}}
\put(580,60){\line(1,0){40}}
\put(580,50){\line(1,0){40}}
\put(580,40){\line(1,0){40}}
\put(580,30){\line(1,0){60}}
\put(600,20){\line(1,0){40}}
\put(600,10){\line(1,0){40}}
\color{black}
\put(580,30){\line(0,1){40}}
\put(590,30){\line(0,1){40}}
\put(600,30){\line(0,1){40}}
\put(610,30){\line(0,1){40}}
\put(620,30){\line(0,1){40}}
\put(580,70){\line(1,0){40}}
\put(580,60){\line(1,0){40}}
\put(580,50){\line(1,0){40}}
\put(580,40){\line(1,0){40}}
\put(580,30){\line(1,0){40}}

\end{picture}
      }}
}}%

\newpage
\thispagestyle{empty}

\begingroup
\hypersetup{linkcolor=black}
\setcounter{tocdepth}{2}
\tableofcontents
\endgroup

\newpage

\section{Introduction}
The spectrum of planar $\mathcal{N}=4$ supersymmetric Yang-Mills theory (SYM) has been intensively studied in the literature during the past 15 years. The high interest in this subject is not surprising as SYM is, on one hand,  the archetypical example of the AdS/CFT correspondence, and it is, on the other hand,  a rare example of a four-dimensional gauge theory that can be explored explicitly at arbitrary coupling thanks to integrability, see \cite{Beisert:2010jr} for a review.

The spectral problem amounts to finding the eigenvalues of the dilatation operator  acting in the space of single-trace operators
\be
\mathbb{D} \,\mathcal{O}(x)=\Delta \, \mathcal{O}(x)\,,\ \ \ \ \mathcal{O}(x)=\Tr\left[\ZZ\DD\Psi\hdots\right]\,.
\ee
After the discovery of integrability in SYM \cite{Minahan:2002ve,Bena:2003wd,Beisert:2003tq}, two important milestones in solving this problem were the asymptotic Bethe Ansatz equations \cite{Beisert:2005fw,Beisert:2006ez} that allowed computing the spectrum of very long operators,  and the thermodynamic Bethe Ansatz equations \cite{Gromov:2009tv,Bombardelli:2009ns,Gromov:2009bc,Arutyunov:2009ur} that were formally suitable for arbitrary operators but whose application was quite limited due to their complexity.

With the formulation of the Quantum Spectral Curve (QSC) \cite{Gromov:2013pga,Gromov:2014caa}, we came to the verge of the  complete practical solution of the spectral problem\footnote{To be fair, we note that the explicit structure of eigenvectors or even the structure of the dilatation operator itself at finite coupling is not provided by the current version of the QSC formalism or any other approach in the literature. It is unclear, though, whether the knowledge of this structure is essential for the computation of quantities such as 3-point functions or if some version of a bootstrap approach based on the spectral data  would be sufficient.}. The analytic and numerical efficiency of QSC has already been successfully demonstrated  in a vast range of specialised scenarios \cite{Gromov:2014bva,Alfimov:2014bwa,Marboe:2014gma,Marboe:2014sya,Gromov:2015wca,Gromov:2015vua,Gromov:2016rrp,Hegedus:2016eop,Marboe:2016igj}, also for  AdS$_4$/CFT$_3$ \cite{Cavaglia:2014exa,Gromov:2014eha,Anselmetti:2015mda,Cavaglia:2016ide,Bombardelli:2017vhk}. Furthermore, QSC was studied in the context of certain integrability-preserving deformations \cite{Kazakov:2015efa,Gromov:2015dfa}. However, in all of the so far tested applications, the explicit solutions were devised for restricted classes of operators, typically belonging to the $\sl(2)$ subsector of the theory or its  generalisation, deformation, or analytic continuation.

This work is the first step towards making the concrete solution of the spectral problem completely general, as well as accessible and automatic. Another motivation is that the spectrum of anomalous dimensions also plays a role in the calculation of quantities beyond the spectrum, in particular structure constants. We aim to provide a user-friendly library for spectral data that is of use in these efforts.


This paper focuses on the solution of QSC at zero value of the coupling constant, with the aim of using the obtained results as the seed for perturbative or numerical QSC-based computations in future developments. The paper is a hybrid of a review and new work. For consistency and to make the material as self-contained as possible, we revisit well-known results from representation theory and its applications in the study of one-loop \cite{Beisert:2003jj,Beisert:2003yb} and asymptotic \cite{Beisert:2005fw} Bethe equations, and recast them in a notation appropriate for QSC. But we also provide new insights and tricks.

The  original contributions are a new way to think of the integrable Q-system underlying non-compact Heisenberg super spin chains, and thus the one-loop spectral problem in  SYM, and a new efficient algorithm to solve this system. This algorithm is a generalisation of our treatment of compact spin chains in \cite{Marboe:2016yyn} to the non-compact case. Our findings systematically use non-compact Young diagrams \cite{Gunaydin:2017lhg}, and we furthermore introduce a notion of extended Young diagrams that, in particular, allow us to count multiplicities in the $\psu(2,2|4)$ spectrum by using compact $\su(N)$ characters only.

In section \ref{sec:rep}, we give a thorough introduction to representation theory of $\psu(2,2|4)$, with an emphasis on understanding the concrete content of the spectrum. In section \ref{sec:Qsys}, we introduce the notion of algebraic Q-systems as they appear in integrable models. We argue that these Q-systems can be naturally  built on Young diagrams, and provide an algorithm to solve them based on polynomial division. We then discuss the connection to the Q-system that appears in the Quantum Spectral Curve and how to fully construct the leading contribution to this system.

The paper is accompanied by a \texttt{Mathematica} notebook, \texttt{spectrum.nb}, with an implementation of the proposed algorithm to solve Q-systems. 



\section{Representation theory} \label{sec:rep}
The Lie superalgebra $\mathfrak{u}(2,2|4)$ is defined by the super Lie-bracket
\be
[E_{mn},E_{kl}\}
&=&\delta_{nk}E_{ml}-(-1)^{(p_m+p_n)(p_k+p_l)}\delta_{ml}E_{kn}\,. \label{psu224}
\ee
Furthermore, the adjoint generators $E_{mn}^\dagger$ are defined by
\be
E_{mn}^\dagger = (-1)^{c_m+c_n}E_{nm}\,.
\ee
The labels $m,n,\ldots$ are elements of the set $\{\mathit{1},\mathit{2},\dot{\mathit{1}},\dot{\mathit{2}},\hat 1,\hat 2,\hat 3,\hat 4\}$, and the value of the grading functions $p,c$ are given by the table
\be
\begin{tabular}[b]{c|c|c|c}
 & $\alpha$ & $\dot\alpha$ & $\hat a$\\
 \hline
 $p$ & 0 & 0 & 1
\\
$c$ & 0 & 1 & 0
\end{tabular}\,.
\ee
The subalgebra of super-traceless elements in $\algu(2,2|4)$ form $\su(2,2|4)$. The projective Lie superalgebra $\psu(2,2|4)$ can effectively be defined by furthermore imposing that the central charge vanishes,
\be
C= \sum_{n} E_{nn}=0\,.
\ee

At finite coupling, $\psu(2,2|4)$ is the symmetry algebra that organises the AdS/CFT spectrum, i.e.\ we decompose the Hilbert space into superconformal multiplets -- irreps of $\psu(2,2|4)$; all states within the same multiplet have identical anomalous part of the conformal dimension. As the rank of the $\psu(2,2|4)$ algebra is six, we can label multiplets using {\it six numbers}\footnote{These can be e.g.\ Cartan charges defined by \eqref{Cartandef}, however we use fundamental weights for this purpose as described in subsection~\ref{sec:qnum}.}.

At zero coupling, however, the symmetry algebra of the spectrum is larger. It is $\mathfrak{pu}(2,2|4)\oplus\mathfrak{u}(1)$, where the extra $\algu(1)$ charge is the length $L$ of single-trace operators\footnote{It is wrong to identify $\mathfrak{pu}(2,2|4)\oplus\mathfrak{u}(1)$ with $\algu(2,2|4)$. For one thing, $L$ is not a combination of  $\algu(2,2|4)$ generators, as opposed to the central element $C$ in $\algu(2,2|4)$ which is a combination of $\algu(2,2|4)$ generators, appears on the r.h.s. of the commutation relations, and should be projected out to get the $\pu$ algebra. In other words, $\algu(2,2|4)$ is the central extension of $\pu(2,2|4)$, not a direct sum.}. The one-loop anomalous part of the dilatation generator commutes with generators of this extended algebra. Therefore the multiplets at zero coupling are labeled by {\it eight numbers}. We will often allow ourselves a loose terminology and refer to all multiplets as $\psu(2,2|4)$ or superconformal multiplets, even if they are effectively representations of the larger algebra at zero coupling.

We define the explicit multiplet labelling at arbitrary and at zero coupling in subsection~\ref{sec:qnum}, and then the discussion of the possible issues follows. Prior to that, we need to introduce the oscillator formalism which allows one to realise the $[\mathfrak{ps}]\mathfrak{u}(2,2|4)$ algebra and  to encode the value of $L$ as well. In effect, we will find that the below-defined eight oscillator numbers is a convenient labelling of the multiplets at zero coupling which we will use throughout the paper.

\subsection{Oscillator formalism} 
The usage of Schwinger oscillators in the description of superconformal algebras goes back to \cite{Bars:1982ep,Gunaydin:1984fk}. They were extensively used in the study of the AdS/CFT spectrum  \cite{Beisert:2003jj,Beisert:2004ry}. Commencing by reviewing the most standard facts about this formalism, we will gradually introduce less-standard notation into the exposition, with the goal to get an appropriate language to describe the QSC solutions in section~\ref{sec:Qsys}.

The following oscillator representations of $\gl(n)$ are used to parametrise the bosonic ($p$-even) generators of $\algu(2,2|4)$:
\begin{subequations}
\be
E_{\a\b}=\aaa^\dagger_\a \aaa_\b\,,&& [\aaa_\a,\aaa_\b^\dagger]=\delta_{\a\b} \label{oa}\,,\ \ \alpha,\beta\in\{\mathit{1},\mathit{2}\}\,,\\
E_{\dot{\a}\dot{\b}}=-\bbb_{\dot{\a}}\bbb^\dagger_{\dot{\b}}\,,&& [\bbb_{\dot{\a}},\bbb_{\dot{\b}}^\dagger]=\delta_{{\dot{\a}}{\dot{\b}}} \label{ob}\,,\quad \dot\alpha,\dot\beta\in\{\dot{\mathit{1}},\dot{\mathit{2}}\}\,,  \\\
E_{\hat a\hat b}=\fff^\dagger_{a} \fff_{b}\,,&& \{\fff_{a},\fff_{b}^\dagger\}=\delta_{ab} \label{of}\,,\quad \hat a,\hat b\in\{\hat 1,\hat 2,\hat 3,\hat 4\}\,.
\ee
\end{subequations}
The fermionic ($p$-odd) generators are parameterised by other bilinear combinations of the oscillators $\aaa,\bbb,\fff$, as precised below in \eqref{allgen}.

The central charge constraint reads
\be
C= -2-n_\bbb+n_\fff+n_\aaa = 0 \,, \label{cccon}
\ee
where $n$ are number operators, e.g.\ $n_{\fff_1}\equiv \fff_1^\dagger \fff_1$, and $n_\fff\equiv \sum_{i=1}^4 n_{\fff_i}$. To avoid notational burden, $\bbb_1$, $\bbb_2$, $\aaa_1$ and $\aaa_2$ will be used to denote, respectively, $\bbb_{\dot{\mathit{1}}}$, $\bbb_{\dot{\mathit{2}}}$, $\aaa_{{\mathit{1}}}$ and $\aaa_{{\mathit{2}}}$.

\subsubsection*{Gradings}
A total order on the  set of 8 labels $\{\mathit{1},\mathit{2},\dot{\mathit{1}},\dot{\mathit{2}},\hat 1,\hat 2,\hat 3,\hat 4\}$ is called a {\it grading}. We always order  the fermionic labels as $\hat 1<\hat 2<\hat 3<\hat 4$ and the bosonic labels as $\dot{\mathit{1}}<\dot{\mathit{2}}<\mathit{1}<\mathit{2}$, while the relative order between bosonic and fermionic labels can vary and will be outlined explicitly each time. The order $\dot{\mathit{1}}<\dot{\mathit{2}}<\mathit{1}<\mathit{2}$ suggests an alternative notation $\{1,2,3,4\}\Leftrightarrow \{\dot{\mathit{1}},\dot{\mathit{2}},\mathit{1},\mathit{2}\}$, which will be used in the questions of grading and in the definition of fundamental weights \eqref{weights2}, and denoted by the indices $i,j,\ldots$.

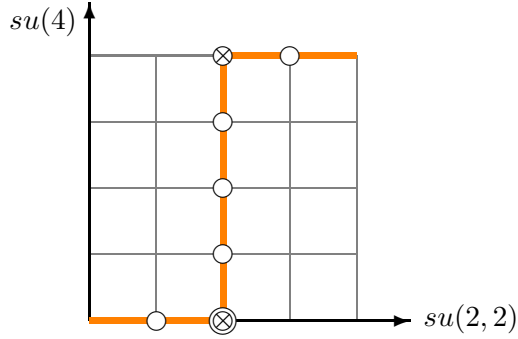
\begin{figure}[t]
\centering
\begin{picture}(150,120)

{\color{gray}
\thinlines
\put(25,25){\line(1,0){100}}
\put(25,50){\line(1,0){100}}
\put(25,75){\line(1,0){100}}
\put(25,100){\line(1,0){100}}
\put(50,0){\line(0,1){100}}
\put(75,0){\line(0,1){100}}
\put(100,0){\line(0,1){100}}
\put(125,0){\line(0,1){100}}
}
\thicklines

\put(25,0){\vector(1,0){120}}
\put(25,0){\vector(0,1){120}}

\linethickness{0.8mm}
\put(25,0){\color{orange}\line(1,0){50}}
\put(75,0){\color{orange}\line(0,1){100}}
\put(75,100){\color{orange}\line(1,0){50}}

\thinlines

{\color{white}
\put(50,0){\circle*{7}}
\put(75,0){\circle*{10}}
\put(75,25){\circle*{7}}
\put(75,50){\circle*{7}}
\put(75,75){\circle*{7}}
\put(75,100){\circle*{7}}
\put(100,100){\circle*{7}}}

\put(50,0){\circle{7}}
\put(75,0){\circle{7}}
\put(75,25){\circle{7}}
\put(75,50){\circle{7}}
\put(75,0){\circle{10}}
\put(75,75){\circle{7}}
\put(75,100){\circle{7}}
\put(100,100){\circle{7}}

\put(70.7,-2.7){$\times$}
\put(70.7,97.3){$\times$}

\put(-5,110){$su(4)$}
\put(150,-2){$su(2,2)$}

\end{picture}
\caption{The compact beauty Dynkin diagram. A cross denotes a change in $p$-grading \cite{KAC19778}, while a double circle denotes a change in the $c$-grading. Dynkin diagrams as two-dimensional paths came into use after \cite{Kazakov:2007fy}.}
\label{fig:beauty}
\end{figure}
\begin{table}[t]
\begin{tabular}{|c|c|c|c|} \hline
Name & grading & short-hand notation & Dynkin diagram \\\hline\hline
\it{compact beauty}&$12\hat{1}\hat{2}\hat{3}\hat{4}34$&2222&
\begin{picture}(140,13)

\thicklines
\put(10,5){\line(1,0){120}}

\thinlines

{\color{white}
\put(10,5){\circle*{7}}
\put(30,5){\circle*{10}}
\put(50,5){\circle*{7}}
\put(70,5){\circle*{7}}
\put(90,5){\circle*{7}}
\put(110,5){\circle*{7}}
\put(130,5){\circle*{7}}}

\put(10,5){\circle{7}}
\put(30,5){\circle{7}}
\put(50,5){\circle{7}}
\put(70,5){\circle{7}}
\put(90,5){\circle{7}}
\put(110,5){\circle{7}}
\put(130,5){\circle{7}}

\put(30,5){\circle{10}}

\put(25.7,2.3){$\times$}
\put(105.7,2.3){$\times$}

\end{picture}
\\\hline
\it{non-compact beauty}&$\hat{1}\hat{2}1234\hat{3}\hat{4}$&0044&
\begin{picture}(140,13)

\thicklines
\put(10,5){\line(1,0){120}}

\thinlines

{\color{white}
\put(10,5){\circle*{7}}
\put(30,5){\circle*{10}}
\put(50,5){\circle*{7}}
\put(70,5){\circle*{10}}
\put(90,5){\circle*{7}}
\put(110,5){\circle*{7}}
\put(130,5){\circle*{7}}}

\put(10,5){\circle{7}}
\put(30,5){\circle{7}}
\put(50,5){\circle{7}}
\put(70,5){\circle{7}}
\put(90,5){\circle{7}}
\put(110,5){\circle{7}}
\put(130,5){\circle{7}}

\put(30,5){\circle{10}}
\put(70,5){\circle{10}}

\put(25.7,2.3){$\times$}
\put(105.7,2.3){$\times$}

\end{picture}
\\\hline
\it{compact ABA}&$\hat{1}12\hat{2}\hat{3}34\hat{4}$&0224&

\begin{picture}(140,13)

\thicklines
\put(10,5){\line(1,0){120}}

\thinlines

{\color{white}
\put(10,5){\circle*{10}}
\put(30,5){\circle*{7}}
\put(50,5){\circle*{10}}
\put(70,5){\circle*{7}}
\put(90,5){\circle*{7}}
\put(110,5){\circle*{7}}
\put(130,5){\circle*{7}}}

\put(10,5){\circle{7}}
\put(30,5){\circle{7}}
\put(50,5){\circle{7}}
\put(70,5){\circle{7}}
\put(90,5){\circle{7}}
\put(110,5){\circle{7}}
\put(130,5){\circle{7}}

\put(10,5){\circle{10}}
\put(50,5){\circle{10}}

\put(5.7,2.3){$\times$}
\put(45.7,2.3){$\times$}
\put(85.7,2.3){$\times$}
\put(125.7,2.3){$\times$}

\end{picture}

\\\hline
\it{non-compact ABA}&$1\hat{1}\hat{2}23\hat{3}\hat{4}4$&1133&
\begin{picture}(140,13)

\thicklines
\put(10,5){\line(1,0){120}}

\thinlines

{\color{white}
\put(10,5){\circle*{10}}
\put(30,5){\circle*{7}}
\put(50,5){\circle*{10}}
\put(70,5){\circle*{10}}
\put(90,5){\circle*{7}}
\put(110,5){\circle*{7}}
\put(130,5){\circle*{7}}}

\put(10,5){\circle{7}}
\put(30,5){\circle{7}}
\put(50,5){\circle{7}}
\put(70,5){\circle{7}}
\put(90,5){\circle{7}}
\put(110,5){\circle{7}}
\put(130,5){\circle{7}}

\put(10,5){\circle{10}}
\put(50,5){\circle{10}}
\put(70,5){\circle{10}}

\put(5.7,2.3){$\times$}
\put(45.7,2.3){$\times$}
\put(85.7,2.3){$\times$}
\put(125.7,2.3){$\times$}

\end{picture}
\\\hline
\end{tabular}
\centering
\caption{Often used gradings.}
\label{table:gradings}
\end{table}

We denote the total order by a sequence of eight elements where the first element is the smallest one, etc. As an example, consider the grading $12\hat{1}\hat{2}\hat{3}\hat{4}34$, which is  referred to as the {\it{compact beauty}} \cite{Beisert:2003yb} grading. The corresponding oscillator parametrisation is
\be\label{allgen}
E_{mn}=
\begin{pmatrix}
-\bbb_{\dot{\a}}\bbb_{\dot{\b}}^\dagger & -\bbb_{\dot{\a}}\fff_\bbb & - \bbb_{\dot{\a}} \aaa_\b \\
\fff_{a}^\dagger \bbb_{\dot{\b}}^\dagger & \fff_{a}^\dagger \fff_b & \fff_{a}^\dagger \aaa_\b \\
\aaa_{{\a}}^\dagger \bbb_{\dot{\b}}^\dagger & \aaa_{{\a}}^\dagger \fff_b & \aaa_{{\a}}^\dagger \aaa_\b
\end{pmatrix}\,.
\ee
To this grading we associate the path and Dynkin diagram depicted in figure \ref{fig:beauty}.

Another short-hand notation for the grading, which is predominantly used in the \texttt{Mathematica} notebook related to this work, is to use four numbers $\delta_1\delta_2\delta_3\delta_4$, where $\delta_a$ is the number of bosonic labels before $\hat a$, e.g.\  2222 stands for the compact beauty grading. Some commonly encountered gradings in the literature are listed in table \ref{table:gradings}.


\subsubsection*{States and field content interpretation}
A highest-weight state (HWS) is defined by the property that it is annihilated by all generators above the diagonal, i.e.
\be
E_{mn}|\text{HWS}\rangle = 0 \quad \text{for } m<n\,.
\ee
We furthermore define the Fock vacuum, $|0\rangle$, by
\be
\aaa_\alpha|0\rangle=\bbb_{\dot{\alpha}}|0\rangle=\fff_a|0\rangle=0\,.
\ee
This state does not satisfy the central charge constraint \eqref{cccon}. The spectrum of the theory is built by acting on $|0\rangle$ with $\aaa_\alpha^\dagger$, $\bbb_{\dot{\alpha}}^\dagger$ and $\fff_a^\dagger$ and the possible states are summarised in table \ref{table:states}.

\begin{table}[h!]
\centering
\begin{tabular}{|c|c|c|c|c|}\hline
\multicolumn{2}{|c|}{Field interpretation} & Content & $\Delta_0$ & Components\\\hline\hline

scalar & $\Phi_{ab}$&$\fff_a^\dagger \fff_b^\dagger \,|0\rangle$& 1 & 6 \\\hline

\multirow{2}{*}{fermion}&$\Psi_{a\alpha}$&$\fff_a^\dagger \aaa_\alpha^\dagger \,|0\rangle$& $\frac{3}{2}$ & 8 \\\cline{2-5}
&$\bar{\Psi}_{a\dot{\alpha}}$&$\epsilon_{abcd}\fff_b^\dagger \fff_c^\dagger \fff_d^\dagger \bbb_{\dot{\alpha}}^\dagger \,|0\rangle$& $\frac{3}{2}$ & 8 \\\hline

\multirow{2}{*}{field strength}&
$\mathcal{F}_{\alpha\beta}$&$\aaa_\a^\dagger \aaa_\b^\dagger \,|0\rangle$& 2 & 3 \\\cline{2-5}
&$\bar{\mathcal{F}}_{\dot{\a}\dot{\beta}}$&$\fff_1^\dagger \fff_2^\dagger \fff_3^\dagger \fff_4^\dagger \bbb_{\dot{\alpha}}^\dagger \bbb_{\dot{\beta}}^\dagger \,|0\rangle$& 2 & 3 \\\hline

covariant derivative &
$\mathcal{D}_{\alpha\dot{\alpha}}$&$\aaa_\alpha^\dagger \bbb_{\dot{\alpha}}^\dagger$& 1 & 4 \\\hline
\end{tabular}
\caption{States satisfying the central charge constraint. Note that $\Phi_{ab}=-\Phi_{ba}$, i.e.\ there are six independent scalars. We denote them by $\mathcal{Z}\equiv \fff_1^\dagger \fff_2^\dagger |0\rangle$, $\mathcal{X}\equiv \fff_1^\dagger \fff_3^\dagger |0\rangle$, $\mathcal{Y}\equiv \fff_1^\dagger \fff_4^\dagger |0\rangle$, $\bar{\mathcal{Y}}\equiv \fff_2^\dagger \fff_3^\dagger |0\rangle$, $\bar{\mathcal{X}}\equiv \fff_2^\dagger \fff_4^\dagger |0\rangle$, $\bar{\mathcal{Z}}\equiv \fff_3^\dagger \fff_4^\dagger |0\rangle$. Note also that $\mathcal{F}_{\alpha\beta}=\mathcal{F}_{\beta\alpha}$ and $\bar{\mathcal{F}}_{\dot{\alpha}\dot{\beta}}=\bar{\mathcal{F}}_{\dot{\beta}\dot{\alpha}}$. A state can contain one of the fundamental fields and an unlimited number of covariant derivatives.}
\label{table:states}
\end{table}

From this single-field representation of $\psu(2,2|4)$ we can build tensor product states, by dressing up $|0\rangle^{\otimes L}$. We interpret these as single-trace operators of length $L$. As an example, consider the length-three state
\be
\text{Tr}[\,\ZZ\,\, \mathcal{D}_{12} \Psi_{11} \,\, \mathcal{F}_{12} \,] = \fff_1^\dagger \fff_2^\dagger |0\rangle \otimes (\aaa_1^\dagger)^2 \bbb_2^\dagger \fff_1^\dagger  |0\rangle   \otimes  \aaa_1^\dagger \aaa_2^\dagger |0\rangle  \,. \no
\ee
For simplicity, we will often leave out the $\text{Tr}[...]$ symbol. Due to the cyclicity of the trace, some states are equivalent since they are related by cyclic permutations, e.g.\ $\ZZ\XX=\XX\ZZ$. As fermions anticommute, this also means that some states involving fermions must vanish, e.g.\ $\Psi_{a\alpha}\Psi_{a\alpha}=-\Psi_{a\alpha}\Psi_{a\alpha}=0$.

The central charge constraint for a tensor product state is
\be
n_\aaa-n_\bbb+n_\fff=2L\,. \label{ccc}
\ee

The fields are assigned a classical conformal dimension, $\Delta_0$, listed in table \ref{table:states}. The total classical dimension of an operator is
\be
\Delta_0 = \frac{n_\fff}{2}+n_\aaa
\,. \label{delta0}
\ee


\subsection{\label{sec:qnum}Quantum numbers}

\subsubsection*{Fundamental weights}
$\algu(2,2|4)$ contains two bosonic subgroups: $\algu(4)$, generated by bilinear combinations of $\fff_a$, and $\algu(2,2)$, generated by bilinear combinations of $\aaa_\alpha$ and $\bbb_{\dot{\alpha}}$.
By {\it{fundamental weights}}, or just {\it{weights}}, we refer to the eigenvalues of the diagonal elements, $E_{nn}$, acting on the HWS. Denote the $\algu(4)$ weights by
\be
\lambda_a = \fff_a^\dagger \fff_a=n_{\fff_a},\quad\quad a=1,...,4\,,\label{weights1}
\ee
and the $\algu(2,2)$ weights by
\be
\nu_i = \{-\bbb_{\dot{\a}} \bbb_{\dot{\a}}^\dagger , \aaa_\a^\dagger \aaa_\a\}_i=
\{-L-n_{\bbb_{\dot{\a}}}, n_{\aaa_\a}\}_i\,,\quad i=1,...,4\,.\label{weights2}
\ee
The central charge constraint reads
\be\label{cccc}
C=\sum_{a=1}^4\lambda_a+\sum_{i=1}^4\nu_i=0\,.
\ee
The six numbers  needed to classify representations of $\psu(2,2|4)$, are the differences $\lambda_{a}-\lambda_{a+1}$ and $\nu_{j}-\nu_{j+1}$. Therefore, the fundamental weights $\{\lambda,\nu\}$ and $\{\lambda+\Lambda,\nu-\Lambda\}$ define {\it the same} irrep of $\psu(2,2|4)$.

\subsubsection*{Oscillator numbers}
As was already mentioned, eight numbers are needed to properly describe a multiplet at zero coupling. The fundamental weights, despite being eight numbers, cannot serve for this goal. Indeed, due to \eqref{cccc}, they define only seven numbers labelling representations of $\pu(2,2|4)$, but they cannot, generically, define the value of the length $L$.

The appropriate eight numbers describing the multiplet are  the oscillator content $n$ used to construct the highest-weight state:
\be\label{oc}
[\,n_{\bbb_1},n_{\bbb_2}\,|\,n_{\fff_1},n_{\fff_2},n_{\fff_3},n_{\fff_4}\,|\,n_{\aaa_1},n_{\aaa_2}\,]\,.
\ee
When not clear from the context, we will imply the grading by a superscript, e.g.\ $n^{2222}$ for the compact beauty grading. Note that the oscillator content $n$ allows one to find the value of fundamental weights using \eqref{weights1} and \eqref{weights2}, and the length $L$ using the central charge constraint \eqref{ccc}.  Relations between $n$ and other conventionally used parametrisations of the quantum numbers in the literature can be found in \ref{ap:qn}.

\subsubsection*{$\mathcal{N}=4$ SYM at $g\neq0$}
The outlined oscillator representation is only a valid description of $\mathcal{N}=4$ SYM at the classical level, i.e.\ at vanishing coupling, $g=0$. We cannot use the oscillator language at finite coupling, and the language of fundamental weights is more appropriate. The weights $\lambda_a$ are the same as they are at zero coupling, while $\nu_i$ receive a contribution from the anomalous dimension at finite coupling, $\gamma\ge 0$, according to 
\be
\nu_i &=& \nu_i|_{g=0} + \frac{\gamma}{2}\, \{-1,-1,1,1\}_i  \,. \label{ano}
\ee

\subsubsection*{Mixing of operators with different length}
At one-loop, the eigenstates of the dilatation operator are linear combination of operators with the same oscillator content, and consequently with the same length. However, the perturbative corrections to the eigenstates of the higher-loop dilatation operator mixes operators of different length, which is a remarkable feature of the AdS/CFT integrable system \cite{Beisert:2003ys}. This mixing can happen for operators for which $\lambda$ and $\nu$ coincide up to a shift by an integer, i.e.\ $\lambda+\Lambda$ and $\nu-\Lambda$.

To see when this occurs, notice that $\lambda_a$ and $\nu_i$ are invariant under the length-changing replacement
\begin{subequations}
\label{bothL}
\be
\{L,\,n_{\bbb_{\dot{\alpha}}}\} \leftrightarrow \{ L-1, \, n_{\bbb_{\dot{\alpha}}}+1  \} \label{L1}\,,
\ee
and, furthermore, the length-changing replacement
\be
\{L,\,n_{\fff_a},\,n_{\aaa_{\alpha}}\} \leftrightarrow \{ L-1, \, n_{\fff_a}-1,n_{\aaa_\alpha}+1  \} \label{L2}\,
\ee
\end{subequations}
takes $\lambda_a\leftrightarrow\lambda_a-1$ and $\nu_i\leftrightarrow\nu_i+1$.

There are no other ways to amend the length than through \eqref{bothL}. Indeed, we cannot devise more than two ways to change the oscillator content and leave the six $\psu(2,2|4)$ quantum numbers unchanged. 
An example of operators related by these transformations and consequently able to mix at higher loops is given in table \ref{table:mixing}.


\begin{table}[h!]
\def\arraystretch{1.15}
\centering
\begin{tabular}{|c|c|c|c|c|}\hline
$n^{2222}$ & $L$ & Field content example & $\lambda_a$ & $\nu_j$\\\hline\hline
$[1,1|2,2,2,2|1,1]$ & 4 & $\Psi_{11}\Psi_{12}\bar{\Psi}_{11}\bar{\Psi}_{12}$ & $\{2,2,2,2\}$ & $\{-5,-5,1,1\}$ \\ \hline
$[0,0|2,2,2,2|1,1]$&5&$\Psi_{11}\Psi_{12}\ZZb\XXb\YYb$&$\{2,2,2,2\}$ & $\{-5,-5,1,1\}$ \\ \hline
$[1,1|3,3,3,3|0,0]$&5&$\bar{\Psi}_{11}\bar{\Psi}_{12}\ZZ\XX\YY$&$\{3,3,3,3\}$&$\{-6,-6,0,0\}$\\ \hline
\end{tabular}
\caption{Three types of operators with differing oscillator content and length that have the same $\psu(2,2|4)$ quantum numbers: $\lambda_a-\lambda_{a+1}=\{0,0,0\}$ and $\nu_j-\nu_{j+1}=\{0,-6,0\}$.}
\label{table:mixing}
\end{table}

\subsection{Multiplets and duality transformations} 
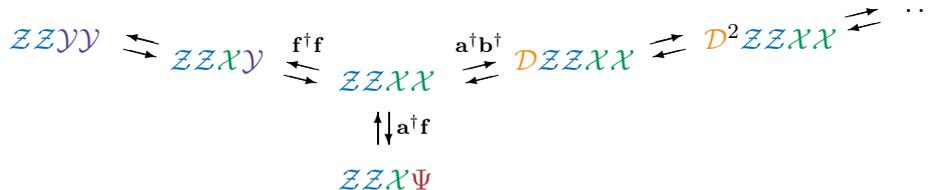
\begin{figure}[t]
\centering
\begin{picture}(250,80)

\put(83,42){$\xZ\xZ\xX\xX$}

\put(98,22){\vector(0,1){12}}
\put(102,34){\vector(0,-1){12}}
\put(83,5){$\xZ\xZ\xX\xPsi$}
\put(105,26){\scriptsize $\aaa^\dagger \fff$}

\put(130,50){\vector(4,1){12}}
\put(143,48){\vector(-4,-1){12}}
\put(127,57){\scriptsize $\aaa^\dagger \bbb^\dagger$}
\put(149,50){$\xD\xZ\xZ\xX\xX$}

\put(200,60){\vector(4,1){12}}
\put(213,58){\vector(-4,-1){12}}
\put(220,58){$\xD^2\xZ\xZ\xX\xX$}

\put(273,70){\vector(4,1){12}}
\put(286,68){\vector(-4,-1){12}}
\put(295,72){$\hdots$}

\put(76,50){\vector(-4,1){12}}
\put(63,48){\vector(4,-1){12}}
\put(66,57){\scriptsize $\fff^\dagger \fff$}
\put(20,50){$\xZ\xZ\xX\xY$}

\put(16,60){\vector(-4,1){12}}
\put(3,58){\vector(4,-1){12}}
\put(-40,58){$\xZ\xZ\xY\xY$}

\end{picture}
\caption{Schematical depiction of part of a multiplet. The action of the symmetry generators results in a different operator. The $R$-symmetry (generated by $\fff^\dagger\fff$) and supersymmetry ($\aaa^\dagger \fff$ and $\bbb\fff$) form compact directions, while the non-compactness comes from the derivatives ($\aaa^\dagger\bbb^\dagger$). Of the shown schematical states, $\ZZ\ZZ\XX\XX$ and $\ZZ\ZZ\XX\Psi$ are highest-weight states in different gradings. $\ZZ\ZZ\YY\YY$ can also be perceived as a highest-weight state, in a label ordering with $4<3$. The rest of the shown states are descendants in any grading.}
\label{fig:mult}
\end{figure}
Operators that are related by the global symmetry, $\psu(2,2|4)$, form superconformal multiplets -- infinite-dimensional unitary irreducible representations of the symmetry algebra, see figure~\ref{fig:mult}. Multiplets contain a finite number of conformal primary operators, i.e.\ those that are annihilated by special conformal transformations (generated by $\aaa_\alpha \bbb_{\dot{\alpha}}$). One of these conformal primaries is also the HWS of the representation. But which one, and consequently the oscillator content \eqref{oc} defining the multiplet, depends on the grading choice.

We could stick, in principle, with one particular choice of the grading. A natural choice is the compact beauty grading whose HWS has the lowest classical dimension \eqref{delta0} among all possible conformal primaries. This is, implicitly, the choice made by Dolan and Osborn \cite{Dolan:2002zh}, though they did not use the same terminology. We indeed routinely choose this grading unless there is a special reason to do otherwise.

However, an understanding of the interplay between different gradings is crucial to understand the properties of the one-loop Q-system, existence of subsectors,  and such effects as multiplet joining. Note also that the asymptotic Bethe Ansatz equations \cite{Beisert:2005fw} paramount for the development of AdS/CFT integrability were formulated in gradings different from the compact beauty grading\footnote{As we understand now \cite{Gromov:2014caa}, asymptotic Bethe Ansatz equations can be written in any grading, though their explicit form will not be given in terms of rational functions of Zhukovsky variables, even if the dressing phase is ignored.}.

The elementary move to modify grading is to  exchange position of two neighbours in the ordered sequence of labels. Moves that permute labels with the same $p$- and $c$-grading only change the HWS up to a relabelling of the fields,  hence they are not interesting and will not be considered. Moves that permute labels with the same $p$- but different $c$-grading obscure the terminology of highest-weight states and their descendants, and they will not be considered either. That is why we focus on orderings that enjoy the properties $1<2<3<4$ and $\hat 1<\hat 2<\hat 3<\hat 4$.

\subsubsection*{Fermionic duality transformations}
Permuting two labels of different $p$-grading changes the field content of the HWS non-trivially. It is known as fermionic duality transformation, and was considered for supersymmetric spin chains in \cite{Woynarovich,tJmodel,Tsuboi:1998ne}. 
Combining such permutations allows one to choose an arbitrary path on the $4\times4$ lattice of figure~\ref{fig:beauty}. There are 70 different choices, and each one, generically, corresponds to a different conformal primary operator being the HWS.

The duality transformation corresponds to the change
\be
\begin{pmatrix} E_{mm} & E_{mn} \\ E_{nm} & E_{nn} \end{pmatrix}\leftrightarrow
\begin{pmatrix} E_{nn} & E_{nm} \\ E_{mn} & E_{mm} \end{pmatrix}\,.
\ee
As $p_m+p_n=1$, $E_{mn}$ and $E_{nm}$ are fermionic. Consider a HWS with respect to the first grading, $|\Omega\rangle$. It satisfies $E_{mn}|\Omega\rangle=0$. As $E_{nm}^2=0$, we can maximally act once with $E_{nm}$ on $|\Omega\rangle$. The case where $E_{nm}|\Omega\rangle=0$ occurs only when $e_m+e_n=0$ (where $e_n$ is the eigenvalue of $E_{nn}$, so one $e$ is a $\lambda$ and one is a $\nu$). Indeed,
\be\label{EE}
E_{mn}E_{nm}|\Omega\rangle = \{E_{mn},E_{nm}\}|\Omega\rangle = (E_{mm}+E_{nn})|\Omega\rangle=(e_m+e_n)|\Omega\rangle\,.
\ee

\subsubsection*{Long representations}
\begin{figure}[h!]
\centering
\begin{picture}(200,68)

{\color{gray}
\thinlines
\put(25,10){\line(1,0){50}}
\put(25,60){\line(1,0){50}}
\put(25,10){\line(0,1){50}}
\put(75,10){\line(0,1){50}}

\put(125,10){\line(1,0){50}}
\put(125,60){\line(1,0){50}}
\put(125,10){\line(0,1){50}}
\put(175,10){\line(0,1){50}}
}

\linethickness{0.7mm}

\put(95,30){$\leftrightarrow$}

\put(25,60){\line(1,0){50}}
\put(25,10){\line(0,1){50}}

\put(125,10){\line(1,0){50}}
\put(175,10){\line(0,1){50}}

\put(16,32){$\lambda$}
\put(48,63){$\nu$}
\put(177,32){$\lambda-1$}
\put(138,0){$\nu+1$}

\end{picture}
\caption{Duality transformation when $\lambda+\nu\neq 0$. The box here is one of the lattice boxes of e.g.\ the lattice in figure~\ref{fig:beauty}.}\label{fig:dual}
\end{figure}
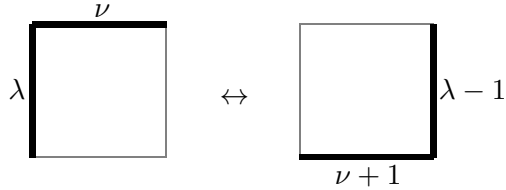

When $e_m+ e_n\neq 0$, which is generally the case for unprotected states at finite coupling, the HWS is $E_{nm}|\Omega\rangle$ with respect to the second grading. The weights of this new HWS can be changed in two ways. If $E_{mn}=\fff_a^\dagger \bbb^\dagger_j$ or $E_{mn}=\fff_a^\dagger \aaa_j$, the new HWS will have the weights $\lambda_a^{E_{nm}|\Omega\rangle}=\lambda_a^{|\Omega\rangle}+1$ and $\nu_j^{E_{nm}|\Omega\rangle}=\nu_j^{|\Omega\rangle}-1$, while the rest are unchanged.
On the other hand, if $E_{mn}=\bbb_j \fff_a$ or $E_{mn}=\aaa^\dagger_j \fff_a$, the new HWS will have the weights $\lambda_a^{E_{nm}|\Omega\rangle}=\lambda_a^{|\Omega\rangle}-1$ and $\nu_j^{E_{nm}|\Omega\rangle}=\nu_j^{|\Omega\rangle}+1$, while the rest are unchanged. The effect of the transformation is summarised in figure \ref{fig:dual}.

\subsubsection*{Short representations}
If $e_m + e_n=0$ then  $E_{mn}E_{nm}|\Omega\rangle=0$,  which implies the shortening condition $E_{nm}|\Omega\rangle=0$ since otherwise the representation would be reducible indecomposable which is impossible for a unitary representation. Therefore $|\Omega\rangle$ is the HWS for both choices of grading, and $\lambda,\nu$ remain unchanged under the duality transformation, so figure~\ref{fig:dual} does not apply.

The phenomenon of shortening occurs at $g=0$, but not at finite coupling, except for the protected chiral primary operators.  We continue its discussion in section~\ref{sec:short}.

\subsection{Non-compact Young diagrams} \label{sec:yd}
For compact algebras, Young diagrams provide an intuitive way of classifying irreducible representations. We generalise this construction to the non-compact case.

\subsubsection*{Definition of non-compact Young diagram\footnote{This builds upon the results of \cite{Gunaydin:2017lhg}. We here restrict to the case when the weights are integers, and some elements of our discussion rely on the vanishing of the central charge. In the general context of \cite{Gunaydin:2017lhg}, these constraints are not needed.}}
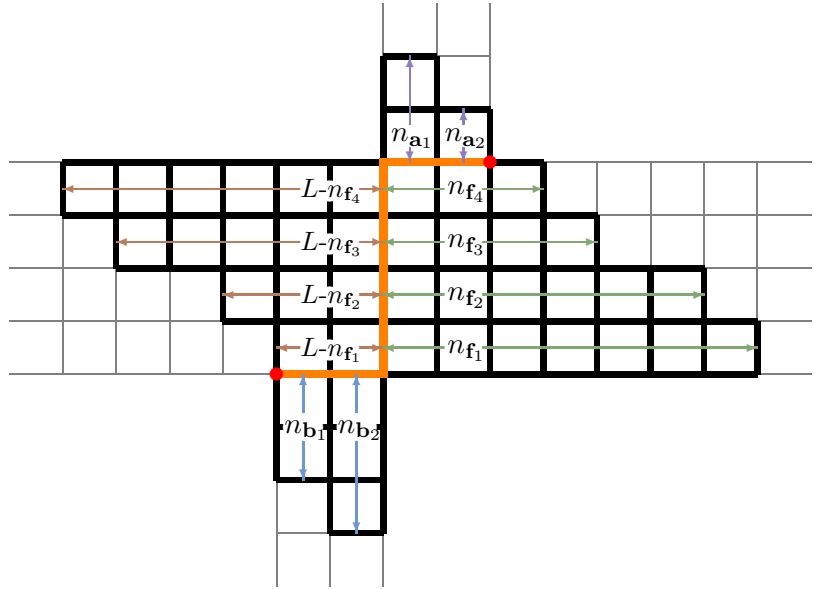
\begin{figure}[h!]
\centering

\begin{picture}(320,220)



\color{gray}
\linethickness{.1mm}

\multiput(10,80)(0,20){5}{\line(1,0){300}}
\multiput(150,180)(0,20){2}{\line(1,0){40}}
\multiput(110,20)(0,20){3}{\line(1,0){40}}

\multiput(30,80)(20,0){4}{\line(0,1){80}}
\multiput(210,80)(20,0){5}{\line(0,1){80}}

\put(110,0){\line(0,1){160}}
\put(130,0){\line(0,1){160}}
\put(150,0){\line(0,1){220}}
\put(170,80){\line(0,1){140}}
\put(190,80){\line(0,1){140}}

\color{black}
\linethickness{0.7mm}

\put(150,200){\line(1,0){20}}
\put(150,180){\line(1,0){40}}
\put(30,160){\line(1,0){180}}
\put(30,140){\line(1,0){200}}
\put(50,120){\line(1,0){220}}
\put(90,100){\line(1,0){200}}
\put(110,80){\line(1,0){180}}
\put(110,60){\line(1,0){40}}
\put(110,40){\line(1,0){40}}
\put(130,20){\line(1,0){20}}

\put(30,140){\line(0,1){20}}
\put(50,120){\line(0,1){40}}
\put(70,120){\line(0,1){40}}
\put(90,100){\line(0,1){60}}
\put(110,40){\line(0,1){120}}
\put(130,20){\line(0,1){140}}
\put(150,20){\line(0,1){180}}
\put(170,80){\line(0,1){120}}
\put(190,80){\line(0,1){100}}
\put(210,80){\line(0,1){80}}
\put(230,80){\line(0,1){60}}
\put(250,80){\line(0,1){40}}
\put(270,80){\line(0,1){40}}
\put(290,80){\line(0,1){20}}


\color{orange}
\linethickness{1mm}
\put(110,80){\line(1,0){41.4}}
\put(150,80){\line(0,1){80}}
\put(148.6,160){\line(1,0){40}}

\color{red}
\put(110,80){\circle*{5}}
\put(190,160){\circle*{5}}

\color{Brown!50}
\linethickness{0.2mm}
\put(30,150){\vector(1,0){120}}\put(150,150){\vector(-1,0){120}}
\put(50,130){\vector(1,0){100}}\put(150,130){\vector(-1,0){100}}
\put(90,110){\vector(1,0){60}}\put(150,110){\vector(-1,0){60}}
\put(110,90){\vector(1,0){40}}\put(150,90){\vector(-1,0){40}}

\color{NavyBlue!60}
\put(120,40){\vector(0,1){40}}\put(120,80){\vector(0,-1){40}}
\put(140,20){\vector(0,1){60}}\put(140,80){\vector(0,-1){60}}

\color{OliveGreen!60}
\put(150,150){\vector(1,0){60}}\put(210,150){\vector(-1,0){60}}
\put(150,130){\vector(1,0){80}}\put(230,130){\vector(-1,0){80}}
\put(150,110){\vector(1,0){120}}\put(270,110){\vector(-1,0){120}}
\put(150,90){\vector(1,0){140}}\put(290,90){\vector(-1,0){140}}

\color{Violet!60}
\put(160,160){\vector(0,1){40}}\put(160,200){\vector(0,-1){40}}
\put(180,160){\vector(0,1){20}}\put(180,180){\vector(0,-1){20}}

\color{black}
\put(180,150){\color{white}\circle*{14}}\put(174,148){$n_{\fff_4}$}
\put(180,130){\color{white}\circle*{14}}\put(174,128){$n_{\fff_3}$}
\put(180,110){\color{white}\circle*{14}}\put(174,108){$n_{\fff_2}$}
\put(180,90){\color{white}\circle*{14}}\put(174,88){$n_{\fff_1}$}

\put(123,150){\color{white}\circle*{8}}\put(130,149){\color{white}\circle*{8}}\put(137,150){\color{white}\circle*{8}}\put(119,147){\small$L {\text-} n_{\fff_4}$}
\put(123,130){\color{white}\circle*{8}}\put(130,129){\color{white}\circle*{8}}\put(137,130){\color{white}\circle*{8}}\put(119,127){\small$L {\text-} n_{\fff_3}$}
\put(123,110){\color{white}\circle*{8}}\put(130,109){\color{white}\circle*{8}}\put(137,110){\color{white}\circle*{8}}\put(119,107){\small$L {\text-} n_{\fff_2}$}
\put(123,90){\color{white}\circle*{8}}\put(130,89){\color{white}\circle*{8}}\put(137,90){\color{white}\circle*{8}}\put(119,87){\small$L {\text-} n_{\fff_1}$}


\put(118,60){\color{white}\circle*{11}}\put(122,60){\color{white}\circle*{11}}\put(113,58){$n_{\bbb_1}$}
\put(138,60){\color{white}\circle*{11}}\put(142,60){\color{white}\circle*{11}}\put(133,58){$n_{\bbb_2}$}

\put(158,170){\color{white}\circle*{9}}\put(162,170){\color{white}\circle*{9}}\put(153,168){$n_{\aaa_1}$}
\put(178,170){\color{white}\circle*{9}}\put(182,170){\color{white}\circle*{9}}\put(173,168){$n_{\aaa_2}$}

\end{picture}
\caption{Young diagram corresponding to the quantum numbers $n_\aaa$, $n_\fff$, $n_\bbb$ with respect to the 2222 grading. Long multiplets correspond to Young diagrams that touch both edge points marked in {\color{red}red}.}
\label{fig:YDdef}
\end{figure}

If the representation is long, choose an arbitrary grading and draw the corresponding Dynkin path. This path consists of eight segments. Draw $n_{\bbb_1}$ boxes below the first horizontal segment, $n_{\bbb_2}$ boxes below the second horizontal segment, $n_{\aaa_1}$ boxes above the third horizontal segment, and $n_{\aaa_2}$ boxes above the fourth horizontal segment. For the $i$'th vertical segment draw $n_{\fff_i}$ boxes to the right and $L-n_{\fff_i}$ boxes to the left. If the representation is short, not all choices of grading produce a meaningful result, but the 2222 grading always guarantees the correct output. See figure \ref{fig:YDdef}.

The oscillator content and the corresponding Young diagram should be admissible ones in the sense that it should be possible to construct the HWS from the given number of oscillators. This is the case if both the left and the right boundary of the non-compact Young diagram is of a ladder shape, i.e.\ the shape of the boundaries of ordinary compact Young diagrams \cite{Gunaydin:2017lhg}. A good example of how things may go wrong is $n^{2222}=[1,1|1,1,1,1|1,1]$. The corresponding HWS is formally $\epsilon^{abcd} \Phi_{ab} \Box \Phi_{cd}$, but it is zero due to the equations of motion. The equations of motion are realised in the oscillator language as $\Box=\epsilon_{\alpha\beta}\epsilon_{\dot\alpha\dot\beta} \mathcal{D}_{\alpha\dot{\alpha}}\mathcal{D}_{\beta\dot{\beta}}=\det\limits_{1\leq \alpha,\dot\alpha\leq 2}\aaa_\alpha^\dagger \bbb_{\dot{\alpha}}^\dagger=0$.

Although we can choose different gradings to explain the drawing procedure, the resulting  diagram does not depend on this choice. The diagram is actually an invariant of the $\pu(2,2|4)\oplus \algu(1)$ multiplet and defines it unambiguously. 
For what concerns the $\psu(2,2|4)$ algebra, admissible diagrams related by the transformations \eqref{bothL}  correspond to isomorphic $\psu(2,2|4)$ representations.

\begin{figure}[h!]
\centering
\begin{picture}(320,220)

\color{gray}
\linethickness{.1mm}

\multiput(10,80)(0,20){5}{\line(1,0){300}}
\multiput(150,180)(0,20){2}{\line(1,0){40}}
\multiput(110,20)(0,20){3}{\line(1,0){40}}

\multiput(30,80)(20,0){4}{\line(0,1){80}}
\multiput(210,80)(20,0){5}{\line(0,1){80}}

\put(110,0){\line(0,1){160}}
\put(130,0){\line(0,1){160}}
\put(150,0){\line(0,1){220}}
\put(170,80){\line(0,1){140}}
\put(190,80){\line(0,1){140}}

\color{black}
\linethickness{0.7mm}

\put(150,200){\line(1,0){20}}
\put(150,180){\line(1,0){40}}
\put(30,160){\line(1,0){180}}
\put(30,140){\line(1,0){200}}
\put(50,120){\line(1,0){220}}
\put(90,100){\line(1,0){200}}
\put(110,80){\line(1,0){180}}
\put(110,60){\line(1,0){40}}
\put(110,40){\line(1,0){40}}
\put(130,20){\line(1,0){20}}

\put(30,140){\line(0,1){20}}
\put(50,120){\line(0,1){40}}
\put(70,120){\line(0,1){40}}
\put(90,100){\line(0,1){60}}
\put(110,40){\line(0,1){120}}
\put(130,20){\line(0,1){140}}
\put(150,20){\line(0,1){180}}
\put(170,80){\line(0,1){120}}
\put(190,80){\line(0,1){100}}
\put(210,80){\line(0,1){80}}
\put(230,80){\line(0,1){60}}
\put(250,80){\line(0,1){40}}
\put(270,80){\line(0,1){40}}
\put(290,80){\line(0,1){20}}

\color{red}
\put(110,80){\circle*{6}}
\put(190,160){\circle*{6}}

\footnotesize


\put(170,190){\color{white}\circle*{8}}\put(168,187){\color{blue}0}
\put(150,190){\color{white}\circle*{8}}\put(148,187){\color{blue}1}

\put(190,170){\color{white}\circle*{8}}\put(188,167){\color{blue}0}
\put(170,170){\color{white}\circle*{8}}\put(168,167){\color{blue}1}
\put(150,170){\color{white}\circle*{8}}\put(148,167){\color{blue}2}

\put(210,150){\color{white}\circle*{8}}\put(208,147){\color{blue}0}
\put(190,150){\color{white}\circle*{8}}\put(188,147){\color{blue}1}
\put(170,150){\color{white}\circle*{8}}\put(168,147){\color{blue}2}
\put(150,150){\color{white}\circle*{8}}\put(148,147){\color{blue}3}
\put(130,150){\color{white}\circle*{8}}\put(128,147){\color{blue}4}
\put(110,150){\color{white}\circle*{8}}\put(108,147){\color{blue}5}
\put(90,150){\color{white}\circle*{8}}\put(88,147){\color{blue}6}
\put(70,150){\color{white}\circle*{8}}\put(68,147){\color{blue}7}
\put(50,150){\color{white}\circle*{8}}\put(48,147){\color{blue}8}
\put(30,150){\color{white}\circle*{8}}\put(28,147){\color{blue}9}

\put(230,130){\color{white}\circle*{8}}\put(228,127){\color{blue}0}
\put(210,130){\color{white}\circle*{8}}\put(208,127){\color{blue}1}
\put(190,130){\color{white}\circle*{8}}\put(188,127){\color{blue}2}
\put(170,130){\color{white}\circle*{8}}\put(168,127){\color{blue}3}
\put(150,130){\color{white}\circle*{8}}\put(148,127){\color{blue}4}
\put(130,130){\color{white}\circle*{8}}\put(128,127){\color{blue}5}
\put(110,130){\color{white}\circle*{8}}\put(108,127){\color{blue}6}
\put(90,130){\color{white}\circle*{8}}\put(88,127){\color{blue}7}
\put(70,130){\color{white}\circle*{8}}\put(68,127){\color{blue}8}
\put(50,130){\color{white}\circle*{8}}\put(48,127){\color{blue}9}

\put(270,110){\color{white}\circle*{8}}\put(268,107){\color{blue}0}
\put(250,110){\color{white}\circle*{8}}\put(248,107){\color{blue}1}
\put(230,110){\color{white}\circle*{8}}\put(228,107){\color{blue}2}
\put(210,110){\color{white}\circle*{8}}\put(208,107){\color{blue}3}
\put(190,110){\color{white}\circle*{8}}\put(188,107){\color{blue}4}
\put(170,110){\color{white}\circle*{8}}\put(168,107){\color{blue}5}
\put(150,110){\color{white}\circle*{8}}\put(148,107){\color{blue}6}
\put(130,110){\color{white}\circle*{8}}\put(128,107){\color{blue}7}
\put(110,110){\color{white}\circle*{8}}\put(108,107){\color{blue}8}
\put(90,110){\color{white}\circle*{8}}\put(88,107){\color{blue}9}

\put(290,90){\color{white}\circle*{8}}\put(288,87){\color{blue}0}
\put(270,90){\color{white}\circle*{8}}\put(268,87){\color{blue}1}
\put(250,90){\color{white}\circle*{8}}\put(248,87){\color{blue}2}
\put(230,90){\color{white}\circle*{8}}\put(228,87){\color{blue}3}
\put(210,90){\color{white}\circle*{8}}\put(208,87){\color{blue}4}
\put(190,90){\color{white}\circle*{8}}\put(188,87){\color{blue}5}
\put(170,90){\color{white}\circle*{8}}\put(168,87){\color{blue}6}
\put(150,90){\color{white}\circle*{8}}\put(148,87){\color{blue}7}
\put(130,90){\color{white}\circle*{8}}\put(128,87){\color{blue}8}
\put(110,90){\color{white}\circle*{8}}\put(108,87){\color{blue}9}

\put(150,70){\color{white}\circle*{8}}\put(148,67){\color{blue}7}
\put(130,70){\color{white}\circle*{8}}\put(128,67){\color{blue}8}
\put(110,70){\color{white}\circle*{8}}\put(108,67){\color{blue}9}

\put(150,50){\color{white}\circle*{8}}\put(148,47){\color{blue}7}
\put(130,50){\color{white}\circle*{8}}\put(128,47){\color{blue}8}
\put(110,50){\color{white}\circle*{8}}\put(108,47){\color{blue}9}

\put(150,30){\color{white}\circle*{8}}\put(148,27){\color{blue}8}
\put(130,30){\color{white}\circle*{8}}\put(128,27){\color{blue}9}


\put(40,160){\color{white}\circle*{13}}\put(34,157){\color{purple}-10}
\put(40,140){\color{white}\circle*{9}}\put(36,137){\color{purple}-9}

\put(60,160){\color{white}\circle*{13}}\put(54,157){\color{purple}-11}
\put(60,140){\color{white}\circle*{13}}\put(54,137){\color{purple}-10}
\put(60,120){\color{white}\circle*{9}}\put(56,117){\color{purple}-9}

\put(80,160){\color{white}\circle*{13}}\put(74,157){\color{purple}-11}
\put(80,140){\color{white}\circle*{13}}\put(74,137){\color{purple}-10}
\put(80,120){\color{white}\circle*{9}}\put(76,117){\color{purple}-9}

\put(100,160){\color{white}\circle*{13}}\put(94,157){\color{purple}-12}
\put(100,140){\color{white}\circle*{13}}\put(94,137){\color{purple}-11}
\put(100,120){\color{white}\circle*{13}}\put(94,117){\color{purple}-10}
\put(100,100){\color{white}\circle*{9}}\put(96,97){\color{purple}-9}

\put(120,160){\color{white}\circle*{13}}\put(114,157){\color{purple}-15}
\put(120,140){\color{white}\circle*{13}}\put(114,137){\color{purple}-14}
\put(120,120){\color{white}\circle*{13}}\put(114,117){\color{purple}-13}
\put(120,100){\color{white}\circle*{13}}\put(114,97){\color{purple}-12}
\put(120,80){\color{white}\circle*{13}}\put(114,77){\color{purple}-11}
\put(120,60){\color{white}\circle*{13}}\put(114,57){\color{purple}-10}
\put(120,40){\color{white}\circle*{9}}\put(116,37){\color{purple}-9}

\put(140,160){\color{white}\circle*{13}}\put(134,157){\color{purple}-16}
\put(140,140){\color{white}\circle*{13}}\put(134,137){\color{purple}-15}
\put(140,120){\color{white}\circle*{13}}\put(134,117){\color{purple}-14}
\put(140,100){\color{white}\circle*{13}}\put(134,97){\color{purple}-13}
\put(140,80){\color{white}\circle*{13}}\put(134,77){\color{purple}-12}
\put(140,60){\color{white}\circle*{13}}\put(134,57){\color{purple}-11}
\put(140,40){\color{white}\circle*{13}}\put(134,37){\color{purple}-10}
\put(140,20){\color{white}\circle*{9}}\put(136,17){\color{purple}-9}

\put(160,200){\color{white}\circle*{8}}\put(157.5,197){\color{purple}0}
\put(160,180){\color{white}\circle*{8}}\put(157.5,177){\color{purple}1}
\put(160,160){\color{white}\circle*{8}}\put(157.5,157){\color{purple}2}
\put(160,140){\color{white}\circle*{8}}\put(157.5,137){\color{purple}3}
\put(160,120){\color{white}\circle*{8}}\put(157.5,117){\color{purple}4}
\put(160,100){\color{white}\circle*{8}}\put(157.5,97){\color{purple}5}
\put(160,80){\color{white}\circle*{8}}\put(157.5,77){\color{purple}6}

\put(180,180){\color{white}\circle*{8}}\put(177.5,177){\color{purple}0}
\put(180,160){\color{white}\circle*{8}}\put(177.5,157){\color{purple}1}
\put(180,140){\color{white}\circle*{8}}\put(177.5,137){\color{purple}2}
\put(180,120){\color{white}\circle*{8}}\put(177.5,117){\color{purple}3}
\put(180,100){\color{white}\circle*{8}}\put(177.5,97){\color{purple}4}
\put(180,80){\color{white}\circle*{8}}\put(177.5,77){\color{purple}5}

\put(200,160){\color{white}\circle*{8}}\put(197.5,157){\color{purple}0}
\put(200,140){\color{white}\circle*{8}}\put(197.5,137){\color{purple}1}
\put(200,120){\color{white}\circle*{8}}\put(197.5,117){\color{purple}2}
\put(200,100){\color{white}\circle*{8}}\put(197.5,97){\color{purple}3}
\put(200,80){\color{white}\circle*{8}}\put(197.5,77){\color{purple}4}

\put(220,140){\color{white}\circle*{8}}\put(217.5,137){\color{purple}0}
\put(220,120){\color{white}\circle*{8}}\put(217.5,117){\color{purple}1}
\put(220,100){\color{white}\circle*{8}}\put(217.5,97){\color{purple}2}
\put(220,80){\color{white}\circle*{8}}\put(217.5,77){\color{purple}3}

\put(240,120){\color{white}\circle*{8}}\put(237.5,117){\color{purple}0}
\put(240,100){\color{white}\circle*{8}}\put(237.5,97){\color{purple}1}
\put(240,80){\color{white}\circle*{8}}\put(237.5,77){\color{purple}2}

\put(260,120){\color{white}\circle*{8}}\put(257.5,117){\color{purple}0}
\put(260,100){\color{white}\circle*{8}}\put(257.5,97){\color{purple}1}
\put(260,80){\color{white}\circle*{8}}\put(257.5,77){\color{purple}2}

\put(280,100){\color{white}\circle*{8}}\put(277.5,97){\color{purple}0}
\put(280,80){\color{white}\circle*{8}}\put(277.5,77){\color{purple}1}

\end{picture}
\caption{Assignment of weights, {\color{blue}$\lambda$} and {\color{magenta}$\nu$}, for the case $n^{2222}=[2,3|7,6,4,3|2,1]$. Any path between the {\color{red}red} circles corresponds to a grading, and the corresponding weights can be read off. Note the way the weights are defined beyond the $4\times4$ square. In the right half of the diagram, the top horizontal lines in each column and the rightmost vertical lines in each row are assigned the weight 0. In the left part of the diagram the bottom horizontal lines in each column are assigned the weight $\nu=-L$, while the leftmost vertical line in each row is assigned the weight $\lambda=+L$.}
\label{fig:YDweights}
\end{figure}
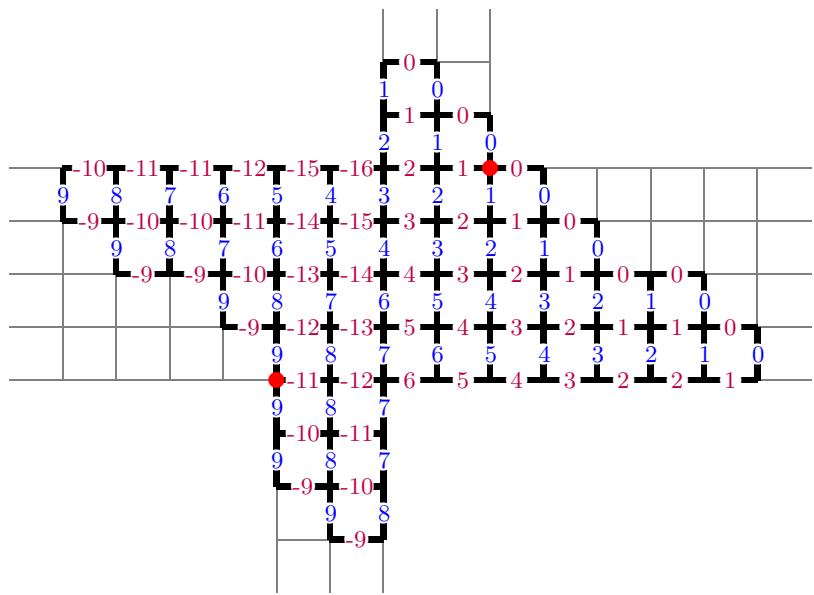

Conversely, the Young diagram makes it easy to read off the weights in a certain grading by simply counting boxes, see figure~\ref{fig:YDweights} for an example.

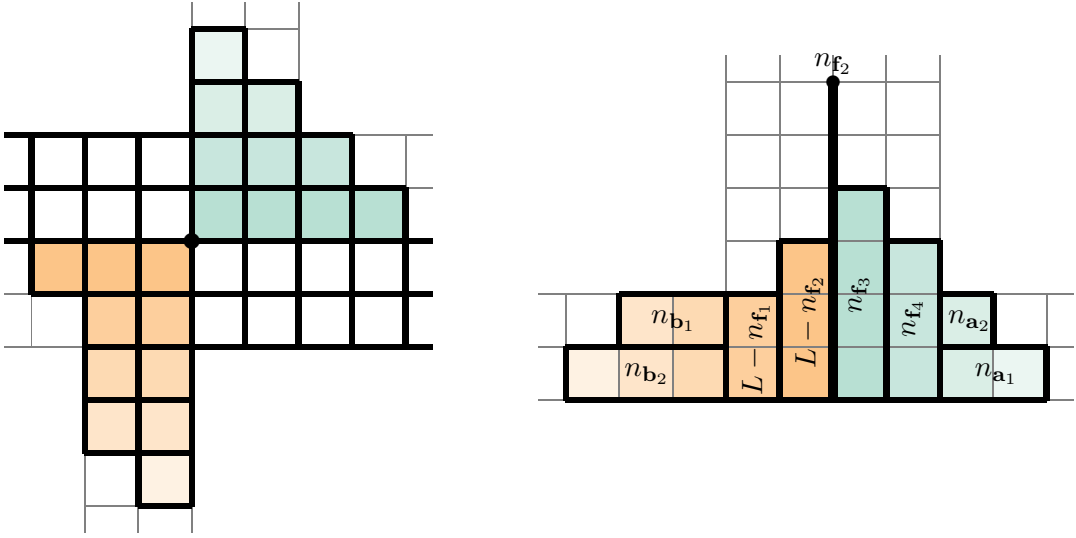
\begin{figure}[h!]
\centering
\begin{picture}(400,200)

\linethickness{0.4mm}
\multiput(10,90)(1,0){60}{\color{BurntOrange!50}\line(0,1){20}}
\multiput(30,70)(1,0){40}{\color{BurntOrange!40}\line(0,1){20}}
\multiput(30,50)(1,0){40}{\color{BurntOrange!30}\line(0,1){20}}
\multiput(30,30)(1,0){40}{\color{BurntOrange!20}\line(0,1){20}}
\multiput(50,10)(1,0){20}{\color{BurntOrange!10}\line(0,1){20}}

\multiput(290,50)(0,1){60}{\color{BurntOrange!50}\line(1,0){20}}
\multiput(270,50)(0,1){40}{\color{BurntOrange!40}\line(1,0){20}}
\multiput(250,50)(0,1){40}{\color{BurntOrange!30}\line(1,0){20}}
\multiput(230,50)(0,1){40}{\color{BurntOrange!20}\line(1,0){20}}
\multiput(210,50)(0,1){20}{\color{BurntOrange!10}\line(1,0){20}}

\multiput(70,110)(1,0){80}{\color{SeaGreen!40}\line(0,1){20}}
\multiput(70,130)(1,0){60}{\color{SeaGreen!30}\line(0,1){20}}
\multiput(70,150)(1,0){40}{\color{SeaGreen!20}\line(0,1){20}}
\multiput(70,170)(1,0){20}{\color{SeaGreen!10}\line(0,1){20}}

\multiput(310,50)(0,1){80}{\color{SeaGreen!40}\line(1,0){20}}
\multiput(330,50)(0,1){60}{\color{SeaGreen!30}\line(1,0){20}}
\multiput(350,50)(0,1){40}{\color{SeaGreen!20}\line(1,0){20}}
\multiput(370,50)(0,1){20}{\color{SeaGreen!10}\line(1,0){20}}

\color{gray}
\linethickness{.1mm}

\multiput(0,70)(0,20){5}{\line(1,0){160}}
\multiput(70,170)(0,20){2}{\line(1,0){40}}
\multiput(30,10)(0,20){3}{\line(1,0){40}}

\multiput(10,70)(20,0){1}{\line(0,1){80}}
\multiput(130,70)(20,0){2}{\line(0,1){80}}

\put(30,0){\line(0,1){150}}
\put(50,0){\line(0,1){150}}
\put(70,0){\line(0,1){200}}
\put(90,70){\line(0,1){130}}
\put(110,70){\line(0,1){130}}

\color{black}
\linethickness{0.7mm}

\put(70,190){\line(1,0){20}}
\put(70,170){\line(1,0){40}}
\put(0,150){\line(1,0){130}}
\put(00,130){\line(1,0){150}}
\put(0,110){\line(1,0){160}}
\put(10,90){\line(1,0){150}}
\put(30,70){\line(1,0){130}}
\put(30,50){\line(1,0){40}}
\put(30,30){\line(1,0){40}}
\put(50,10){\line(1,0){20}}

\put(10,90){\line(0,1){60}}
\put(30,30){\line(0,1){120}}
\put(50,10){\line(0,1){140}}
\put(70,10){\line(0,1){180}}
\put(90,70){\line(0,1){120}}
\put(110,70){\line(0,1){100}}
\put(130,70){\line(0,1){80}}
\put(150,70){\line(0,1){60}}

\put(70,110){\circle*{6}}



\color{gray}

\linethickness{0.1mm}
\put(200,50){\line(1,0){200}}
\put(200,70){\line(1,0){200}}
\put(200,90){\line(1,0){200}}
\put(270,110){\line(1,0){80}}
\put(270,130){\line(1,0){80}}
\put(270,150){\line(1,0){80}}
\put(270,170){\line(1,0){80}}

\put(210,50){\line(0,1){40}}
\put(230,50){\line(0,1){40}}
\put(250,50){\line(0,1){40}}
\put(270,50){\line(0,1){130}}
\put(290,50){\line(0,1){130}}
\put(310,50){\line(0,1){130}}
\put(330,50){\line(0,1){130}}
\put(350,50){\line(0,1){130}}
\put(370,50){\line(0,1){40}}
\put(390,50){\line(0,1){40}}

\color{black}
\linethickness{0.7mm}

\put(210,50){\line(1,0){180}}
\put(210,70){\line(1,0){60}}
\put(230,90){\line(1,0){60}}
\put(290,110){\line(1,0){20}}
\put(310,130){\line(1,0){20}}
\put(330,110){\line(1,0){20}}
\put(350,90){\line(1,0){20}}
\put(350,70){\line(1,0){40}}

\put(210,50){\line(0,1){20}}
\put(230,70){\line(0,1){20}}
\put(270,50){\line(0,1){40}}
\put(290,50){\line(0,1){60}}
\put(310,50){\line(0,1){80}}
\put(330,50){\line(0,1){80}}
\put(350,50){\line(0,1){60}}
\put(370,70){\line(0,1){20}}
\put(390,50){\line(0,1){20}}

\put(232,59){$n_{\bbb_2}$}
\put(242,79){$n_{\bbb_1}$}

\put(276,52){\begin{turn}{90} $L-n_{\fff_1}$ \end{turn}}
\put(296,62){\begin{turn}{90} $L-n_{\fff_2}$\end{turn}}
\put(316,83){\begin{turn}{90}$n_{\fff_3}$\end{turn}}
\put(336,74){\begin{turn}{90}$n_{\fff_4}$\end{turn}}

\put(363,59){$n_{\aaa_1}$}
\put(353,79){$n_{\aaa_2}$}

\linethickness{1.2mm}
\put(310,50){\line(0,1){120}}
\put(303,176){$n_{\fff_2}$}
\put(310,170){\circle*{5}}


\end{picture}
\caption{(left) Central charge constraint: the number of boxes in the upper-right and lower-left quadrants must be the same. (right) T-hook diagram corresponding to the quantum numbers $n_\aaa$, $n_\fff$, $n_\bbb$ in the $2222$ grading. It is the upper-right and lower-left quadrants of the Young diagram plus a line of height $n_{\fff_2}$.}
\label{fig:cccon}
\end{figure}

The central charge constraint \eqref{ccc}
has a natural interpretation in terms of Young diagrams: It  states that the number of boxes in the upper-right and lower-left quadrants must be the same, see figure \ref{fig:cccon}.

\subsubsection*{Relation to T-hook}
If the value of $L$ is known, the lower-left quadrant and the upper-right quadrant of the Young diagram are enough to restore all the information about the multiplet.
A T-hook diagram, as shown in figure \ref{fig:cccon}, is the combination of these two quadrants plus an additional line specifying $n_{\fff_2}$ (it allows to read off the missing charge, $L$ in our context). The T-hook diagrams were proposed in \cite{Gromov:2010vb},  studied further in \cite{Volin:2010xz}, and generalised to the form presented here in \cite{Tsuboi:2011iz,Gunaydin:2017lhg}.

\subsubsection*{Infinite extension of Young diagram}
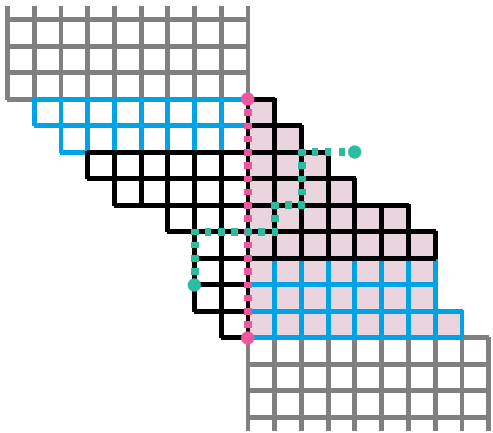
\begin{figure}[h!]
\centering
\begin{picture}(190,160)

\color{VioletRed!50!gray!30}
\linethickness{0.4mm}
\multiput(95,35)(0,1){90}{\line(1,0){10}}
\multiput(105,35)(0,1){80}{\line(1,0){10}}
\multiput(115,35)(0,1){70}{\line(1,0){10}}
\multiput(125,35)(0,1){60}{\line(1,0){10}}
\multiput(135,35)(0,1){50}{\line(1,0){10}}
\multiput(145,35)(0,1){50}{\line(1,0){10}}
\multiput(155,35)(0,1){40}{\line(1,0){10}}
\multiput(165,35)(0,1){10}{\line(1,0){10}}


\color{gray}
\linethickness{0.5mm}
\put(95,155){\line(-1,0){90}}
\put(95,145){\line(-1,0){90}}
\put(95,135){\line(-1,0){90}}

\put(95,25){\line(1,0){90}}
\put(95,15){\line(1,0){90}}
\put(95,5){\line(1,0){90}}

\put(5,125){\line(0,1){35}}
\put(15,115){\line(0,1){45}}
\put(25,105){\line(0,1){55}}
\put(35,105){\line(0,1){55}}
\put(45,105){\line(0,1){55}}
\put(55,105){\line(0,1){55}}
\put(65,105){\line(0,1){55}}
\put(75,105){\line(0,1){55}}
\put(85,105){\line(0,1){55}}
\put(95,0){\line(0,1){160}}
\put(105,0){\line(0,1){60}}
\put(115,0){\line(0,1){60}}
\put(125,0){\line(0,1){60}}
\put(135,0){\line(0,1){60}}
\put(145,0){\line(0,1){60}}
\put(155,0){\line(0,1){60}}
\put(165,0){\line(0,1){60}}
\put(175,0){\line(0,1){45}}
\put(185,0){\line(0,1){35}}

\put(95,0){\line(0,1){160}}

\put(95,125){\line(-1,0){90}}
\put(175,35){\line(1,0){10}}


\color{Cerulean}
\put(95,125){\line(-1,0){80}}
\put(95,115){\line(-1,0){80}}
\put(95,105){\line(-1,0){70}}
\put(95,55){\line(1,0){70}}
\put(95,45){\line(1,0){80}}
\put(95,35){\line(1,0){80}}

\put(15,115){\line(0,1){10}}
\put(25,105){\line(0,1){20}}
\put(35,105){\line(0,1){20}}
\put(45,105){\line(0,1){20}}
\put(55,105){\line(0,1){20}}
\put(65,105){\line(0,1){20}}
\put(75,105){\line(0,1){20}}
\put(85,105){\line(0,1){20}}
\put(105,35){\line(0,1){30}}
\put(115,35){\line(0,1){30}}
\put(125,35){\line(0,1){30}}
\put(135,35){\line(0,1){30}}
\put(145,35){\line(0,1){30}}
\put(155,35){\line(0,1){30}}
\put(165,35){\line(0,1){30}}
\put(175,35){\line(0,1){10}}


\color{black}
\put(95,125){\line(1,0){10}}
\put(95,115){\line(1,0){20}}
\put(35,105){\line(1,0){90}}
\put(35,95){\line(1,0){100}}
\put(45,85){\line(1,0){110}}
\put(65,75){\line(1,0){100}}
\put(75,65){\line(1,0){90}}
\put(75,55){\line(1,0){20}}
\put(75,45){\line(1,0){20}}
\put(85,35){\line(1,0){10}}

\put(35,95){\line(0,1){10}}
\put(45,85){\line(0,1){20}}
\put(55,85){\line(0,1){20}}
\put(65,75){\line(0,1){30}}
\put(75,45){\line(0,1){60}}
\put(85,35){\line(0,1){70}}
\put(95,35){\line(0,1){90}}
\put(105,65){\line(0,1){60}}
\put(115,65){\line(0,1){50}}
\put(125,65){\line(0,1){40}}
\put(135,65){\line(0,1){30}}
\put(145,65){\line(0,1){20}}
\put(155,65){\line(0,1){20}}
\put(165,65){\line(0,1){10}}


\color{VioletRed}
\linethickness{0.9mm}
\put(95,35){\circle*{5}}
\put(95,125){\circle*{5}}
\multiput(95,39)(0,5){18}{\line(0,1){2}}

\color{SeaGreen}
\put(75,55){\circle*{5}}
\put(135,105){\circle*{5}}
\multiput(75,59)(0,5){4}{\line(0,1){2}}
\multiput(79,75)(5,0){6}{\line(1,0){2}}
\multiput(105,79)(0,5){2}{\line(0,1){2}}
\multiput(109,85)(5,0){2}{\line(1,0){2}}
\multiput(115,89)(0,5){4}{\line(0,1){2}}
\multiput(119,105)(5,0){4}{\line(1,0){2}}


\end{picture}
\caption{Extension of the Young diagram for $n^{2222}=[2,3|7,6,4,3|2,1]$. We call the part marked in {\color{Cerulean}blue} the {\it non-trivial extension}, while the {\color{gray}grey} part is called the {\it trivial extension}. The trivial part continues infinitely in the vertical direction without any horizontal shifts. The two {\color{VioletRed}red} points correspond to the algebra $\su(9)$, and the shaded boxes show the corresponding $\su(9)$ Young diagram. The two {\color{SeaGreen}green} points correspond to the algebra $su(2,4|5)$. As one of the points is outside the diagram, the corresponding multiplet is short.
}
\label{fig:YDext}
\end{figure}

The Young diagrams can be thought of as being part of an infinite column that has a bend corresponding to the diagram, see figure \ref{fig:YDext} for an example. All rows have length $L$ and the rows that do not intersect with the original diagram are completely aligned to the left or right of the central vertical line. Weights can be assigned in the same way as explained in figure \ref{fig:YDweights}.

A diagram is not related to a specific algebra. It can denote a representation of any $[\mathfrak{ps}]\algu(N,M|K)$ algebra whose rank is high enough to accommodate the representation. If one chooses two points on the $\mathbb{Z}^2$ lattice such that the extended Young diagram is between them and that the right point is not lower than the left point, then the relative position of these points to one another and to the central vertical line specify the algebra and the Young diagram then defines a representation of this algebra. If both points are on the boundary of the extended diagram then this representation is a long multiplet, otherwise it is a short multiplet.

\subsection{Shortening and joining} \label{sec:short}
All unprotected multiplets must be long at finite coupling because the shortening condition $\lambda+\nu=0$ cannot be satisfied for generic value of the anomalous dimension, cf.\ \eqref{ano}. However, shortenings may re-emerge at zero coupling. If that happens, one long multiplet splits into several short ones. Taking the opposite view, we  can hence ask which groups of short multiplets join to form a long one; or we can split this question into the following two: First, given a short multiplet at zero coupling, which long multiplet will it be a member of at finite coupling. Second, which other short multiplets will be members of the same long multiplet.

To answer these questions, we take a closer look on the phenomena related to shortenings.

\subsubsection*{Shortening and unitarity bounds}
Shortenings happen when certain unitarity bounds are reached for a representation. Indeed, since the generators $E_{mn}$ and $E_{nm}$ are conjugate to one another up to a sign, we can use \eqref{EE} to conclude that for fermionic $E_{mn}$
\be
\langle \Omega'|\Omega' \rangle \propto (e_n+e_m) \langle \Omega|\Omega \rangle\,,
\ee
where $|\Omega'\rangle \propto E_{nm}|\Omega\rangle$. Hence the relative sign of the norms of $|\Omega'\rangle$ and $|\Omega\rangle$ changes when the value of $e_n+e_m$ crosses zero, while all states should have positive norm in a unitary representation. When $e_n+e_m=0$, the states $|\Omega'\rangle$ and $|\Omega\rangle$ decouple and become parts of two different short multiplets.

\begin{subequations}\label{bothU1}
For what concerns $\psu(2,2|4)$, there are only two non-trivial unitarity restrictions\footnote{The "trivial" ones are that $\lambda_a-\lambda_{a+1}$, $\nu_3-\nu_4$, and $\nu_1-\nu_2$ are non-negative integers (in any grading). They originate from studying the action of the compact subalgebra $\su(2)\oplus \su(4)\oplus \su(2)$ on the HWS, whereas \eqref{bothU1} ensure that unitarity is present when non-compactness  and supersymmetry are properly accounted. See e.g \cite{Gromov:2014caa} that translates the unitarity bounds of \cite{Dobrev:1985qv} to the language we use here.}:
\be
\lambda_1 + \nu_1 &\le&  0 \label{U4a1}\,, \\
\lambda_4 + \nu_4 &\ge&  0 \label{U5a1}\,.
\ee
\end{subequations}
One requires \eqref{U4a1}  for a grading of type $\{1\hat 1\}...$, and \eqref{U5a1} for a grading of type $...\{4\hat 4\}$, where the notation $\{ab\}$ means that $a$ and $b$ can be in either order. If the inequalities \eqref{bothU1} are true for the mentioned gradings, they will be true in any grading.

\subsubsection*{Shortening and restriction of oscillator numbers}
Saturation of either of the unitarity bounds \eqref{bothU1} imposes one constraint on the values of fundamental weights. However, the oscillator content is restricted more severely. In the oscillator language, the shortening conditions read:
\begin{subequations}
\label{bothU}
\be
\lambda_1 + \nu_1 \quad =& n_{\fff_1}-L-n_{\bbb_1}&=\quad 0  \label{U4a}\,, \\
\lambda_4 + \nu_4 \quad =& n_{\fff_4}+n_{\aaa_2} &=\quad 0 \label{U5a}\,.
\ee
\end{subequations}
Consider first \eqref{U4a}. Given that the number of fermions can not be larger than the length of the operator, \eqref{U4a} can be realised only if $n_{\fff_1}=L$ and $n_{\bbb_1}=0$. Therefore the shortening \eqref{U4a} implies not one but two constraints on the oscillator content. What happens, is that the transformation \eqref{L1} is not allowed anymore.

Likewise, \eqref{U5a} implies two constraints $n_{\fff_4}=n_{\aaa_2}=0$, and forbids the transformation \eqref{L2}. Note that if both \eqref{U4a} and \eqref{U5a} are satisfied, we are in a special situation where it is possible to read off the oscillator content, in particular $L$, directly from six Cartan charges.

Switching off the transformations \eqref{bothL} restricts the possibility of mixing, but, as we shall see, it opens the possibility of short multiplets joining at finite coupling.


\subsubsection*{Joining  and duality transformations}
At $g=0$, the HWS remains unchanged in a short multiplet under certain fermionic duality transformations. However, at finite coupling each HWS only remains a HWS in one particular grading. Thus, we need to supply the short multiplet with a grading to specify the long multiplet which it joins into at finite coupling. This clarifies the answer to our first question about the fate of a short multiplet, however it should still be understood how strongly the grading choice affects the result, and whether each grading choice will result in multiplet joining at all. This can be understood only after a careful analysis.

\begin{subequations}
\label{sm12}
Consider first the shortening \eqref{U4a}. It restricts the oscillator content to the values
\be\label{sm1}
[0,n_{\bbb_2}|L,\bullet,\bullet,\bullet|\bullet,\bullet]^{L}_{1\hat{1}...}\,.
\ee 
In superscript, we outlined the operator's length (it is not a new piece of information, but it is to avoid confusion in the discussion). The subscript outlines the grading choice to properly process the joining mechanism.

At zero coupling, the oscillator content would be unchanged after the duality transformation to the grading $\hat 1 1...$. However, if we switch on the coupling, the multiplet becomes long and the duality transformation would have nontrivial effect on the weights: $\{\lambda_1,\nu_1\}\rightarrow \{\lambda_1+1,\nu_1-1\}$. Going back to zero coupling we formally get the inadmissible oscillator content $[1,n_{\bbb_2}|L+1,\bullet,\bullet,\bullet|\bullet,0]^{L}_{\hat{1}1...}$, but this can be countered by the length-changing transformation \eqref{L1} to get
\be\label{sm2}
[0,n_{\bbb_2}-1|L+1,\bullet,\bullet,\bullet|\bullet,\bullet]^{L+1}_{\hat{1}1...}\,.
\ee
\end{subequations}
One has two different multiplets \eqref{sm1} and \eqref{sm2} that join into one long multiplet at finite coupling.

\begin{subequations}
Performing a similar analysis for the shortening \eqref{U5a} we conclude that the following two multiplets should join into one 
\be
&&\label{sm3} [\bullet,\bullet|\bullet,\bullet,\bullet,0|n_{\aaa_1},0]^L_{...\hat 4 4}\,,
\\
&&\label{sm4} [\bullet,\bullet|\bullet+1,\bullet+1,\bullet+1,0|n_{\aaa_1}-1,0]^{L+1}_{...4\hat 4}\,.
\ee
\end{subequations}
Finally, if both shortenings \eqref{bothU} are applied simultaneously, the following four multiplets join:
\be
&&[0,\bullet|L,\bullet,\bullet,0|\bullet,0]^L_{1...4}\,,\nonumber\\
&&[0,\bullet-1|L+1,\bullet,\bullet,0|\bullet,0]^{L+1}_{\hat 1...4}\,,\nonumber\\
&&[0,\bullet|L+1,\bullet+1,\bullet+1,0|\bullet-1,0]^{L+1}_{1...\hat 4}\,,\nonumber\\
&&[0,\bullet-1|L+2,\bullet+1,\bullet+1,0|\bullet-1,0]^{L+2}_{\hat 1...\hat 4}\,.\label{sms}
\ee
This accomplishes the answer to the second question that we posed. Returning to our first question, we now see that it only matters in the choice of grading whether the first label is $1$ or $\hat 1$, for the shortening \eqref{U4a}, and whether the last label is $4$ or $\hat 4$, for the shortening \eqref{U5a}.  Finally, there are cases when only special choices of grading makes joining possible. We postpone this discussion to the very end of this section.

An example of the presented analysis is provided in table \ref{table:konmul} which assembles detailed information about sample members of the Konishi multiplet.

\begin{table}[h!]
\def\arraystretch{1.15}
\centering
\begin{tabular}{|c|c|c|c|c|c|c|} \hline
\multicolumn{2}{|c|}{Grading} & $n$ & Possible field content & $L$ & $\Delta_0$ & Young diagram \\\hline \hline

\multirow{3}{*}{$1...4$} &2222 & $[0,0|1,1,1,1|0,0]$ & $\ZZ\ZZb$, $\XX\XXb$, $\YY\YYb$ & \multirow{3}{*}{2}& 2 &
\\\cline{2-4}\cline{6-6}
&1222 & $[0,1|2,1,1,1|0,0]$ & $\ZZ\bar{\Psi}_{2,2}$, $\XX\bar{\Psi}_{32}$, $\YY\bar{\Psi}_{42}$ & & $\frac{5}{2}$&\\\cline{2-4}\cline{6-6}
&1133 & $[0,2|2,2,0,0|2,0]$ &$\DD_{12}^2\ZZ^2$& &4&
\begin{picture}(36,0)
\color{gray}
\linethickness{.1mm}
\multiput(-8,6)(0,6){5}{\line(1,0){52}}
\put(6,0){\line(1,0){12}}
\put(18,36){\line(1,0){12}}
\put(-6,6){\line(0,1){24}}
\put(0,6){\line(0,1){24}}
\put(6,-2){\line(0,1){32}}
\put(12,-2){\line(0,1){32}}
\put(18,-2){\line(0,1){40}}
\put(24,6){\line(0,1){32}}
\put(30,6){\line(0,1){32}}
\put(36,6){\line(0,1){24}}
\put(42,6){\line(0,1){24}}
\linethickness{0.4mm}\color{orange}
\put(5.8,6){\line(1,0){6.2}}
\put(24,30){\line(1,0){6.2}}
\linethickness{0.4mm}\color{black}
\put(12,6){\line(1,0){12}}
\put(12,12){\line(1,0){12}}
\put(12,18){\line(1,0){12}}
\put(12,24){\line(1,0){12}}
\put(12,30){\line(1,0){12}}
\put(12,6){\line(0,1){24}}
\put(18,6){\line(0,1){24}}
\put(24,6){\line(0,1){24}}

\end{picture}

\\\hline\hline

\multirow{3}{*}{$\hat{1}...4$}
&0222& $[0,0|3,1,1,1|0,0]$ &$\ZZ\XX\YY$&\multirow{3}{*}{3}&3&
\\\cline{2-4}\cline{6-6}

&0233& $[0,0|3,1,0,0|2,0]$ &$\ZZ\Psi_{11}^2$&&4&
\\\cline{2-4}\cline{6-6}


&0000& $[3,3|3,3,3,3|0,0]$ &$\FFb_{12}^3$, $\FFb_{11}\FFb_{12}\FFb_{21}$&&6&
\begin{picture}(36,0)
\color{gray}
\linethickness{.1mm}
\multiput(-8,6)(0,6){5}{\line(1,0){52}}
\put(6,0){\line(1,0){12}}
\put(18,36){\line(1,0){12}}
\put(-6,6){\line(0,1){24}}
\put(0,6){\line(0,1){24}}
\put(6,-2){\line(0,1){32}}
\put(12,-2){\line(0,1){32}}
\put(18,-2){\line(0,1){40}}
\put(24,6){\line(0,1){32}}
\put(30,6){\line(0,1){32}}
\put(36,6){\line(0,1){24}}
\put(42,6){\line(0,1){24}}
\linethickness{0.4mm}\color{orange}
\put(6,5.8){\line(0,1){6.2}}
\put(24,30){\line(1,0){6.2}}
\linethickness{0.4mm}\color{black}
\put(18,6){\line(0,1){6}}
\put(24,6){\line(0,1){6}}
\put(30,6){\line(0,1){6}}
\put(36,6){\line(0,1){6}}
\put(24,12){\line(1,0){12}}
\put(18,6){\line(1,0){18}}
\color{black}
\put(6,12){\line(1,0){18}}
\put(6,18){\line(1,0){18}}
\put(6,24){\line(1,0){18}}
\put(6,30){\line(1,0){18}}
\put(6,12){\line(0,1){18}}
\put(12,12){\line(0,1){18}}
\put(18,12){\line(0,1){18}}
\put(24,12){\line(0,1){18}}
\end{picture}
 \\\hline\hline

\multirow{3}{*}{$1...\hat{4}$} & 2224& $[0,0|2,2,2,0|0,0]$ &$\ZZ\XX\YYb$&\multirow{3}{*}{3}&3&
\\\cline{2-4}\cline{6-6}

& 1124& $[0,2|3,3,2,0|0,0]$ &$\ZZ\bar{\Psi}_{42}^2$&&4&
\\\cline{2-4}\cline{6-6}
&4444& $[0,0|0,0,0,0|3,3]$ &$\FF_{12}^3$, $\FF_{11}\FF_{12}\FF_{21}$&&6&
\begin{picture}(36,0)
\color{gray}
\linethickness{.1mm}
\multiput(-8,6)(0,6){5}{\line(1,0){52}}
\put(6,0){\line(1,0){12}}
\put(18,36){\line(1,0){12}}
\put(-6,6){\line(0,1){24}}
\put(0,6){\line(0,1){24}}
\put(6,-2){\line(0,1){32}}
\put(12,-2){\line(0,1){32}}
\put(18,-2){\line(0,1){40}}
\put(24,6){\line(0,1){32}}
\put(30,6){\line(0,1){32}}
\put(36,6){\line(0,1){24}}
\put(42,6){\line(0,1){24}}
\linethickness{0.4mm}\color{orange}
\put(5.8,6){\line(1,0){6.2}}
\put(30,24){\line(0,1){6.2}}
\linethickness{0.4mm}\color{black}
\put(0,24){\line(0,1){6}}
\put(6,24){\line(0,1){6}}
\put(12,24){\line(0,1){6}}
\put(18,24){\line(0,1){6}}
\put(0,30){\line(1,0){18}}
\put(0,24){\line(1,0){12}}
\color{black}
\put(12,6){\line(1,0){18}}
\put(12,12){\line(1,0){18}}
\put(12,18){\line(1,0){18}}
\put(12,24){\line(1,0){18}}
\put(12,6){\line(0,1){18}}
\put(18,6){\line(0,1){18}}
\put(24,6){\line(0,1){18}}
\put(30,6){\line(0,1){18}}
\end{picture}
\\\hline\hline

\multirow{3}{*}{$\hat{1}...\hat{4}$}&0224& $[0,0|4,2,2,0|0,0]$ & $\ZZ^2\XX^2$&&4&\\\cline{2-4}\cline{6-6}
&0044& $[1,1|4,4,0,0|1,1]$ & $\DD_{11}\DD_{22}\ZZ^4$, $\DD_{12}\DD_{21}\ZZ^4$&4&6&\\\cline{2-4}\cline{6-6}
&0004& $[2,2|4,4,4,0|0,0]$ & $\bar{\Psi}_{41}^2\bar{\Psi}_{42}^2$&&6&
\begin{picture}(36,0)
\color{gray}
\linethickness{.1mm}
\multiput(-8,6)(0,6){5}{\line(1,0){52}}
\put(6,0){\line(1,0){12}}
\put(18,36){\line(1,0){12}}
\put(-6,6){\line(0,1){24}}
\put(0,6){\line(0,1){24}}
\put(6,-2){\line(0,1){32}}
\put(12,-2){\line(0,1){32}}
\put(18,-2){\line(0,1){40}}
\put(24,6){\line(0,1){32}}
\put(30,6){\line(0,1){32}}
\put(36,6){\line(0,1){24}}
\put(42,6){\line(0,1){24}}
\linethickness{0.4mm}\color{orange}
\put(6,5.8){\line(0,1){6.2}}
\put(30,24){\line(0,1){6.2}}
\linethickness{0.4mm}\color{black}
\put(-6,30){\line(1,0){24}}
\put(-6,24){\line(1,0){12}}
\put(30,12){\line(1,0){12}}
\put(18,6){\line(1,0){24}}
\put(-6,24){\line(0,1){6}}
\put(0,24){\line(0,1){6}}
\put(6,24){\line(0,1){6}}
\put(12,24){\line(0,1){6}}
\put(18,24){\line(0,1){6}}
\put(18,6){\line(0,1){6}}
\put(24,6){\line(0,1){6}}
\put(30,6){\line(0,1){6}}
\put(36,6){\line(0,1){6}}
\put(42,6){\line(0,1){6}}
\color{black}
\put(6,12){\line(1,0){24}}
\put(6,18){\line(1,0){24}}
\put(6,24){\line(1,0){24}}
\put(6,12){\line(0,1){12}}
\put(12,12){\line(0,1){12}}
\put(18,12){\line(0,1){12}}
\put(24,12){\line(0,1){12}}
\put(30,12){\line(0,1){12}}
\end{picture}
\\\hline
\end{tabular}
\caption{Selected components of the Konishi multiplet, which splits into four short multiplets at $g=0$.}
\label{table:konmul}
\end{table}

\subsubsection*{Joining and Young diagrams}
The above-made conclusions have a natural interpretation on the level of Young diagrams. If to generalise our statements in section~\ref{sec:yd} about $\psu(2,2|4)$ diagrams to an arbitrary rank: $\su(N,M|K)$ diagrams can be defined for arbitrary $N,M,K$, and the diagrams bijectively define isomorphism classes of $\algu(N,M|K)\oplus \algu(1)$ irreps that can be constructed using the outlined oscillator formalism. But there are, generically, different diagrams that correspond to the same isomorphism class of an $\su(N,M|K)$ irrep. The diagrams of isomorphic representations are related to one another through two moves \eqref{bothL}, which are generically permitted {\it if the representation in question is long}.

If the shortening \eqref{U4a} is present, we can think of the corresponding Young diagram as an extension of the $\su(1,2|3)$ diagram. In contrast to the $\psu(2,2|4)$ algebra, the $\su(1,2|3)$ Young diagram defines a long representation. Transition from \eqref{sm1} to \eqref{sm2} is the move \eqref{L1} realised on the $\su(1,2|3)$  diagram.

Likewise, transition from \eqref{sm3} to \eqref{sm4} is the move \eqref{L2} on an $\su(2,1|3)$ diagram, and the four oscillator contents in \eqref{sms} originate from both moves \eqref{bothL} on an $\su(1,1|2)$ diagram. See figure \ref{fig:short} for an example.

\begin{figure}[h!]
\centering
\begin{picture}(390,100)

\color{gray!40}
\linethickness{0.4mm}
\multiput(10,50)(0,1){10}{\line(1,0){10}}
\multiput(20,50)(0,1){10}{\line(1,0){10}}
\multiput(30,50)(0,1){10}{\line(1,0){10}}
\multiput(40,20)(0,1){10}{\line(1,0){10}}
\multiput(50,20)(0,1){10}{\line(1,0){10}}
\multiput(60,20)(0,1){10}{\line(1,0){10}}

\multiput(100,50)(0,1){10}{\line(1,0){10}}
\multiput(110,50)(0,1){10}{\line(1,0){10}}
\multiput(120,50)(0,1){10}{\line(1,0){10}}
\multiput(130,50)(0,1){10}{\line(1,0){10}}
\multiput(140,20)(0,1){10}{\line(1,0){10}}
\multiput(150,20)(0,1){10}{\line(1,0){10}}
\multiput(160,20)(0,1){10}{\line(1,0){10}}
\multiput(170,20)(0,1){10}{\line(1,0){10}}

\multiput(200,50)(0,1){10}{\line(1,0){10}}
\multiput(210,50)(0,1){10}{\line(1,0){10}}
\multiput(220,50)(0,1){10}{\line(1,0){10}}
\multiput(230,50)(0,1){10}{\line(1,0){10}}
\multiput(240,20)(0,1){10}{\line(1,0){10}}
\multiput(250,20)(0,1){10}{\line(1,0){10}}
\multiput(260,20)(0,1){10}{\line(1,0){10}}
\multiput(270,20)(0,1){10}{\line(1,0){10}}

\multiput(290,50)(0,1){10}{\line(1,0){10}}
\multiput(300,50)(0,1){10}{\line(1,0){10}}
\multiput(310,50)(0,1){10}{\line(1,0){10}}
\multiput(320,50)(0,1){10}{\line(1,0){10}}
\multiput(330,50)(0,1){10}{\line(1,0){10}}
\multiput(340,20)(0,1){10}{\line(1,0){10}}
\multiput(350,20)(0,1){10}{\line(1,0){10}}
\multiput(360,20)(0,1){10}{\line(1,0){10}}
\multiput(370,20)(0,1){10}{\line(1,0){10}}
\multiput(380,20)(0,1){10}{\line(1,0){10}}

\color{orange!50}
\linethickness{0.4mm}
\multiput(130,0)(0,1){10}{\line(1,0){10}}
\multiput(240,70)(0,1){10}{\line(1,0){10}}
\multiput(330,0)(0,1){10}{\line(1,0){10}}
\multiput(340,70)(0,1){10}{\line(1,0){10}}

\color{cyan!50}
\linethickness{0.4mm}
\multiput(110,30)(0,1){10}{\line(1,0){10}}
\multiput(110,40)(0,1){10}{\line(1,0){10}}
\multiput(260,30)(0,1){10}{\line(1,0){10}}
\multiput(260,40)(0,1){10}{\line(1,0){10}}
\multiput(310,30)(0,1){10}{\line(1,0){10}}
\multiput(310,40)(0,1){10}{\line(1,0){10}}
\multiput(360,30)(0,1){10}{\line(1,0){10}}
\multiput(360,40)(0,1){10}{\line(1,0){10}}

\color{black}
\linethickness{0.5mm}

\put(30,0){\line(1,0){10}}
\put(30,10){\line(1,0){10}}
\put(30,20){\line(1,0){40}}
\put(20,30){\line(1,0){50}}
\put(20,40){\line(1,0){40}}
\put(10,50){\line(1,0){50}}
\put(10,60){\line(1,0){40}}
\put(40,70){\line(1,0){10}}
\put(40,80){\line(1,0){10}}
\put(10,50){\line(0,1){10}}
\put(20,30){\line(0,1){30}}
\put(30,0){\line(0,1){60}}
\put(40,0){\line(0,1){80}}
\put(50,20){\line(0,1){60}}
\put(60,20){\line(0,1){30}}
\put(70,20){\line(0,1){10}}

\put(130,10){\line(1,0){10}}
\put(130,20){\line(1,0){50}}
\put(110,30){\line(1,0){70}}
\put(110,40){\line(1,0){50}}
\put(100,50){\line(1,0){60}}
\put(100,60){\line(1,0){50}}
\put(140,70){\line(1,0){10}}
\put(140,80){\line(1,0){10}}
\put(100,50){\line(0,1){10}}
\put(110,30){\line(0,1){30}}
\put(120,30){\line(0,1){30}}
\put(130,10){\line(0,1){50}}
\put(140,10){\line(0,1){70}}
\put(150,20){\line(0,1){60}}
\put(160,20){\line(0,1){30}}
\put(170,20){\line(0,1){10}}
\put(180,20){\line(0,1){10}}

\put(230,0){\line(1,0){10}}
\put(230,10){\line(1,0){10}}
\put(230,20){\line(1,0){50}}
\put(220,30){\line(1,0){60}}
\put(220,40){\line(1,0){50}}
\put(200,50){\line(1,0){70}}
\put(200,60){\line(1,0){50}}
\put(240,70){\line(1,0){10}}
\put(200,50){\line(0,1){10}}
\put(210,50){\line(0,1){10}}
\put(220,30){\line(0,1){30}}
\put(230,0){\line(0,1){60}}
\put(240,0){\line(0,1){70}}
\put(250,20){\line(0,1){50}}
\put(260,20){\line(0,1){30}}
\put(270,20){\line(0,1){30}}
\put(280,20){\line(0,1){10}}

\put(330,10){\line(1,0){10}}
\put(330,20){\line(1,0){60}}
\put(310,30){\line(1,0){80}}
\put(310,40){\line(1,0){60}}
\put(290,50){\line(1,0){80}}
\put(290,60){\line(1,0){60}}
\put(340,70){\line(1,0){10}}
\put(290,50){\line(0,1){10}}
\put(300,50){\line(0,1){10}}
\put(310,30){\line(0,1){30}}
\put(320,30){\line(0,1){30}}
\put(330,10){\line(0,1){50}}
\put(340,10){\line(0,1){60}}
\put(350,20){\line(0,1){50}}
\put(360,20){\line(0,1){30}}
\put(370,20){\line(0,1){30}}
\put(380,20){\line(0,1){10}}
\put(390,20){\line(0,1){10}}

\put(30,90){$1...4$}
\put(130,90){$\hat{1}...4$}
\put(230,90){$1...\hat{4}$}
\put(330,90){$\hat{1}...\hat{4}$}

\end{picture}
\caption{An example of four different $\psu(2,2|4)$ Young diagrams for short multiplets that join into one at finite coupling. Removal of the grey boxes defines $\su(1,1|2)$ diagrams. Conversely, the collection of grey boxes is the uniquely defined extension of an $\su(1,1|2)$ diagram to a $\psu(2,2|4)$ diagram, cf.\ figure~\ref{fig:YDext}. The changes of the $\su(1,1|2)$ diagrams with respect to the diagram on the left are highlighted.
}
\label{fig:short}
\end{figure}

\subsubsection*{Stronger shortenings}
In the above-described joining mechanisms, the participating short multiplets should correspond to admissible Young diagrams. Otherwise, the joining mechanism cannot be realised. 
An example of this restriction is $n_{\bbb_2}=0$ for the short multiplet \eqref{sm1}. Then the state \eqref{sm2} is inadmissible.
Consequently, the state with $n_{\bbb_2}=0$ is permitted to join only in one way, while being  a HWS  in the $\hat 1...$ grading, i.e.\ of the type \eqref{sm2}, at finite coupling.


The condition $n_{\bbb_2}=0$ corresponds to the extra shortening $\lambda_1+\nu_2=0$, in addition to $\lambda_1+\nu_1=0$, which at zero coupling should hold in all $\{1\hat 12\}...$ gradings, i.e.\ the HWS is unchanged within these gradings. For joining to be possible, the state can only remain a HWS in the $\hat{1}12...$ grading at finite coupling.
Likewise, the situation $n_{\fff_2}=L$ allows the joining as described in previous subsections if the state remains a HWS of the type \eqref{sm1} at finite coupling, i.e.\ in the $1...$ grading. This case corresponds to the shortenings $\lambda_2+\nu_1=\lambda_1+\nu_1=0$ in all $\{1\hat 1\hat 2\}...$ gradings. 
Two analogous restrictions exist for enhancement of the $\lambda_4+\nu_4=0$ shortening.
We discuss the finite coupling consequences of the strong shortenings in more detail in section \ref{sec:sectors}.

There is also a situation where we have both $n_{\bbb_2}=0$ in the $\{1\hat 12\}...$ gradings and $n_{\fff_2}=L$ in the $\{1\hat 1\hat 2\}...$ gradings. Then neither of the joining mechanisms, through $\hat 1...$ or $1...$ gradings, is permitted, so the corresponding state remains short at any coupling. Hence its conformal dimension is protected from quantum corrections. 
Because of the central charge constraint, there are only two multiplets of length $L$ that have this property. The first one has the oscillator content
\be
n^{2222}=[0,0|L,L-1,1,0|0,0]\,.
\ee
This multiplet cannot be realised using single-trace operators because the required  antisymmetrisations of oscillator labels are incompatible  with cyclicity of the trace.

The second one is the 1/2-BPS multiplet, with
\be
n^{2222}=[0,0|L,L,0,0|0,0]\,.
\ee
Its HWS in the compact beauty grading is the BMN vacuum $\Tr \ZZ^L$.

\subsection{Counting the spectrum}
The spectrum of multiplets, or irreducible representations, can be understood using {\it character theory}. For AdS$_5$/CFT$_4$, this was previously done in \cite{Bianchi:2003wx,Beisert:2003te} by an {\it ad hoc} use of $\mathfrak{so}(N)$ characters. Here, we propose a trick to avoid the complications of non-compactness and supersymmetry by mapping the $\psu(2,2|4)$ representations to representations of $\su(N)$. The trade-off is that the rank $N$ is unbounded, but it can be truncated if a cut-off is made in the classical dimension.

\subsubsection*{From $\psu(2,2|4)$ to $\su(N)$ representations}
In section \ref{sec:yd} we explained that an extended Young diagram characterises an irreducible representation of any algebra that can accommodate it.
All $\psu(2,2|4)$ diagrams can in fact be seen as diagrams of $\su(4+n_{\bbb_2}+n_{\aaa_1})$.
The corresponding representations can be thought of as tensor products of single-particle states built from the action of $2+n_{\bbb_2}$ fermionic oscillators on the Fock vacuum, i.e.
\be
\Phi_{i_1...i_{2+n_{\bbb_2}}}=\fff_{i_1}^\dagger \cdots \fff_{i_{2+n_{\bbb_2}}}^\dagger |0\rangle \,,
\ee
where the indices are antisymmetric. An example was given in figure \ref{fig:YDext}, where the {\color{VioletRed}red} path corresponds to the algebra $\su(9)$ with a single-site representation made of five antisymmetric indices ($\Phi_{ijklm}= \fff_i^\dagger \fff_j^\dagger \fff_k^\dagger \fff_l^\dagger \fff_m^\dagger | 0 \rangle$). In this algebra, the shown diagram corresponds to the weights $\lambda=n_{\fff_i}=\{8,7,7,7,6,4,3,2,1\}$, and the shaded boxes show the corresponding compact $\su(9)$ Young diagram.

To account for all multiplets up to a given maximal classical dimension $\Delta_{\text{max}}$, it is sufficient to consider $\su(2\Delta_{\text{max}})$ representations where the one-site state is a completely antisymmetric representation with $\Delta_{\text{max}}$ indices.

\subsubsection*{$\gl(N)$ characters and tensor product multiplicity}
The {\it character} for a $\gl(N)$ representation with weights $\lambda=\{\lambda_1,...,\lambda_N\}$ is given by the Schur polynomial
\be
\chi_\lambda = \frac{\det_{1\le i,j\le N} x_i^{\lambda_j+N-j}}{\det_{1\le i,j\le N} x_i^{N-j}}\equiv \frac{W_\lambda}{\Delta_V}\,,
\ee
where the denominator $\Delta_V=\prod_{i<j} (x_i - x_j)$ is the  Vandermonde determinant. 
For a given representation $\lambda$, the polynomial $W_\lambda$ will contain a term of the kind
\be
x_1^{\lambda_1+N-1}x_2^{\lambda_2+N-2} \cdots x_{N-1}^{\lambda_{N-1}+1}x_N^{\lambda_N}\,,
\ee
which we will call the {\it dominant term}.

The tensor product of $L$ representations decomposes into a direct sum of its irreducible representations, and the multiplicity of these irreducible representations, $c_{\lambda'}$, can be read off from the corresponding character decomposition,
\be
\chi_\lambda^L =  \sum_{\lambda'} c_{\lambda'} \chi_{\lambda'}\,. \label{tens}
\ee
The Vandermonde determinant is a common denominator on the right-hand side and can be factored out:
\be
\frac{W_{\lambda}^L}{\Delta_V^{L-1}} =  \sum_{\lambda'} c_{\lambda'} W_{\lambda'}\,.
\ee
The multiplicities $c_{\lambda'}$ can be found by calculating the coefficient of the dominant term of $W_{\lambda'}$ on the left-hand side. The advantage of $W$ compared to $\chi$ is that the dominant terms are unique to $W_\lambda$, i.e.\ they do not appear in $W_{\lambda'}$ if $\lambda\neq \lambda'$.

The multiplicities tell us exactly how many representations of a given type appear in the tensor product of $L$ single-site representations. However, this does not account for the cyclicity of the trace which reduces the Hilbert space. Furthermore, we have to keep in mind that shortening means that short multiplets combine into long.

\subsubsection*{Imposing cyclicity - Polya theory}
The tensor product character \eqref{tens} is a sum of all the states in the spectrum with length $L$. The Polya theorem \cite{Polya,Polyakov:2001af} provides a way to account for states that are related by equivalence relations, in our case by the cyclic group $\mathbb{Z}_L$. The corresponding sum of states $Z$ is given by
\be
Z=\sum_{L=2}^\infty \sum_{d|L} \frac{\phi(d)}{L} \chi_1(x_1^d,...,x_N^d)^{\frac{L}{d}}\,,
\ee
where $d|L$ means all divisors of $L$, $\phi(d)$ is the Euler totient function\footnote{The Euler totient function, $\phi(n)$, is given by the number of integers between $1$ and $n$ that are mutually prime with $n$.}, and $\chi_1$ is the character of the single-site $\su(N)$ representation. We refer to \cite{Bianchi:2003wx} for a more detailed explanation.

We will consider tensor products of up to $\Delta_{\text{max}}$ single-site representations of $\su(2\Delta_{\text{max}})$ with $\Delta_{\text{max}}$ antisymmetric indices, i.e.
\be
\chi_1(x_1,...,x_{\Delta_{\text{max}}})=\sum_{1\le i_1<...<i_{\Delta_{\text{max}}}\le 2\Delta_{\text{max}}} \prod_{n=1}^{\Delta_{\text{max}}} x_{i_n}\,.
\ee
Our goal is to evaluate the left-hand side of
\be
\Delta_{\text{V}} \, Z = \sum_\lambda c_\lambda W_\lambda \label{DelZ} \,,
\ee
to be able to read off $c_{\lambda}$. The combinatorics of this task grow factorially with $\Delta_{\text{max}}$, and already for $\su(10)$ it is challenging to expand the polynomial expression. However this can be overcome by using the fact that all information is captured by the dominant terms, and by a sort of reverse engineering procedure which we describe in \ref{ap:char}. In this way, we are able to produce the full decomposition of multiplets for $\Delta_0\le 8$, and further results can be produced if necessary. We find complete agreement with the results ($\Delta_0\le \frac{13}{2}$) provided in Appendix D of \cite{Beisert:2003te}. The multiplets with $\Delta_0\le\frac{11}{2}$ are listed in table \ref{table:simplestop}. 
We provide the full list of multiplets with $\Delta_0\le 8$ in \ref{ap:spec} for future reference. 

\subsubsection*{Example: $\Delta_0\le4$}
To describe the spectrum up to $\Delta_0\le4$, we should consider $\su(8)$ representations. The single site representation has a character of the form
\be
\chi_1(x_1,...,x_8)=x_1x_2x_3x_4 + x_1x_2x_3x_5 + ...
\ee
and the Vandermonde determinant looks like
\be
\Delta_V= x_1^3x_2^2x_3-x_1^2x_2^3x_3\pm...
\ee
The truncated sum of states is
\be
Z&=&
\frac{\phi(1)}{2}\chi_1(x_i)^2+\frac{\phi(2)}{2}\chi_1(x_i^2)
+\frac{\phi(1)}{3}\chi_1(x_i)^3+\frac{\phi(3)}{3}\chi_1(x_i^3)\\\nonumber
&&+\frac{\phi(1)}{4}\chi_1(x_i)^4+\frac{\phi(2)}{4}\chi_1(x_i^2)^2+\frac{\phi(4)}{4}\chi_1(x_i^4)\,.
\ee
It is straightforward, though already computationally demanding, to expand the product $\Delta_V Z$. The dominant terms in the expression corresponding to $\Delta_0\le4$ are
\be
\frac{\left. \Delta_V Z \right|_{\text{dominant},\Delta_0\le 4}}{x_1^7x_2^6x_3^5x_4^4x_5^3x_6^2x_7} &=&
{\color{gray}x_1^{2}x_2^{2}x_3^{2}x_4^{2}}
+ {\color{Cerulean}x_1^{2}x_2^{2}x_3^{1}x_4^{1}x_5^{1}x_6^{1}}
\no\\&&\no\\&&
+ {\color{gray}x_1^{3}x_2^{3}x_3^{3}x_4^{3}}
+ {\color{Cerulean} x_1^{3}x_2^{3}x_3^{2}x_4^{2}x_5^{2}}
+ {\color{Cerulean}x_1^{3}x_2^{3}x_3^{3}x_4^{1}x_5^{1}x_6^{1}}
+ {\color{Fuchsia} x_1^{3}x_2^{3}x_3^{2}x_4^{2}x_5^{1}x_6^{1}}
\no\\&&\no\\&&
+ {\color{gray}x_1^{4}x_2^{4}x_3^{4}x_4^{4}}
+ {\color{Cerulean}x_1^{4}x_2^{4}x_3^{4}x_4^{2}x_5^{2}}
+ {\color{Fuchsia} x_1^{4}x_2^{4}x_3^{3}x_4^{3}x_5^{2}}
+{\color{Fuchsia}  x_1^{4}x_2^{4}x_3^{4}x_4^{2}x_5^{1}x_6^{1}}
\no\\&&
+ x_1^{4}x_2^{4}x_3^{3}x_4^{2}x_5^{2}x_6^{1}
+ 2 x_1^{3}x_2^{2}x_3^{2}x_4^{2}x_5^{1}x_6^{1}x_7^{1}
+ 2 x_1^{4}x_2^{4}x_3^{2}x_4^{2}x_5^{2}x_6^{2}
\no\\&&
+ 2 x_1^{4}x_2^{4}x_3^{3}x_4^{3}x_5^{1}x_6^{1}
+ x_1^{1}x_2^{1}x_3^{1}x_4^{1}x_5^{1}x_6^{1}x_7^{1}x_8^{1}
\no\\&&
+ x_1^{2}x_2^{2}x_3^{2}x_4^{2}x_5^{2}x_6^{2}
+ x_1^{3}x_2^{3}x_3^{1}x_4^{1}x_5^{1}x_6^{1}x_7^{1}x_8^{1}\,.
\ee
The dominant terms corresponding to chiral primaries (members of 1/2-BPS multiplets) are marked in {\color{gray}grey}, while the short representations that make up the Konishi multiplet are marked in {\color{Cerulean}blue}, and those that make up the $n^{2222}=[0,0|2,2,1,1|0,0]$ multiplet (containing the lowest twist-3 operator) are marked in {\color{Fuchsia}purple} (one term with $\Delta_0=5$ is missing due to the truncation).

By combining short multiplets into long ones, shifting to oscillator number notation, and leaving out chiral primaries, the spectrum of unprotected multiplets with $\Delta_0\le 4$ is
\be
&&[0,0|1,1,1,1|0,0]+[0,0|2,2,1,1|0,0]+[0,0|1,1,1,1|2,0]\no\\
&&+[0,0|3,2,2,1|0,0]+[0,2|1,1,1,1|2,0]+[0,2|2,2,2,2|0,0]\no\\
&&+2\cdot[0,0|2,2,2,2|0,0]+2\cdot[0,0|3,3,1,1|0,0]+2\cdot[0,1|2,2,1,1|1,0]\,.
\ee
Note that for higher $\Delta_{\text{max}}$ the $\su(2\Delta_{\text{max}})$ sum of states \eqref{DelZ} will contain representations that lie outside the spectrum of $\psu(2,2|4)$, i.e.\ where 
the Young diagram does not fit inside the cross-shaped region defined by $\psu(2,2|4)$, see figure \ref{fig:YDdef}. These can simply be dropped from the sum. 


\newpage

\subsection{Sectors}\label{sec:sectors}
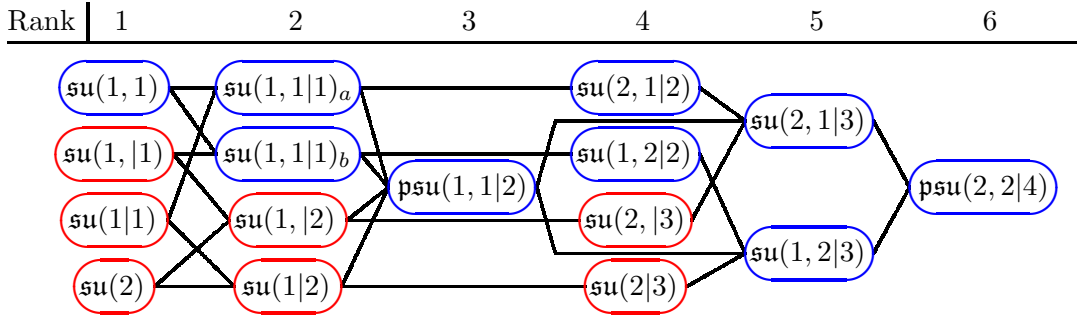
\begin{figure}[h!]\centering\begin{picture}(400,105)
\thicklines

\put(0,100){Rank}
\put(40,100){1}
\put(105,100){2}
\put(170,100){3}
\put(235,100){4}
\put(300,100){5}
\put(365,100){6}
\put(0,95){\line(1,0){400}}
\put(30,95){\line(0,1){15}}

\qbezier(60.5,78)(60.5,78)(78,78)
\qbezier(60.5,78)(60.5,78)(78,53)
\qbezier(62,53)(62,53)(78,53)
\qbezier(62,53)(62,53)(83,28)
\qbezier(60,28)(60,28)(78,78)
\qbezier(60,28)(60,28)(85,3)
\qbezier(55,3)(55,3)(83,28)
\qbezier(55,3)(55,3)(85,3)

\qbezier(132,78)(132,78)(142.5,40.5)
\qbezier(132,53)(132,53)(142.5,40.5)
\qbezier(127,28)(127,28)(142.5,40.5)
\qbezier(125,3)(125,3)(142.5,40.5)


\qbezier(198,40.5)(198,40.5)(205,65.5)
\qbezier(198,40.5)(198,40.5)(205,15.5)
\qbezier(205,65.5)(276,65.5)(276,65.5)
\qbezier(205,15.5)(276,15.5)(276,15.5)

\qbezier(132,78)(132,78)(211,78)
\qbezier(132,53)(132,53)(211,53)
\qbezier(127,28)(127,28)(214,28)
\qbezier(125,3)(125,3)(216,3)


\qbezier(259,78)(259,78)(276,65.5)
\qbezier(259,53)(259,53)(276,15.5)
\qbezier(256,28)(256,28)(276,65.5)
\qbezier(254,3)(254,3)(276,15.5)

\qbezier(324,15.5)(324,15.5)(337.5,40.5)
\qbezier(324,65.5)(324,65.5)(337.5,40.5)

\put(27,0){$\su(2)$}\put(40,3){\color{red} \oval(30,20)}
\put(23,25){$\su(1|1)$}\put(40,28){\color{red} \oval(40,20)}
\put(21,50){$\su(1,|1)$}\put(40,53){\color{red} \oval(44,20)}
\put(22,75){$\su(1,1)$}\put(40,78){\color{blue} \oval(41,20)}

\put(88,0){$\su(1|2)$}\put(105,3){\color{red} \oval(40,20)}
\put(85,25){$\su(1,|2)$}\put(105,28){\color{red} \oval(44,20)}
\put(81,50){$\su(1,1|1)_b$}\put(105,53){\color{blue} \oval(54,20)}
\put(81,75){$\su(1,1|1)_a$}\put(105,78){\color{blue} \oval(54,20)}

\put(146,37.5){$\psu(1,1|2)$}\put(170,40.5){\color{blue} \oval(55,20)}

\put(218,0){$\su(2|3)$}\put(235,3){\color{red} \oval(38,20)}
\put(216,25){$\su(2,|3)$}\put(235,28){\color{red} \oval(42,20)}
\put(213,50){$\su(1,2|2)$}\put(235,53){\color{blue} \oval(48,20)}
\put(213,75){$\su(2,1|2)$}\put(235,78){\color{blue} \oval(48,20)}

\put(278,12.5){$\su(1,2|3)$}\put(300,15.5){\color{blue} \oval(48,20)}
\put(278,62.5){$\su(2,1|3)$}\put(300,65.5){\color{blue} \oval(48,20)}

\put(341,37.5){$\psu(2,2|4)$}\put(365,40.5){\color{blue} \oval(55,20)}

\end{picture}
\caption{Closed sectors. \suii compact \suici non-compact.} 
\label{fig:sectors}
\end{figure}

\noindent By a {\it closed sector}, we mean operators with restricted field content that do not mix with operators with different field content, i.e.\ the finite coupling eigenstates of the dilatation operator will be a linear combination of single trace operators of only this kind. This is ensured by the fact that it is impossible to construct other types of operators from the $\psu(2,2|4)$ quantum numbers, e.g.\ the six numbers $\lambda_a-\lambda_{a+1}$ and $\nu_j-\nu_{j+1}$, corresponding to the sector. 

In a closed sector, some oscillators are passive in the sense that they are either saturated (e.g.\ $n_{\fff_1}=L$) or not excited at all (e.g.\ $n_{\fff_4}=n_{\aaa_1}=0$), while some oscillators are excited and will be part of a set of non-trivial raising operators. 
Furthermore, the length changing transformations \eqref{L1} and \eqref{L2} do not excite the passive oscillators.

\begin{table}[b!]
\def\arraystretch{1.7}
\centering
\begin{tabular}{|c|c|c|c|} \cline{2-4}
\multicolumn{1}{c|}{}&$[n_b|n_f|n_a]$&Grading & {\footnotesize$\begin{matrix} \text{Empty} \\[-3mm] \text{cells} \end{matrix}$} \\\cline{2-4}\hline
\multirow{2}{*}{simple}&$[0,\bullet| L, \bullet, \bullet, \bullet | \bullet, \bullet ]$ & $\{1\hat{1}\}...$ &
\begin{picture}(0,0)
\color{orange!70}
\linethickness{1.8mm}
\put(-9.8,-3.1){\line(1,0){5}}
\color{black}
\linethickness{0.1mm}
\multiput(-9.8,-5.6)(0,5){5}{\line(1,0){20}}
\multiput(-9.8,-5.6)(5,0){5}{\line(0,1){20}}
\end{picture}
%
\\\cline{2-4}
&$[\bullet,\bullet| \bullet, \bullet, \bullet, 0 | \bullet, 0 ]$ & $...\{4\hat{4}\}$ &
\begin{picture}(0,0)
\color{orange!70}
\linethickness{1.8mm}
\put(5.2,11.9){\line(1,0){5}}
\color{black}
\linethickness{0.1mm}
\multiput(-9.8,-5.6)(0,5){5}{\line(1,0){20}}
\multiput(-9.8,-5.6)(5,0){5}{\line(0,1){20}}
\end{picture}
\\\hline\hline
\multirow{4}{*}{strong}&$[0,0| L, \bullet, \bullet, \bullet | \bullet, \bullet ]$ & $\hat{1}12...$ &
\begin{picture}(0,0)
\color{orange!70}
\linethickness{1.8mm}
\put(-9.8,-3.1){\line(1,0){10}}
\color{black}
\linethickness{0.1mm}
\multiput(-9.8,-5.6)(0,5){5}{\line(1,0){20}}
\multiput(-9.8,-5.6)(5,0){5}{\line(0,1){20}}
\end{picture}
\\\cline{2-4}
&$[0,\bullet| L, L, \bullet, \bullet | \bullet, \bullet ]$ & $1\hat{1}\hat{2}...$ &
\begin{picture}(0,0)
\color{orange!70}
\linethickness{3.6mm}
\put(-9.8,-0.6){\line(1,0){5}}
\color{black}
\linethickness{0.1mm}
\multiput(-9.8,-5.6)(0,5){5}{\line(1,0){20}}
\multiput(-9.8,-5.6)(5,0){5}{\line(0,1){20}}
\end{picture}
\\\cline{2-4}
&$[\bullet,\bullet| \bullet, \bullet, \bullet, 0 | 0, 0 ]$ & $...34\hat{4}$ &
\begin{picture}(0,0)
\color{orange!70}
\linethickness{1.8mm}
\put(0.2,11.9){\line(1,0){10}}
\color{black}
\linethickness{0.1mm}
\multiput(-9.8,-5.6)(0,5){5}{\line(1,0){20}}
\multiput(-9.8,-5.6)(5,0){5}{\line(0,1){20}}
\end{picture}
\\\cline{2-4}
&$[\bullet,\bullet| \bullet, \bullet, 0, 0 | \bullet, 0 ]$ & $...\hat{3}\hat{4}4$ &
\begin{picture}(0,0)
\color{orange!70}
\linethickness{3.6mm}
\put(5.2,9.4){\line(1,0){5}}
\color{black}
\linethickness{0.1mm}
\multiput(-9.8,-5.6)(0,5){5}{\line(1,0){20}}
\multiput(-9.8,-5.6)(5,0){5}{\line(0,1){20}}
\end{picture}
\\\hline
\end{tabular}
\caption{Types of shortening and resulting grading for the state that remains a HWS at finite coupling. Brackets $\{\}$ denote interchangeable gradings. The rightmost column shows which cells on the 4$\times$4 square that are not covered by the corresponding Young diagram. Note that joining can combine short multiplets with a strong shortening in one grading and a simple one in another.}
\label{table:shtype}
\end{table}

Closed sectors combine highest-weight states of certain type, and in certain gradings, and are closely related to shortening. Two scenarios can be used to characterise sectors. Shortening happens when $n_{\fff_1}-L = n_{\bbb_1}=0$ or $n_{\fff_4}=n_{a_2}=0$. Due to the unitarity constraints, the gradings where this can happen must, in the first case, go through the point $(1,1)$ on the $4\times 4$ square, i.e.\ start as $1\hat{1}...$ or $\hat{1}1...$, or, in the second case, go through $(3,3)$, i.e.\ end as $...4\hat{4}$ or $...\hat{4}4$. We here refer to these shortenings as {\it simple}. If additionally $\fff_2$ is passive, $n_{\fff_2}=L$, the unitarity constraint $\lambda_1\ge \lambda_2$ forces the grading to start from $1\hat{1}\hat{2}$ for the finite coupling HWS. Likewise, $n_{\bbb_2}=0$ forces the grading to be $\hat{1}12...$, $n_{\fff_3}=0$ implies $...\hat{3}\hat{4}4$, and $n_{\aaa_1}=0$ implies $...34\hat{4}$. We refer to these four cases as {\it strong} shortenings. The six different possibilities are summarised in table \ref{table:shtype}.

%

\begin{table}[t]
\def\arraystretch{1.3}
\centering
\begin{picture}(250,80)
\thicklines

\small\bf
\put(53,45){strong $|$ simple $|$ none}

\put(75,-30){\circle*{6}}
\put(75,-10){\circle*{6}}
\put(75,10){\circle*{6}}
\put(75,30){\circle*{6}}

\put(110,-20){\circle*{6}}
\put(110,20){\circle*{6}}

\put(145,0){\circle*{6}}

\qbezier(75,-30)(110,-20)(110,-20)
\qbezier(75,-10)(110,-20)(110,-20)
\qbezier(75,10)(110,20)(110,20)
\qbezier(75,30)(110,20)(110,20)

\qbezier(110,-20)(110,-20)(145,0)
\qbezier(110,20)(145,0)(145,0)

\footnotesize
\put(150,-2){$[\bullet,\bullet|\bullet,\bullet,\bullet,\bullet|\bullet,\bullet]$}

\put(112,26){$[0,\bullet|L,\bullet,\bullet,\bullet|\bullet,\bullet]$}
\put(112,-30){$[\bullet,\bullet|\bullet,\bullet,\bullet,0|\bullet,0]$}

\put(1,27){$[0,0|L,\bullet,\bullet,\bullet|\bullet,\bullet]$}
\put(0,7){$[0,\bullet|L,L,\bullet,\bullet|\bullet,\bullet]$}
\put(1,-13){$[\bullet,\bullet|\bullet,\bullet,0,0|\bullet,0]$}
\put(1,-33){$[\bullet,\bullet|\bullet,\bullet,\bullet,0|0,0]$}

\end{picture}
\begin{tabular}{|c|c|c|} \hline
Rank & 
Type & \# Sectors \\\hline\hline
6 & 
\begin{picture}(36,0)
\thicklines
\put(33,4){\circle*{4}}
\end{picture}
& 1 \\\hline
5 &
\begin{picture}(36,0)
\thicklines
\put(33,4){\circle*{4}}
\put(20,4){\circle*{4}}
\qbezier(33,4)(33,4)(20,4)
\end{picture}
& 2 \\\hline
4 &
\begin{picture}(36,0)
\thicklines
\put(33,4){\circle*{4}}
\put(20,4){\circle*{4}}
\put(5,4){\circle*{4}}
\qbezier(33,4)(33,4)(20,4)
\qbezier(20,4)(20,4)(5,4)
\end{picture}
& 4 \\\hline
3 &
\begin{picture}(36,0)
\thicklines
\put(33,4){\circle*{4}}
\put(20,0){\circle*{4}}
\put(20,8){\circle*{4}}
\qbezier(33,4)(33,4)(20,8)
\qbezier(33,4)(33,4)(20,0)
\end{picture}
 & 1 \\\hline
2 & 
\begin{picture}(36,0)
\thicklines
\put(33,4){\circle*{4}}
\put(20,0){\circle*{4}}
\put(20,8){\circle*{4}}
\put(5,8){\circle*{4}}
\qbezier(33,4)(33,4)(20,8)
\qbezier(33,4)(33,4)(20,0)
\qbezier(20,8)(20,8)(5,8)
\end{picture}
& 4 \\\hline
1 &
\begin{picture}(36,0)
\thicklines
\put(33,4){\circle*{4}}
\put(20,0){\circle*{4}}
\put(20,8){\circle*{4}}
\put(5,8){\circle*{4}}
\put(5,0){\circle*{4}}
\qbezier(33,4)(33,4)(20,8)
\qbezier(33,4)(33,4)(20,0)
\qbezier(20,8)(20,8)(5,8)
\qbezier(20,0)(20,0)(5,0)
\end{picture}
& 4 \\\hline
\end{tabular}
\caption{Overview of the possible shortenings. The strong shortenings are special cases of the simple shortenings. The shortenings due to $\l_1$, $\l_2$, $\nu_1$ and $\nu_2$ are independent of those due to $\l_3$, $\l_4$, $\nu_3$ and $\nu_4$, and all sectors can be understood as combinations of such shortenings.}
\label{table:secshor}
\end{table}


All closed sectors can be understood in terms of these types of shortening. For example, the $\su(1,1)=\mathfrak{sl}(2)$ sector combines the fourth and sixth type: $[0,\bullet| L, L, 0, 0 | \bullet, 0 ]$, while the $\su(2,1|3)$ sector is of the second type: $[\bullet,\bullet| \bullet, \bullet, \bullet, 0 | \bullet, 0 ]$. 
The rank of a sector is one lower than the number of active oscillators. 
Table \ref{table:secshor} provides an overview of the ways in which the simple and strong types of shortenings combine and create subsectors of different rank. 
All closed sectors are listed in table \ref{table:allsectors}. 
Sectors of low rank are subsectors of higher rank sectors, and the relationship between sectors is summarised in figure \ref{fig:sectors}.
\begin{figure}[b]
\centering
\begin{picture}(335,120)

\color{gray}
\linethickness{.1mm}

\multiput(0,30)(0,15){5}{\line(1,0){120}}
\put(30,15){\line(1,0){15}}
\put(75,105){\line(1,0){15}}
\put(15,30){\line(0,1){60}}
\put(30,0){\line(0,1){60}}
\put(90,60){\line(0,1){60}}
\put(105,30){\line(0,1){60}}

\multiput(185,30)(0,15){5}{\line(1,0){150}}
\put(230,15){\line(1,0){30}}
\put(260,105){\line(1,0){30}}
\put(200,30){\line(0,1){30}}
\put(215,30){\line(0,1){15}}
\put(230,0){\line(0,1){45}}
\put(245,0){\line(0,1){30}}
\put(260,0){\line(0,1){120}}
\put(275,90){\line(0,1){30}}
\put(290,75){\line(0,1){45}}
\put(305,75){\line(0,1){15}}
\put(320,45){\line(0,1){45}}

\color{black}
\linethickness{0.7mm}

\put(45,0){\line(1,0){15}}
\put(45,30){\line(1,0){60}}
\put(45,45){\line(1,0){60}}
\put(15,60){\line(1,0){90}}
\put(15,75){\line(1,0){60}}
\put(15,90){\line(1,0){60}}
\put(60,120){\line(1,0){15}}

\put(15,60){\line(0,1){30}}
\put(45,0){\line(0,1){60}}
\put(60,0){\line(0,1){120}}
\put(75,60){\line(0,1){60}}
\put(105,30){\line(0,1){30}}

\put(245,30){\line(1,0){90}}
\put(215,45){\line(1,0){120}}
\put(200,60){\line(1,0){120}}
\put(185,75){\line(1,0){120}}
\put(185,90){\line(1,0){90}}

\put(185,75){\line(0,1){15}}
\put(200,60){\line(0,1){15}}
\put(215,45){\line(0,1){15}}
\put(245,30){\line(0,1){15}}
\put(260,30){\line(0,1){60}}
\put(275,75){\line(0,1){15}}
\put(305,60){\line(0,1){15}}
\put(320,45){\line(0,1){15}}
\put(335,30){\line(0,1){15}}

\put(15,105){$\sl(2)$}
\put(210,105){$\su(2)$}

\footnotesize
\put(71,50){$L-1$}
\put(71,35){$L-1$}
\put(26,80){$L-1$}
\put(26,65){$L-1$}
\put(49,3){\begin{turn}{90} $S-2$ \end{turn}}
\put(64,93){\begin{turn}{90} $S-2$ \end{turn}}

\put(285,35){$L-3$}
\put(269,50){$L-M-1$}
\put(271,65){$M-1$}

\put(225,50){$M-1$}
\put(208,65){$L-M-1$}
\put(212,80){$L-3$}

\end{picture}
\caption{Young diagrams ($1...4$ grading) for multiplets containing $\sl(2)$, $\DD_{12}^S\ZZ^L$, and $\su(2)$, $\ZZ^{L-M}\XX^M$, operators.}
\label{fig:secYD}
\end{figure}

Duality transformations involving the active oscillators will simply change the HWS within the sector. For the two simple shortenings, it is furthermore possible to do the duality transformations $1\hat{1}\leftrightarrow \hat{1}1$ or $4\hat{4}\leftrightarrow \hat{4}4$. These operations result in a new HWS that is still inside the sector, but of a different length. 

\begin{table}[t!]
\small
\centering
\def\arraystretch{1.15}

\colorlet{morange}{black}
\colorlet{mgray}{gray}
\begin{tabular}{|c|c|c|c|c|} \hline
Rank & Sector & Field content & $[n_\bbb|n_\fff|n_\aaa]$ & Grading(s)  \\\hline \hline

\multirow{4}{*}{1}

& $\su(1,1)$ & $\mathcal{D}_{12}^S \ZZ^L$ &$[\,{\color{mgray}0}\,, \, {\color{morange}S} \,| \, {\color{mgray}L}\, , \, {\color{mgray}L}\, , \, {\color{mgray}0}\, , \,{\color{mgray}0}\, | \, {\color{morange}S} \, ,\, {\color{mgray}0}\, ]$ & $1\hat{1}\hat{2}23\hat{3}\hat{4}$4 \\\cline{2-5}

&$\su(1,|1)$ & $\ZZ^{L-N}\bar{\Psi}_{42}^N$&$[\, {\color{mgray}0}\,,\, {\color{morange}N} \,|\, {\color{mgray}L} \,,\, {\color{mgray}L} \,,\, {\color{morange}N} \,,\, {\color{mgray}0} \,|\, {\color{mgray}0} \,,\, {\color{mgray}0} \,]$&$1\hat{1}\hat{2}[2\hat{3}]34\hat{4}$\\\cline{2-5}

&$\su(1|1)$ & $\ZZ^{L-N}\Psi_{11}^N$&$[\, {\color{mgray}0} \,,\, {\color{mgray}0} \,|\, {\color{mgray}L} \,,\, {\color{morange}L\!-\!N} \,,\, {\color{mgray}0} \,,\, {\color{mgray}0} \,|\, {\color{morange}N} \,,\, {\color{mgray}0} \,]$&$\hat{1}12[3\hat{2}]\hat{3}\hat{4}4$\\\cline{2-5}

&$\su(2)$ & $\ZZ^{L-M}\XX^M$&$[\, {\color{mgray}0}\,,\, {\color{mgray}0} \,|\, {\color{mgray}L} \,,\, {\color{morange}L\!-\!M} \,,\, {\color{morange}M} \,,\, {\color{mgray}0} \,|\,{\color{mgray}0} \,,\,{\color{mgray}0} \,]$&$\hat{1}12\hat{2}\hat{3}34\hat{4}$\\\hline\hline

\multirow{4}{*}{2}

&$\su(1,1|1)_a$ & $\mathcal{D}_{12}^S \ZZ^{L-N} \Psi_{11}^N$&$[\, {\color{mgray}0} , {\color{morange}S} \,|\, {\color{mgray}L} \,,\, {\color{morange}L\!-\!N} \,,\,{\color{mgray}0} \,,\, {\color{mgray}0} \,|\, {\color{morange}S\!+\!N} , {\color{mgray}0} \,]$&$\{1\hat{1}\}[23\hat{2}]\hat{3}\hat{4}4$\\\cline{2-5}

&$\su(1,1|1)_b$ & $\mathcal{D}_{12}^S \ZZ^{L-N} \bar{\Psi}_{42}^N$ &$[\, {\color{mgray}0} , {\color{morange}S\!+\!N} \,|\, {\color{mgray}L} \,,\, {\color{mgray}L} \,,\, {\color{morange}N} \,,\, {\color{mgray}0} \,|\, {\color{morange}S} , {\color{mgray}0} \,]$&$1\hat{1}\hat{2}[23\hat{3}]\{4\hat{4}\}$\\\cline{2-5}

&$\su(1,|2)$ & $\ZZ^{L-M-N}\XX^M\bar{\Psi}_{42}^{N}$&
$[\, {\color{mgray}0} , {\color{morange}N} \,|\, {\color{mgray}L} , {\color{morange}L\!-\!M} , {\color{morange}M\!+\!N} , {\color{mgray}0} \,|\, {\color{mgray}0} , {\color{mgray}0} \,]$&$\{1\hat{1}\}[2\hat{2}\hat{3}]34\hat{4}$\\\cline{2-5}

&$\su(1|2)$ &$\ZZ^{L-M-N} \XX^M \Psi_{11}^N$ &$[\, {\color{mgray}0} , {\color{mgray}0} \,|\, {\color{mgray}L} , {\color{morange}L\!-\!M\!-\!N} , {\color{morange}M} ,{\color{mgray}0} \,|\, {\color{morange}N} ,0 \,]$&$\hat{1}12[3\hat{2}\hat{3}]\{4\hat{4}\}$\\\hline\hline


3&$\psu(1,1|2)$ & $\begin{matrix}\mathcal{D}_{12}^n \ZZ^{L-M-N_1-N_2+n}\\ \XX^{M+n} \Psi_{11}^{N_1-n}\bar{\Psi}_{42}^{N_2-n}\end{matrix}$&$\begin{matrix}[\, {\color{mgray}0} \,,\, {\color{morange}N_2} \,|\,{\color{mgray}L} \,,\, {\color{morange}L\!-\!M\!-\!N_1} \,,\, \\ {\color{morange}M\!+\!N_2} \,,\, {\color{mgray}0} \,|\, {N_1} \,,\, {\color{mgray}0} \,]\end{matrix}$&$\!\!\{1\hat{1}\}[23\hat{2}\hat{3}]\{4\hat{4}\}\!\!$\\\hline\hline

\multirow{7}{*}{4}

&$\su(2,1|2)$ & $\begin{matrix}\DD_{11}^{S_1}\DD_{12}^{S_2} \ZZ^{L-N_1-N_2+n}\\ {\Psi}_{11}^{N_1-n}{\Psi}_{21}^{N_2-n}\mathcal{F}_{11}^n\end{matrix}$&$\begin{matrix}[\, {\color{morange}S_1} \,,\, {\color{morange}S_2} \,|\, {\color{morange}L\!-\!N_2} \,,\, {\color{morange}L\!-\!N_1} \,,\, \\ {\color{mgray}0} \,,\, {\color{mgray}0} \,|\, {\color{morange}N_1\!+\!N_2\!+\!S_1\!+\!S_2} \,,\, {\color{mgray}0} \,]\end{matrix}$&$[123\hat{1}\hat{2}]\hat{3}\hat{4}4$\\\cline{2-5}

&$\su(2,|3)$ & $\begin{matrix}\ZZ^{L-M_1-M_2-N_1-N_2}\\ \XX^{M_1}\YYb^{M_2}\bar{\Psi}_{42}^{N_1}\bar{\Psi}_{41}^{N_2}\end{matrix}$&$\begin{matrix}[\, {\color{morange}N_2} \,,\, {\color{morange}N_1} \,|\, {\color{morange}L\!-\!M_2} \,,\, {\color{morange}L\!-\!M_1} \,,\,\\ {\color{morange}M_1\!+\!M_2\!+\!N_1\!+\!N_2} \,,\, {\color{mgray}0} \,|\, {\color{mgray}0} \,,\, {\color{mgray}0} \,]\end{matrix}$&$[12\hat{1}\hat{2}\hat{3}]34\hat{4}$\\\cline{2-5}

&$\su(1,2|2)$ & $\begin{matrix} \DD_{12}^{S_1}\DD_{22}^{S_2} \ZZ^{L-N_1-N_2+n}\\ \bar{\Psi}_{42}^{N_1-n}\bar{\Psi}_{32}^{N_2-n}\bar{\mathcal{F}}_{22}^n\end{matrix}$&$\begin{matrix}[\, {\color{mgray}0} \,,\, {\color{morange}S_1\!+\!S_2\!+\!N_1\!+\!N_2} \,|\, {\color{mgray}L} \,,\, {\color{mgray}L} \,,\, \\ {\color{morange}N_1} \,,\, {\color{morange}N_2} \,|\, {\color{morange}S_1} \,,\, {\color{morange}S_2} \,]\end{matrix}$&$1\hat{1}\hat{2}[234\hat{3}\hat{4}]$\\\cline{2-5}

&$\su(2|3)$ & $\begin{matrix}\ZZ^{L-M_1-M_2-N_1-N_2}\\ \XX^{M_1}\YY^{M_2}\Psi_{11}^{N_1}\Psi_{12}^{N_2}\end{matrix}$&
$\begin{matrix}[\, {\color{mgray}0} , {\color{mgray}0} \,|\, {\color{mgray}L} , {\color{morange}L\!-\!M_1\!-\!M_2\!-\!N_1\!-\!N_2} \,,\, \\ {\color{morange}M_1} \,,\, {\color{morange}M_2} \,|\, {\color{morange}N_1} \,,\, {\color{morange}N_2} \,]\end{matrix}$&$\hat{1}12[34\hat{2}\hat{3}\hat{4}]$\\\hline\hline

\multirow{3}{*}{5}

&$\su(1,2|3)$ 
&
$\begin{matrix} 
\DD_{12}^\bullet\DD_{22}^{\bullet} \XX^\bullet\YY^\bullet \ZZ^\bullet \\
\Psi_{11}^{\bullet}\Psi_{12}^{\bullet} \bar{\Psi}_{42}^\bullet\bar{\Psi}_{32}^\bullet\bar{\Psi}_{22}^\bullet\FFb_{22}^\bullet
\end{matrix}$
&
$[{\color{mgray}0},\bullet|{\color{mgray}L},\bullet,\bullet,\bullet|\bullet,\bullet]
$&$[123\hat{1}\hat{2}\hat{3}]\{4\hat{4}\}$\\\cline{2-5}

&$\su(2,1|3)$ 
& 
$\begin{matrix} \DD_{11}^\bullet\DD_{12}^\bullet \XX^\bullet\YYb^\bullet\ZZ^\bullet\\\bar{\Psi}_{42}^\bullet\bar{\Psi}_{41}^\bullet{\Psi}_{11}^\bullet{\Psi}_{21}^\bullet{\Psi}_{31}^\bullet\FF_{11}^\bullet\end{matrix}$
&
$[\bullet,\bullet|\bullet,\bullet,\bullet,{\color{mgray}0}|\bullet,{\color{mgray}0}]$
&$\{1\hat{1}\} [234\hat{2}\hat{3}\hat{4}]$\\\hline

\end{tabular}
\caption{Closed sectors. Passive oscillators are marked in {\color{mgray}grey}. A bracket $[\,]$ denotes duality transformations that shuffle active oscillators and change the HWS within the sector while preserving the length. A bracket $\{\}$ denotes a duality transformation that shuffles passive oscillators and changes the HWS within the sector while changing $L$. Note that the field content in the $\psu(1,1|2)$, $\su(2,1|2)$, $\su(1,2|2)$, $\su(2,1|3)$ and $\su(1,2|3)$ sectors is not completely fixed by the oscillator numbers. For the first three cases, this ambiguity appears in the above field content through the integer $n$, while in the last two cases the ambiguity is larger.} 
\label{table:allsectors}
\end{table} 

Note that in sectors where only one of the shortening conditions is satisfied, one of the length changing transformations \eqref{L1} and \eqref{L2} is allowed and results in a state with the same $\lambda$ and $\nu$ (up to a shift of $\pm1$), but with a different length. As discussed, this means that at higher loops the eigenstates of the dilatation operator can mix operators of different length.

Young diagrams for multiplets containing a HWS within the $\sl(2)$ and $\su(2)$ sectors are depicted in figure \ref{fig:secYD}. Note that a certain class of finite-coupling multiplets contain operators belonging to all rank 1 sectors, namely those for which
\be
n^{2222} = [0,0|L-1,L-1,1,1|0,0]\,.
\ee
%

\subsection{Table of simplest multiplets}
\begin{table}[h!]
\centering
\def\arraystretch{1.07}
{\small
\begin{tabular}{|c||c|c|c|c|c|c|} \hline
$\Delta_0^{2222}$&$L$&$[\,n_\bbb\,| \, n_\fff \, | \, n_\aaa \,]^{2222}$& U$_1$ & U$_2$ & Sector & Multiplicity 
\\\hline\hline\hline

2&2-4&$[0,0|1,1,1,1|0,0]$&\ding{52}&\ding{52}& all & 1\\\hline\hline

3&3-5&$[0,0|2,2,1,1|0,0]$&\ding{52}&\ding{52}& all & 1\\\hline\hline

\multirow{7}{*}{4}&4-6&$[0,0|3,3,1,1|0,0]$&\ding{52}&\ding{52}& all &2\\\cline{2-7}

&3-5& $[0,1|2,2,1,1|1,0]$&\ding{52}&\ding{52}& $\sl(2)$&2\\\cline{2-7}

&2-4&$[0,2|1,1,1,1|2,0]$&\ding{52}&\ding{52}& $\sl(2)$&1\\\cline{2-7}

&4-6&$[0,0|3,2,2,1|0,0]$&\ding{52}&\ding{52}&$\su(2)$&1\\\cline{2-7}

&\multirow{2}{*}{3-4}&$[0,0|1,1,1,1|2,0]$&\ding{56}&\ding{52}& $\su(2,1|2)$ &1
\\
&&$[0,2|2,2,2,2|0,0]$&\ding{52}&\ding{56}& $\su(1,2|2)$&1
\\\cline{2-7}

&4&$[0,0|2,2,2,2|0,0]$&\ding{56}&\ding{56}&$\psu(2,2|4)$&2\\\hline\hline

\multirow{13}{*}{5}&5-7&$[0,0|4,4,1,1|0,0]$&\ding{52}&\ding{52}&all&2\\\cline{2-7}

&4-6& $[0,1|3,3,1,1|1,0]$ &\ding{52}&\ding{52}&$\sl(2)$&2\\\cline{2-7}

&3-5&$[0,2|2,2,1,1|2,0]$&\ding{52}&\ding{52}&$\sl(2)$&1\\\cline{2-7}

&5-7&$[0,0|4,3,2,1|0,0]$&\ding{52}&\ding{52}&$\su(2)$&2\\\cline{2-7}

&\multirow{2}{*}{4-6}&$[0,0|3,1,1,1|2,0]$&\multirow{2}{*}{\ding{52}}&\multirow{2}{*}{\ding{52}}&$\su(1|1)$&1
\\
&&$[0,2|3,3,3,1|0,0]$&&& $\su(1,|1)$&1\\\cline{2-7}

&4-6&$[0,1|3,2,2,1|1,0]$&\ding{52}&\ding{52}&$\psu(1,1|2)$&4\\\cline{2-7}

&\multirow{2}{*}{5-6}&$[0,0|3,3,3,1|0,0]$
&\ding{56}&\ding{52}& $\su(2,|3)$&2
\\
&&$[0,0|4,2,2,2|0,0]$
&\ding{52}&\ding{56}& $\su(2|3)$&2
\\\cline{2-7}

&\multirow{2}{*}{4-5}&$[0,0|2,2,1,1|2,0]$
&\ding{56}&\ding{52}&$\su(2,1|2)$&2\\
&&$[0,2|3,3,2,2|0,0]$&\ding{52}&\ding{56}&$\su(1,2|2)$&2\\\cline{2-7}

&5&$[0,0|3,3,2,2|0,0]$&\ding{56}&\ding{56}&$\psu(2,2|4)$&4\\\cline{2-7}

&4&$[0,1|2,2,2,2|1,0]$&\ding{56}&\ding{56}&$\psu(2,2|4)$&2\\\hline\hline

\multirow{12}{*}{$\frac{11}{2}$}&\multirow{2}{*}{5-7}&$[0,0|4,3,1,1|1,0]$&\multirow{2}{*}{\ding{52}}&\multirow{2}{*}{\ding{52}}&$\su(1|1)$&2\\
&&$[0,1|4,4,2,1|0,0]$&&&$\su(1,|1)$&2\\\cline{2-7}

&\multirow{2}{*}{4-6}&$[0,1|3,2,1,1|2,0]$
&\multirow{2}{*}{\ding{52}}&\multirow{2}{*}{\ding{52}}&$\su(1,1|1)_a$&2
\\
&&$[0,2|3,3,2,1|1,0]$&&&$\su(1,1|1)_b$&2
\\\cline{2-7}

&\multirow{2}{*}{3-5}&$[0,2|2,1,1,1|3,0]$
&\multirow{2}{*}{\ding{52}}&\multirow{2}{*}{\ding{52}}&$\su(1,1|1)_a$&2
\\
&&$[0,3|2,2,2,1|2,0]$
&&&$\su(1,1|1)_b$&2
\\\cline{2-7}

&\multirow{2}{*}{5-6}&$[0,0|3,3,2,1|1,0]$
&\ding{56}&\ding{52}& $\su(2,1|3)$&4
\\
&&$[0,1|4,3,2,2|0,0]$
&\ding{52}&\ding{56}& $\su(1,2|3)$&4
\\\cline{2-7}

&\multirow{2}{*}{4-5}&$[0,1|2,2,2,1|2,0]$
&\ding{56}&\ding{52}&$\su(2,1|3)$&4
\\
&&$[0,2|3,2,2,2|1,0]$
&\ding{52}&\ding{56}&$\su(1,2|3)$&4
\\\cline{2-7}

&\multirow{2}{*}{5}&$[0,0|3,2,2,2|1,0]$
&\multirow{2}{*}{\ding{56}}&\multirow{2}{*}{\ding{56}}&\multirow{2}{*}{$\psu(2,2|4)$}&4
\\
&&$[0,1|3,3,3,2|0,0]$&&&&4
\\\hline
\end{tabular}
}
\caption{All non-protected primaries with $\Delta_0^{2222}\le\frac{11}{2}$. The columns U$_1$ and U$_2$ denote whether the two unitarity bounds (shortening conditions), \eqref{U4a1} and \eqref{U5a1}, are saturated, respectively. A larger and more detailed table can be found in the ancillary \texttt{Mathematica} notebook.}\label{table:simplestop}
\end{table}
\newpage
\section{One-loop Q-systems} \label{sec:Qsys}\label{sec:Qsys}
The eigenstates of integrable spin chains are in one-to-one correspondence to solutions of a corresponding Q-system with particular analytic structure. 
For representations of compact algebras, the whole Q-system is polynomial and the involved Q-functions carry Bethe roots as zeros, while a class of non-rational functions appears in the non-compact case. We here propose a new way to think of these Q-systems: as living on the Young diagrams of the irreducible representation in question, instead of being associated to the symmetry algebra. However, we will end up making the transition back to the $\psu(2,2|4)$ Q-system, which is the basic structure in the quantum spectral curve.

The discussion in sections \ref{sec:QQ}-\ref{sec:alg} is valid for any $\su(N,M|K)$ spin chain, but, for clarity of the exposition, we work with homogeneous spin chains and keep the spin chain nodes in irreducible representations  that have only one non-zero Dynkin label in some grading\footnote{They are often called rectangular representations, due to the shape of their Young diagram on the T-hook. For the $\su(N)$ case, they are also known as  Kirillov-Reshetikhin modules.}. The corresponding Dynkin node is called {\it momentum-carrying}. Starting from section~\ref{sec:Qshort}, we focus on the particular features of the $\psu(2,2|4)$ system and the consequences of the cyclicity of trace-operators.

\subsection{Q-systems} \label{sec:QQ}
Supersymmetric Q-systems are made up of a set of Q-functions of the spectral parameter $u$,
\be
Q_{A|I}(u)=Q_{a_1a_2\hdots|i_1i_2\hdots}(u)\,,
\ee
with $A$ and $I$ denoting multi-indices. The Q-functions are separately antisymmetric in the two types of indices.

\subsubsection*{QQ-relations}
The Q-functions are related by three types of finite difference equations,
\begin{subequations}
\be
Q_{A|I}Q_{Aab|I}&=&Q^+_{Aa|I}Q^-_{Ab|I}-Q^-_{Aa|I}Q^+_{Ab|I}\label{QQ1}\\
Q_{A|I}Q_{A|Iij}&=&Q^+_{A|Ii}Q^-_{A|Ij}-Q^-_{A|Ii}Q^+_{A|Ij}\label{QQ2}\\
Q_{Aa|I}Q_{A|Ii}&=&Q^+_{Aa|Ii}Q^-_{A|I}-Q^-_{Aa|Ii}Q^+_{A|I} \label{QQ3}\,.
\ee
\end{subequations}
We use the standard notation $Q^\pm=Q(u\pm\frac{i}{2})$ and $Q^{[n]}=Q(u+\frac{in}{2})$ for shifts in the spectral parameter.

The overall normalisation of Q-functions is irrelevant for us (unless it is zero), hence all the QQ-relations should be understood in this projective sense. If a Q-function is a polynomial, we normalise it to be a monic polynomial for convenience.

\subsubsection*{Distinguished Q-functions}
By {\it distinguished} Q-functions we refer to those where the indices take the lowest possible values, i.e.
\be
\dQ_{a,s}\equiv Q_{12...(a-1)a|12...(s-1)s}\,. \label{dqdef}
\ee

\subsubsection*{Traditional way of thinking: Q-system belongs to an algebra}
\begin{figure}[h!]
\centering
\begin{picture}(400,100)

\linethickness{0.4mm}

\put(15,50){\line(1,0){80}}
\put(15,50){\circle*{4}}
\put(55,50){\color{NavyBlue}\circle*{6}}
\put(95,50){\circle*{4}}
\put(-3,48){$Q_\emptyset$}
\put(50,57){$Q_{1}$}
\put(50,38){$Q_{2}$}
\put(100,48){$Q_{12}$}

\put(160,10){\line(1,0){40}}
\put(160,50){\line(1,0){40}}
\put(160,90){\line(1,0){40}}
\put(160,10){\line(0,1){80}}
\put(200,10){\line(0,1){80}}
\put(160,50){\color{NavyBlue}\circle*{6}}
\put(200,50){\circle*{4}}
\put(160,10){\circle*{4}}
\put(200,10){\circle*{4}}
\put(160,90){\circle*{4}}
\put(200,90){\circle*{4}}
\put(138,5){$Q_{\emptyset|\emptyset}$}
\put(203,5){$Q_{\emptyset|1}$}
\put(138,56){$Q_{1|\emptyset}$}
\put(138,43){$Q_{2|\emptyset}$}
\put(203,56){$Q_{1|1}$}
\put(203,43){$Q_{2|1}$}
\put(203,90){$Q_{12|1}$}
\put(133,90){$Q_{12|\emptyset}$}

\put(290,10){\line(1,0){80}}
\put(290,50){\line(1,0){80}}
\put(290,90){\line(1,0){80}}
\put(290,10){\line(0,1){80}}
\put(330,10){\line(0,1){80}}
\put(370,10){\line(0,1){80}}
\put(290,50){\circle*{4}}
\put(330,50){\color{NavyBlue}\circle*{6}}
\put(370,50){\circle*{4}}
\put(290,10){\circle*{4}}
\put(330,10){\circle*{4}}
\put(370,10){\circle*{4}}
\put(290,90){\circle*{4}}
\put(330,90){\circle*{4}}
\put(370,90){\circle*{4}}
\put(268,5){$Q_{\emptyset|\emptyset}$}
\put(268,56){$Q_{1|\emptyset}$}
\put(268,43){$Q_{2|\emptyset}$}
\put(263,90){$Q_{12|\emptyset}$}
\put(332,1){$Q_{\emptyset|2}$}
\put(332,55){$Q_{1|2}$}
\put(332,41){$Q_{2|2}$}
\put(332,94){$Q_{12|2}$}
\put(310,1){$Q_{\emptyset|1}$}
\put(309,55){$Q_{1|1}$}
\put(309,41){$Q_{2|1}$}
\put(306,94){$Q_{12|1}$}
\put(373,5){$Q_{\emptyset|12}$}
\put(373,56){$Q_{1|12}$}
\put(373,43){$Q_{2|12}$}
\put(373,90){$Q_{12|12}$}

\end{picture}
\caption{Examples of Q-systems: $\sl(2)$ (left), $\su(1|2)$ (middle), and $\psu(1,1|2)$ (right). The momentum-carrying node is marked in {\color{NavyBlue}blue}. Note that there is one distinguished Q-function at each node.}\label{fig:QQalg}
\end{figure}
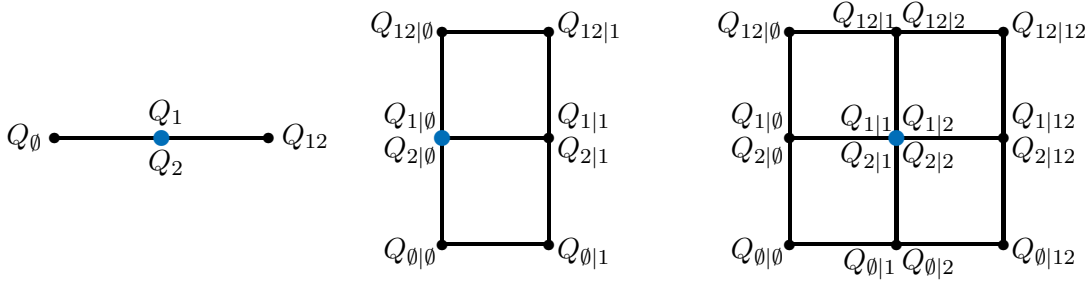

Traditionally, Q-systems are associated with algebras \cite{Tsuboi:1997iq,Kazakov:2007fy,Tsuboi:2009ud,Tsuboi:2011iz,Kazakov:2015efa}. For example, the $\sl(2)$ Q-system consists of four Q-functions, $Q_{\emptyset}$, $Q_{1}$, $Q_2$ and $Q_{12}$ related by a single QQ-relation of the type \eqref{QQ2}, while the $\su(1|2)$ Q-system contains eight Q-functions related by two QQ-relations of type \eqref{QQ1} and four of type \eqref{QQ3}, see figure \ref{fig:QQalg}. The algebra and representation impose certain restrictions on the analytic structure of the Q-functions, and we return to this in section \ref{sec:Qstruc}.

\subsubsection*{New approach: Q-system belongs to a Young diagram}
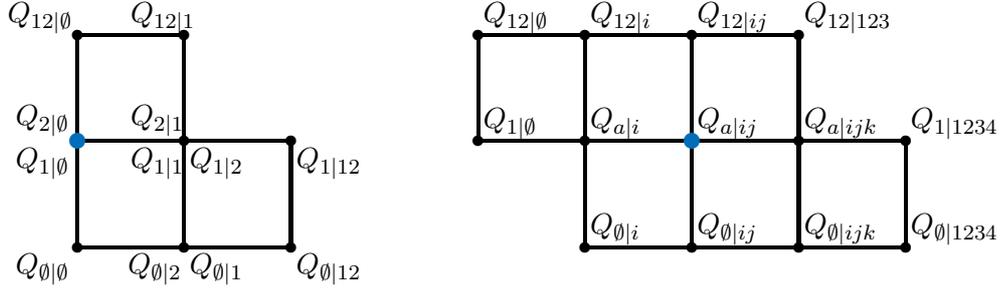
\begin{figure}[h!]
\centering
\begin{picture}(340,100)

\linethickness{0.4mm}

\put(20,90){\line(1,0){40}}
\put(20,50){\line(1,0){80}}
\put(20,10){\line(1,0){80}}
\put(20,10){\line(0,1){80}}
\put(60,10){\line(0,1){80}}
\put(100,10){\line(0,1){40}}
\put(20,10){\circle*{4}}
\put(60,10){\circle*{4}}
\put(100,10){\circle*{4}}
\put(20,50){\color{NavyBlue}\circle*{6}}
\put(60,50){\circle*{4}}
\put(100,50){\circle*{4}}
\put(20,90){\circle*{4}}
\put(60,90){\circle*{4}}

\put(210,10){\line(1,0){120}}
\put(170,50){\line(1,0){160}}
\put(170,90){\line(1,0){120}}
\put(170,50){\line(0,1){40}}
\put(210,10){\line(0,1){80}}
\put(250,10){\line(0,1){80}}
\put(290,10){\line(0,1){80}}
\put(330,10){\line(0,1){40}}
\put(170,50){\circle*{4}}
\put(170,90){\circle*{4}}
\put(210,10){\circle*{4}}
\put(210,50){\circle*{4}}
\put(210,90){\circle*{4}}
\put(250,10){\circle*{4}}
\put(250,50){\circle*{4}}
\put(250,90){\circle*{4}}
\put(290,10){\circle*{4}}
\put(290,50){\circle*{4}}
\put(290,90){\circle*{4}}
\put(330,10){\circle*{4}}
\put(330,50){\circle*{4}}
\put(250,50){\color{NavyBlue}\circle*{6}}


\put(-3,0){$Q_{\emptyset|\emptyset}$}
\put(39,0){$Q_{\emptyset|2}$}
\put(62,0){$Q_{\emptyset|1}$}
\put(102,0){$Q_{\emptyset|12}$}
\put(-3,56){$Q_{2|\emptyset}$}
\put(-3,40){$Q_{1|\emptyset}$}
\put(-6,95){$Q_{12|\emptyset}$}
\put(40,56){$Q_{2|1}$}
\put(40,40){$Q_{1|1}$}
\put(40,95){$Q_{12|1}$}
\put(62,40){$Q_{1|2}$}
\put(102,40){$Q_{1|12}$}

\put(212,15){$Q_{\emptyset|i}$}
\put(252,15){$Q_{\emptyset|ij}$}
\put(292,15){$Q_{\emptyset|ijk}$}
\put(332,15){$Q_{\emptyset|1234}$}
\put(172,55){$Q_{1|\emptyset}$}
\put(212,55){$Q_{a|i}$}
\put(252,55){$Q_{a|ij}$}
\put(292,55){$Q_{a|ijk}$}
\put(332,55){$Q_{1|1234}$}
\put(172,95){$Q_{12|\emptyset}$}
\put(212,95){$Q_{12|i}$}
\put(252,95){$Q_{12|ij}$}
\put(292,95){$Q_{12|123}$}

\end{picture}
\caption{Q-systems on a compact Young diagram (left) and on a non-compact diagram (right). The leftmost diagram corresponds to a long representation of $\su(2|1)$ and $\su(1|2)$ (and subalgebras thereof) and the Young diagram Q-system thus contains the Q-functions belonging to both the $\su(2|1)$ and $\su(1|2)$ Q-systems.}\label{fig:QQYD}
\end{figure}

We will take another point of view: that the Q-system lives on the Young diagram of the considered irreducible representation, see figure \ref{fig:QQYD}. Extended diagrams may be considered if needed, and this provides an algebra-independent description. The extended Young diagram corresponds to a representation in any $\su(N,M|K)$ algebra where it fits inside the cross-shaped area defined by the algebra, see e.g.\ figure \ref{fig:YDdef} for $\psu(2,2|4)$. The Q-system on the Young diagram consists of all Q-functions that are part of an $\su(N,M|K)$ Q-system for which the Young diagram defines a long representation\footnote{For a more detailed discussion of Young diagram Q-systems, see the thesis \cite{Marboe:2017zdv}.}. We will mostly focus on the set of distinguished Q-functions \eqref{dqdef}. There is one $\dQ$ at each node, and this subset is related solely by QQ-relations of the type \eqref{QQ3}.

\subsection{Structure of Q-functions} \label{sec:Qstruc}
We here review the structure of Q-functions for homogeneous rational spin chains with in principle arbitrary symmetry algebra, though we tailor our notation to the $\psu(2,2|4)$ case.

\subsubsection*{Distinguished Q-functions}
All distinguished Q-functions are rational, and we here give their specific structure.

\noindent {\bf Asymptotic power}. We choose the origin on the $\mathbb{Z}^2$ lattice such that the momentum-carrying node has coordinates $(2,2)$. Then $\dQ_{2,2}$ is, by default, the distinguished Q-function at the momentum-carrying node. The large $u$ asymptotic power of  $\dQ_{n,n}$ with minimal $n$ (such that the node $(n,n)$ lies on the left boundary of the Young diagram) is $u^0$. Starting from this point, the asymptotic power of any other distinguished Q-function can be found by summing the weights $\lambda$ and $\nu$ that are encountered on the way from $\dQ_{n,n}$ to $\dQ_{a,s}$,
\be
\dQ_{a,s}\simeq u^{-\sum_{b=n}^a \lambda_b-\sum_{k=n}^s \nu_k}\,. \label{powercount}
\ee

\noindent {\bf Full structure}. The full structure of the Q-functions in a non-compact Q-system is
\be
\dQ_{a.s}=\Phi_{a,s}^L\,\, q_{a,s}\,, \label{Qstruc}
\ee
where $\Phi_{a,s}$ is the fusion factor given by
\be\label{fusion}
\Phi_{a,s}(u)=\left\{ \begin{matrix} u^{[s-a]_{\rm D}}\,, & s\le 1\vee (s=2\wedge a\le 2)  \\  1 & \quad\text{otherwise} \end{matrix} \right.\,,
\ee
where we define $f^{[r]_{\rm D}}\equiv \prod\limits_{k=\frac{-|r|+1}2}^{\frac{|r|-1}2} f\left( u+i\,k \right)^{\text{sign}(r)}\,,$ for $r\in\mathbb{Z}$.

\noindent {\bf Bethe roots}.
The degree of the polynomial $q_{a,s}$ can be found as the difference of the asymptotic power and the power coming from $\Phi_{a,s}^L$. Young diagrams provide an intuitive way of counting Bethe roots, see figure \ref{fig:BRex}. In the right half of the diagram, the Bethe roots in each $\dQ$ equals the number of boxes to the right and above its position. In the left half of the diagram, the number of Bethe roots equals the number of boxes below and to the left. On the central vertical line ($\dQ_{a,2}$), the counting towards the right should be used above the central point ($\dQ_{2,2}$), while the counting towards the left should be used below the central point. 

\begin{figure}[t!]
\centering
\begin{picture}(320,220)

\color{Green}
\put(140,130){\line(1,-1){20}}
\put(139,130){\line(1,-1){20}}\put(139.5,130){\line(1,-1){20}}\put(140.5,130){\line(1,-1){20}}\put(141,130){\line(1,-1){20}}

\linethickness{.8mm}
\put(160,10){\line(0,1){100}}
\put(140,130){\line(0,1){80}}

\color{gray}
\linethickness{.1mm}

\multiput(10,80)(0,20){5}{\line(1,0){300}}
\multiput(150,180)(0,20){2}{\line(1,0){40}}
\multiput(110,20)(0,20){3}{\line(1,0){40}}

\multiput(30,80)(20,0){4}{\line(0,1){80}}
\multiput(210,80)(20,0){5}{\line(0,1){80}}

\put(110,0){\line(0,1){160}}
\put(130,0){\line(0,1){160}}
\put(150,0){\line(0,1){220}}
\put(170,80){\line(0,1){140}}
\put(190,80){\line(0,1){140}}

\color{black}
\linethickness{0.7mm}

\put(150,200){\line(1,0){20}}
\put(150,180){\line(1,0){40}}
\put(30,160){\line(1,0){180}}
\put(30,140){\line(1,0){200}}
\put(50,120){\line(1,0){220}}
\put(90,100){\line(1,0){200}}
\put(110,80){\line(1,0){180}}
\put(110,60){\line(1,0){40}}
\put(110,40){\line(1,0){40}}
\put(130,20){\line(1,0){20}}

\put(30,140){\line(0,1){20}}
\put(50,120){\line(0,1){40}}
\put(70,120){\line(0,1){40}}
\put(90,100){\line(0,1){60}}
\put(110,40){\line(0,1){120}}
\put(130,20){\line(0,1){140}}
\put(150,20){\line(0,1){180}}
\put(170,80){\line(0,1){120}}
\put(190,80){\line(0,1){100}}
\put(210,80){\line(0,1){80}}
\put(230,80){\line(0,1){60}}
\put(250,80){\line(0,1){40}}
\put(270,80){\line(0,1){40}}
\put(290,80){\line(0,1){20}}

\put(150,120){\circle*{6}}

\color{blue}
\footnotesize

\put(170,200){\color{white}\circle*{8}}\put(168,197){\color{gray}0}
\put(150,200){\color{white}\circle*{8}}\put(148,197){\color{gray}0}

\put(190,180){\color{white}\circle*{8}}\put(188,177){\color{gray}0}
\put(170,180){\color{white}\circle*{8}}\put(168,177){\color{gray}0}
\put(150,180){\color{white}\circle*{8}}\put(148,177){1}

\put(210,160){\color{white}\circle*{8}}\put(208,157){\color{gray}0}
\put(190,160){\color{white}\circle*{8}}\put(188,157){\color{gray}0}
\put(170,160){\color{white}\circle*{8}}\put(168,157){1}
\put(150,160){\color{white}\circle*{8}}\put(148,157){3}
\put(130,160){\color{white}\circle*{10}}\put(125,157){14}
\put(110,160){\color{white}\circle*{8}}\put(108,157){8}
\put(90,160){\color{white}\circle*{8}}\put(88,157){5}
\put(70,160){\color{white}\circle*{8}}\put(68,157){3}
\put(50,160){\color{white}\circle*{8}}\put(48,157){1}
\put(30,160){\color{white}\circle*{8}}\put(28,157){\color{gray}0}

\put(230,140){\color{white}\circle*{8}}\put(228,137){\color{gray}0}
\put(210,140){\color{white}\circle*{8}}\put(208,137){\color{gray}0}
\put(190,140){\color{white}\circle*{8}}\put(188,137){1}
\put(170,140){\color{white}\circle*{8}}\put(168,137){3}
\put(150,140){\color{white}\circle*{8}}\put(148,137){6}
\put(130,140){\color{white}\circle*{8}}\put(128,137){9}
\put(110,140){\color{white}\circle*{8}}\put(108,137){4}
\put(90,140){\color{white}\circle*{8}}\put(88,137){2}
\put(70,140){\color{white}\circle*{8}}\put(68,137){1}
\put(50,140){\color{white}\circle*{8}}\put(48,137){\color{gray}0}
\put(30,140){\color{white}\circle*{8}}\put(28,137){\color{gray}0}

\put(270,120){\color{white}\circle*{8}}\put(268,117){\color{gray}0}
\put(250,120){\color{white}\circle*{8}}\put(248,117){\color{gray}0}
\put(230,120){\color{white}\circle*{8}}\put(228,117){\color{gray}0}
\put(210,120){\color{white}\circle*{8}}\put(208,117){1}
\put(190,120){\color{white}\circle*{8}}\put(188,117){3}
\put(170,120){\color{white}\circle*{8}}\put(168,117){6}
\put(150,120){\color{white}\circle*{10}}\put(145,117){10}
\put(130,120){\color{white}\circle*{8}}\put(128,117){5}
\put(110,120){\color{white}\circle*{8}}\put(108,117){1}
\put(90,120){\color{white}\circle*{8}}\put(88,117){\color{gray}0}
\put(70,120){\color{white}\circle*{8}}\put(68,117){\color{gray}0}
\put(50,120){\color{white}\circle*{8}}\put(48,117){\color{gray}0}

\put(290,100){\color{white}\circle*{8}}\put(288,97){\color{gray}0}
\put(270,100){\color{white}\circle*{8}}\put(268,97){\color{gray}0}
\put(250,100){\color{white}\circle*{8}}\put(248,97){1}
\put(230,100){\color{white}\circle*{8}}\put(228,97){2}
\put(210,100){\color{white}\circle*{8}}\put(208,97){4}
\put(190,100){\color{white}\circle*{8}}\put(188,97){7}
\put(170,100){\color{white}\circle*{10}}\put(165,97){11}
\put(150,100){\color{white}\circle*{8}}\put(148,97){7}
\put(130,100){\color{white}\circle*{8}}\put(128,97){3}
\put(110,100){\color{white}\circle*{8}}\put(108,97){\color{gray}0}
\put(90,100){\color{white}\circle*{8}}\put(88,97){\color{gray}0}

\put(290,80){\color{white}\circle*{8}}\put(288,77){\color{gray}0}
\put(270,80){\color{white}\circle*{8}}\put(268,77){1}
\put(250,80){\color{white}\circle*{8}}\put(248,77){3}
\put(230,80){\color{white}\circle*{8}}\put(228,77){5}
\put(210,80){\color{white}\circle*{8}}\put(208,77){8}
\put(190,80){\color{white}\circle*{10}}\put(185,77){12}
\put(170,80){\color{white}\circle*{10}}\put(165,77){17}
\put(150,80){\color{white}\circle*{8}}\put(148,77){5}
\put(130,80){\color{white}\circle*{8}}\put(128,77){2}
\put(110,80){\color{white}\circle*{8}}\put(108,77){\color{gray}0}

\put(150,60){\color{white}\circle*{8}}\put(148,57){3}
\put(130,60){\color{white}\circle*{8}}\put(128,57){1}
\put(110,60){\color{white}\circle*{8}}\put(108,57){\color{gray}0}

\put(150,40){\color{white}\circle*{8}}\put(148,37){1}
\put(130,40){\color{white}\circle*{8}}\put(128,37){\color{gray}0}
\put(110,40){\color{white}\circle*{8}}\put(108,37){\color{gray}0}

\put(150,20){\color{white}\circle*{8}}\put(148,17){\color{gray}0}
\put(130,20){\color{white}\circle*{8}}\put(128,17){\color{gray}0}

\end{picture}
\caption{Number of Bethe roots in the distinguished Q-functions on the Young diagram corresponding to the $\psu(2,2|4)$ multiplet $n^{2222}=[2,3|7,6,4,3|2,1]$. To the right of the {\color{Green}green} line, the number of roots equals the boxes to the right and above the location. To the left of the green line, the number of roots equals the boxes below and to the left of the location.}
\label{fig:BRex}
\end{figure}

\subsubsection*{Rational and non-rational Q-functions}
For non-compact algebras, the Q-system contains non-rational functions. These functions are however restricted to only have poles at $i\mathbb{Z}$, and they can in general be written in terms of $\eta$-functions \cite{Leurent:2012ab,Leurent:2013mr},
\be\label{etadef}
\eta_{k}(u)\equiv\sum_{n=0}^\infty \frac{1}{(u+in)^k}\,,\quad \eta_{k_1,k_2,...}(u)\equiv\sum_{n=0}^\infty \frac{\eta_{k_2,...}(u+i+in)}{(u+in)^{k_1}}\,.
\ee

To understand which Q-functions are non-rational, note that the Young diagram can be seen as two compact diagrams glued together (one of them upside down). The members of the Q-system on the left compact diagram are $Q_{A|J}$, where $a\leq 2$ for any $a\in A$. The members of the Q-system on the right compact diagram are $Q_{\bar B|J}$, where  $b>2$ for all $b\in B$ and $\bar B$ is the complementary set. All these Q-functions are rational, with the only non-polynomial part stemming from the fusion factor. In particular, this includes all the distinguished functions $\dQ$.

The full Young diagram extends these two subsets to a bigger Q-system that includes non-rational Q-functions with poles allowed at $i\mathbb{Z}$. We treat the full Q-system in section \ref{sec:psuQsys}.

\subsubsection*{Good solutions of the full Q-system}
Bethe equations have solutions that do not correspond to spin chain states, and identifying these solutions is quite non-trivial \cite{Hao:2013jqa,Hao:2013rza,Nepomechie:2014hma}. As we discussed in \cite{Marboe:2016yyn}, these unphysical solutions are allowed because the Bethe equations do not guarantee that the full Q-system, but only a small subset, has the right structure. We furthermore proved that for compact Young diagrams polynomiality of the distinguished Q-functions in the Young diagram Q-system implies that all Q-functions are polynomial, which is the requirement for a solution to be physical.

This argument generalises immediately to the rational subset of non-compact Q-systems. If all distinguished Q-functions on the non-compact Young diagram are rational expressions as described above (polynomial times a factor of fused $u^{\pm L}$), then the remaining rational Q-functions also have this structure (note that the symmetries of the Q-system discussed in section \ref{sec:sym} allow to shift the fusion factors completely to one side of the Young diagram).

In practice we observe that this always leads to well-behaved non-rational Q-functions as well, i.e. containing only $\eta$-functions \eqref{etadef} and rational functions in $u$. Although we could not prove this property solely based on QQ-relations, there is a simple counting argument. We can always think about an extended Young diagram as the extension of a compact Young diagram, i.e.\ there is always a bijection between a non-compact and compact Q-systems, and therefore there is simply no room for non-well-behaved solutions. Therefore we conjecture that polynomiality of $q_{a,s}$ in \eqref{Qstruc} for all distinguished Q-functions is a necessary and sufficient condition for a solution of the Q-system to correspond to a physical spin chain multiplet.

\subsection{Algorithm: Distinguished Q-functions on Young diagrams} \label{sec:alg}
We now propose an efficient algorithm that exactly imposes the structure \eqref{Qstruc} on all distinguished Q-functions and finds the corresponding solutions analytically.

\subsubsection*{Step 1: choice of path and ansatz}
Choose a path from the left side of the diagram to the right. On this path, write an arbitrary ansatz for the polynomial parts of the Q-functions:
\be
q_{a,s}= u^{p_{a,s}} + \sum_{k=0}^{p_{a,s}-1} c_{a,s}^{(k)} u^k\,,
\ee
where $p_{a,s}$ is the number of roots in $q_{a,s}$.

It is often advantageous to look for the path from the left boundary to the right boundary of the extended Young diagram on which the least number of roots is encountered. 
See figure \ref{fig:outside} for an example.
\begin{figure}[h!]
\centering
\begin{picture}(320,220)

\color{gray}
\linethickness{.1mm}

\multiput(10,80)(0,20){5}{\line(1,0){300}}
\multiput(150,180)(0,20){2}{\line(1,0){40}}
\multiput(110,20)(0,20){3}{\line(1,0){40}}

\multiput(30,80)(20,0){4}{\line(0,1){80}}
\multiput(210,80)(20,0){5}{\line(0,1){80}}

\put(110,0){\line(0,1){160}}
\put(130,0){\line(0,1){160}}
\put(150,0){\line(0,1){220}}
\put(170,80){\line(0,1){140}}
\put(190,80){\line(0,1){140}}

\color{yellow}
\linethickness{3mm}

\put(110,180){\line(1,0){40}}

\color{orange}
\linethickness{0.7mm}

\put(70,200){\line(1,0){80}}
\put(70,180){\line(1,0){80}}
\put(190,80){\line(1,0){20}}
\put(150,60){\line(1,0){60}}
\put(150,40){\line(1,0){80}}
\put(150,20){\line(1,0){80}}

\put(70,180){\line(0,1){40}}
\put(90,180){\line(0,1){40}}
\put(110,160){\line(0,1){60}}
\put(130,160){\line(0,1){60}}
\put(150,180){\line(0,1){40}}
\put(150,0){\line(0,1){40}}
\put(170,0){\line(0,1){80}}
\put(190,0){\line(0,1){80}}
\put(210,0){\line(0,1){80}}
\put(230,0){\line(0,1){40}}

\color{black}
\linethickness{0.7mm}

\put(150,180){\line(1,0){40}}
\put(110,160){\line(1,0){80}}
\put(110,140){\line(1,0){80}}
\put(110,120){\line(1,0){80}}
\put(110,100){\line(1,0){80}}
\put(110,80){\line(1,0){80}}
\put(130,60){\line(1,0){20}}
\put(130,40){\line(1,0){20}}

\put(110,80){\line(0,1){80}}
\put(130,40){\line(0,1){120}}
\put(150,40){\line(0,1){140}}
\put(170,80){\line(0,1){100}}
\put(190,80){\line(0,1){100}}

\color{blue}
\footnotesize


\put(70,220){\color{white}\circle*{8}}\put(68,217){\color{gray}0}
\put(90,220){\color{white}\circle*{8}}\put(88,217){2}
\put(110,220){\color{white}\circle*{8}}\put(108,217){4}
\put(130,220){\color{white}\circle*{10}}\put(125,217){11}
\put(150,220){\color{white}\circle*{8}}\put(148,217){\color{gray}0}

\put(70,200){\color{white}\circle*{8}}\put(68,197){\color{gray}0}
\put(90,200){\color{white}\circle*{8}}\put(88,197){1}
\put(110,200){\color{white}\circle*{8}}\put(108,197){2}
\put(130,200){\color{white}\circle*{8}}\put(128,197){8}
\put(150,200){\color{white}\circle*{8}}\put(148,197){\color{gray}0}

\put(190,180){\color{white}\circle*{8}}\put(188,177){\color{gray}0}
\put(170,180){\color{white}\circle*{8}}\put(168,177){\color{gray}0}
\put(150,180){\color{white}\circle*{8}}\put(148,177){\color{gray}0}
\put(130,180){\color{white}\circle*{8}}\put(128,177){5}
\put(110,180){\color{white}\circle*{8}}\put(108,177){\color{gray}0}
\put(90,180){\color{white}\circle*{8}}\put(88,177){\color{gray}0}
\put(70,180){\color{white}\circle*{8}}\put(68,177){\color{gray}0}

\put(190,160){\color{white}\circle*{8}}\put(188,157){\color{gray}0}
\put(170,160){\color{white}\circle*{8}}\put(168,157){1}
\put(150,160){\color{white}\circle*{8}}\put(148,157){2}
\put(130,160){\color{white}\circle*{8}}\put(128,157){4}
\put(110,160){\color{white}\circle*{8}}\put(108,157){\color{gray}0}

\put(190,140){\color{white}\circle*{8}}\put(188,137){\color{gray}0}
\put(170,140){\color{white}\circle*{8}}\put(168,137){2}
\put(150,140){\color{white}\circle*{8}}\put(148,137){4}
\put(130,140){\color{white}\circle*{8}}\put(128,137){3}
\put(110,140){\color{white}\circle*{8}}\put(108,137){\color{gray}0}

\put(190,120){\color{white}\circle*{8}}\put(188,117){\color{gray}0}
\put(170,120){\color{white}\circle*{8}}\put(168,117){3}
\put(150,120){\color{white}\circle*{8}}\put(148,117){6}
\put(130,120){\color{white}\circle*{8}}\put(128,117){2}
\put(110,120){\color{white}\circle*{8}}\put(108,117){\color{gray}0}

\put(190,100){\color{white}\circle*{8}}\put(188,97){\color{gray}0}
\put(170,100){\color{white}\circle*{8}}\put(168,97){4}
\put(150,100){\color{white}\circle*{8}}\put(148,97){4}
\put(130,100){\color{white}\circle*{8}}\put(128,97){1}
\put(110,100){\color{white}\circle*{8}}\put(108,97){\color{gray}0}

\put(210,80){\color{white}\circle*{8}}\put(208,77){\color{gray}0}
\put(190,80){\color{white}\circle*{8}}\put(188,77){\color{gray}0}
\put(170,80){\color{white}\circle*{8}}\put(168,77){5}
\put(150,80){\color{white}\circle*{8}}\put(148,77){2}
\put(130,80){\color{white}\circle*{8}}\put(128,77){\color{gray}0}
\put(110,80){\color{white}\circle*{8}}\put(108,77){\color{gray}0}

\put(210,60){\color{white}\circle*{8}}\put(208,57){\color{gray}0}
\put(190,60){\color{white}\circle*{8}}\put(188,57){1}
\put(170,60){\color{white}\circle*{8}}\put(168,57){7}
\put(150,60){\color{white}\circle*{8}}\put(148,57){1}
\put(130,60){\color{white}\circle*{8}}\put(128,57){\color{gray}0}

\put(230,40){\color{white}\circle*{8}}\put(228,37){\color{gray}0}
\put(210,40){\color{white}\circle*{8}}\put(208,37){\color{gray}0}
\put(190,40){\color{white}\circle*{8}}\put(188,37){2}
\put(170,40){\color{white}\circle*{8}}\put(168,37){9}
\put(150,40){\color{white}\circle*{8}}\put(148,37){\color{gray}0}
\put(130,40){\color{white}\circle*{8}}\put(128,37){\color{gray}0}

\put(230,20){\color{white}\circle*{8}}\put(228,17){\color{gray}0}
\put(210,20){\color{white}\circle*{8}}\put(208,17){1}
\put(190,20){\color{white}\circle*{8}}\put(188,17){4}
\put(170,20){\color{white}\circle*{10}}\put(165,17){12}
\put(150,20){\color{white}\circle*{8}}\put(148,17){\color{gray}0}

\put(230,0){\color{white}\circle*{8}}\put(228,-3){\color{gray}0}
\put(210,0){\color{white}\circle*{8}}\put(208,-3){2}
\put(190,0){\color{white}\circle*{8}}\put(188,-3){6}
\put(170,0){\color{white}\circle*{10}}\put(165,-3){15}
\put(150,0){\color{white}\circle*{8}}\put(148,-3){\color{gray}0}

\end{picture}
\caption{For $n^{2222}=[0,2|2,2,2,2|1,1]$ the path with the minimal number of roots ({\color{black}yellow}) lies outside the Young diagram.}
\label{fig:outside}
\end{figure}
The number of roots increases monotonically in the infinite extensions once the columns are aligned completely to the left or right of the central vertical line, and thus the path with least roots is always within the non-trivial extension.

\subsubsection*{Step 2: remaining $\dQ$ from fermionic QQ-relations}

Generate the remaining $\dQ$ from fermionic QQ-relations \eqref{QQ3}:
\begin{eqnarray}
q_{a,s} \propto \frac{1}{f_{a,s}} \frac{{\dQ}_{a\pm1,s}^+{\dQ}_{a,s\mp 1}^--{\dQ}_{a\pm 1,s}^-{\dQ}_{a,s\mp 1}^+}{{\dQ}_{a\pm 1,s\mp 1}}\,.
\end{eqnarray}
The unknown Q-function is a ratio of two polynomials, but is required to be a polynomial itself, and thus it can be assigned the quotient of the polynomial division of the numerator by the denominator. The remainder of this polynomial division 
should vanish, but it is not necessary to impose this yet. 

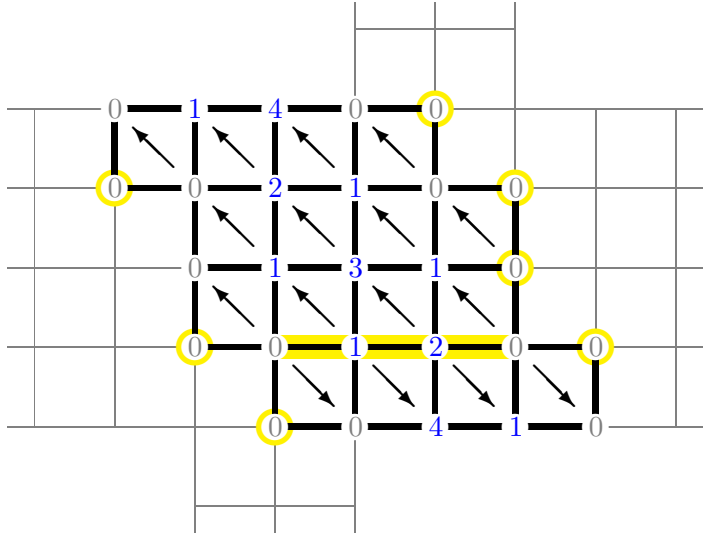
\begin{figure}[t!]
\centering
\begin{picture}(260,200)

\color{gray}
\linethickness{.1mm}

\multiput(0,40)(0,30){5}{\line(1,0){260}}
\multiput(10,40)(30,0){9}{\line(0,1){120}}

\put(130,190){\line(1,0){60}}
\put(70,10){\line(1,0){60}}

\put(70,0){\line(0,1){40}}
\put(100,0){\line(0,1){40}}
\put(130,0){\line(0,1){40}}

\put(130,160){\line(0,1){40}}
\put(160,160){\line(0,1){40}}
\put(190,160){\line(0,1){40}}

\color{yellow}
\linethickness{3mm}

\put(100,70){\line(1,0){90}}

\put(220,70){\color{yellow}\circle*{14}}
\put(100,40){\color{yellow}\circle*{14}}
\put(70,70){\color{yellow}\circle*{14}}
\put(40,130){\color{yellow}\circle*{14}}
\put(190,100){\color{yellow}\circle*{14}}
\put(190,130){\color{yellow}\circle*{14}}
\put(160,160){\color{yellow}\circle*{14}}

\color{black}
\linethickness{0.7mm}

\put(40,160){\line(1,0){120}}
\put(40,130){\line(1,0){150}}
\put(70,100){\line(1,0){120}}
\put(70,70){\line(1,0){150}}
\put(100,40){\line(1,0){120}}

\put(40,130){\line(0,1){30}}
\put(70,70){\line(0,1){90}}
\put(100,40){\line(0,1){120}}
\put(130,40){\line(0,1){120}}
\put(160,40){\line(0,1){120}}
\put(190,40){\line(0,1){90}}
\put(220,40){\line(0,1){30}}

\put(40,130){\line(0,1){30}}
\put(40,130){\line(0,1){30}}

\color{blue}
\put(40,160){\color{white}\circle*{10}}\put(37.5,156.5){\color{gray}0}
\put(40,130){\color{white}\circle*{10}}\put(37.5,126.5){\color{gray}0}

\put(70,160){\color{white}\circle*{10}}\put(67.5,156.5){1}
\put(70,130){\color{white}\circle*{10}}\put(67.5,126.5){\color{gray}0}
\put(70,100){\color{white}\circle*{10}}\put(67.5,96.5){\color{gray}0}
\put(70,70){\color{white}\circle*{10}}\put(67.5,66.5){\color{gray}0}

\put(100,160){\color{white}\circle*{10}}\put(97.5,156.5){4}
\put(100,130){\color{white}\circle*{10}}\put(97.5,126.5){2}
\put(100,100){\color{white}\circle*{10}}\put(97.5,96.5){1}
\put(100,70){\color{white}\circle*{10}}\put(97.5,66.5){\color{gray}0}
\put(100,40){\color{white}\circle*{10}}\put(97.5,36.5){\color{gray}0}

\put(130,160){\color{white}\circle*{10}}\put(127.5,156.5){\color{gray}0}
\put(130,130){\color{white}\circle*{10}}\put(127.5,126.5){1}
\put(130,100){\color{white}\circle*{10}}\put(127.5,96.5){3}
\put(130,70){\color{white}\circle*{10}}\put(127.5,66.5){1}
\put(130,40){\color{white}\circle*{10}}\put(127.5,36.5){\color{gray}0}

\put(160,160){\color{white}\circle*{10}}\put(157.5,156.5){\color{gray}0}
\put(160,130){\color{white}\circle*{10}}\put(157.5,126.5){\color{gray}0}
\put(160,100){\color{white}\circle*{10}}\put(157.5,96.5){1}
\put(160,70){\color{white}\circle*{10}}\put(157.5,66.5){2}
\put(160,40){\color{white}\circle*{10}}\put(157.5,36.5){4}

\put(190,130){\color{white}\circle*{10}}\put(187.5,126.5){\color{gray}0}
\put(190,100){\color{white}\circle*{10}}\put(187.5,96.5){\color{gray}0}
\put(190,70){\color{white}\circle*{10}}\put(187.5,66.5){\color{gray}0}
\put(190,40){\color{white}\circle*{10}}\put(187.5,36.5){1}

\put(220,70){\color{white}\circle*{10}}\put(217.5,66.5){\color{gray}0}
\put(220,40){\color{white}\circle*{10}}\put(217.5,36.5){\color{gray}0}

\color{black}
\thicklines
\put(107,63){\vector(1,-1){15}}
\put(137,63){\vector(1,-1){15}}
\put(167,63){\vector(1,-1){15}}
\put(197,63){\vector(1,-1){15}}

\put(92,78){\vector(-1,1){15}}
\put(122,78){\vector(-1,1){15}}
\put(152,78){\vector(-1,1){15}}
\put(182,78){\vector(-1,1){15}}

\put(92,108){\vector(-1,1){15}}
\put(122,108){\vector(-1,1){15}}
\put(152,108){\vector(-1,1){15}}
\put(182,108){\vector(-1,1){15}}

\put(92,138){\vector(-1,1){15}}
\put(122,138){\vector(-1,1){15}}
\put(152,138){\vector(-1,1){15}}
\put(62,138){\vector(-1,1){15}}

\end{picture}
\caption{Example: generating all $\dQ$ on the Young diagram corresponding to the $\psu(2,2|4)$ multiplet $n^{2222}=[0,0|3,2,2,1|0,0]$ from a path that minimises the number of Bethe roots. The $\dQ$'s at the positions encircled in yellow are set to 1 by default.}
\label{fig:allDQ}
\end{figure}

In this way all distinguished Q-functions are generated in terms of the coefficients $c_{a,s}^{(k)}$ that were introduced on the path. See figure \ref{fig:allDQ} for an example. All the remainders of the polynomial divisions are collected.


\subsubsection*{Step 3: solve polynomiality constraints}
The final step of the algorithm is to simultaneously impose that all remainders of the polynomial divisions vanish. This completely fixes $c_{a,s}^{(k)}$. This set of algebraic equations can be solved efficiently in most symbolical programming languages when the total number of unknown Bethe roots is not too high. In practice, the solution is usually effortless when the number of Bethe roots is less than ten. Note that the algorithm finds {\it exactly} the expected number of solutions, i.e.\ it is not necessary to discard any of the obtained solutions in contrast to the solutions of Bethe equations. In table \ref{tab:solfound} and \ref{ap:spec} we mark the multiplets with $\Delta_0^{2222}\le 8$ for which we were able to generate the Q-system in less than 15 minutes on a standard laptop with our general implementation of the algorithm. This includes all 495 multiplets with $\Delta_0^{2222}\le \frac{13}{2}$. 



\begin{table}[t!]
\centering
\begin{tabular}{|l||c|c||c|c|} \hline
\multirow{2}{*}{$\Delta_0^{2222}$} & \multicolumn{2}{c||}{Diagrams}  & \multicolumn{2}{c|}{Solutions} \\ 
& solved & total & found & total\\\hline\hline
2 & {\color{black}1} & 1 & {\color{black}1} & 1 \\\hline
3 & {\color{black}1} & 1 & {\color{black}1} & 1 \\\hline
4 & {\color{black}7} & 7 & {\color{black}10} & 10 \\\hline
5 & {\color{black}13} & 13 & {\color{black}27} & 27 \\\hline
5.5& {\color{black}12} & 12 & {\color{black}36} & 36 \\\hline
6& {\color{black}39} & 39 & {\color{black}144} & 144 \\\hline
6.5& {\color{black}36} & 36 & {\color{black}276} & 276 \\\hline
7 & {\color{black}68} & 77 & {\color{black}600} & 918  \\\hline
7.5& {\color{black}54} & 84 & {\color{black}694} & 2204 \\\hline
8& {\color{black}107} & 180 & {\color{black}1395} & 6918 \\\hline
\end{tabular}
\caption{Solutions found with our \texttt{Mathematica}-implementation of the algorithm. 15 minutes of computation time (on a 1.8GHz laptop with 4GB memory) were allowed per diagram. The explicit multiplet content for each value of $\Delta_0$ is given in \ref{ap:spec}.}
\label{tab:solfound}
\end{table}

\subsubsection*{Remark: paths without roots}
As a side remark, note that in specific cases there are paths without any Bethe roots, and the Q-system is completely fixed without solving any equations.

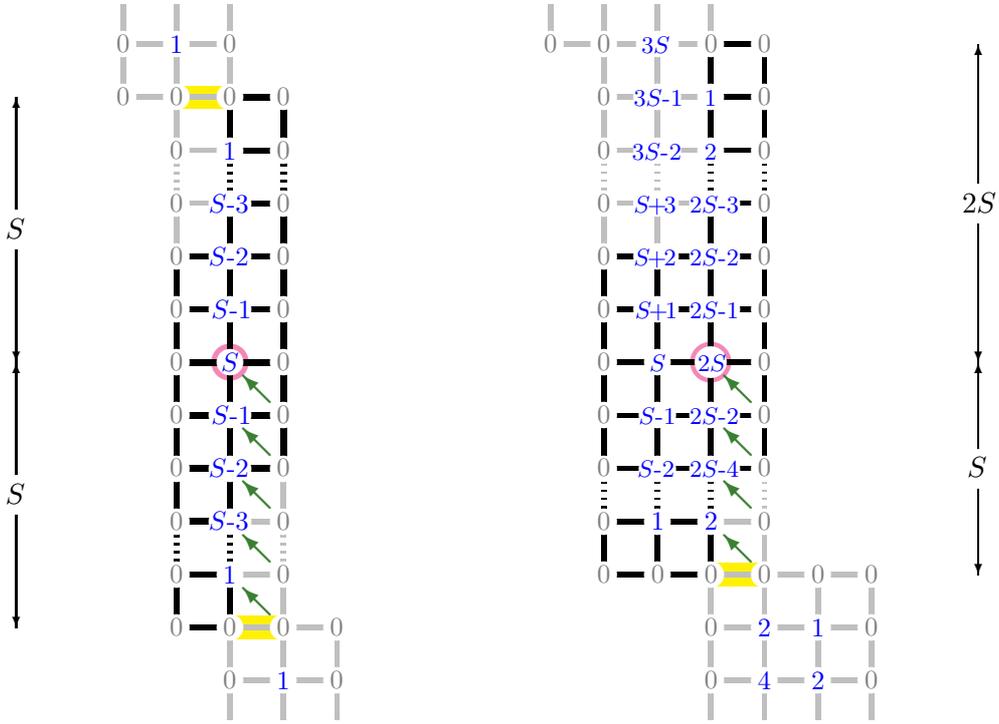
\begin{figure}[t!]
\centering
\begin{picture}(300,275)

\color{magenta!60}
\put(50,140){\circle*{14}}
\put(230,140){\circle*{15}}
\put(230,140){\circle{15.5}}

\color{black}
\put(-30,40){\vector(0,1){100}}
\put(-30,140){\vector(0,-1){100}}
\put(-30,140){\vector(0,1){100}}
\put(-30,240){\vector(0,-1){100}}

\put(330,60){\vector(0,1){80}}
\put(330,140){\vector(0,-1){80}}
\put(330,140){\vector(0,1){120}}
\put(330,260){\vector(0,-1){120}}

\put(-30,90){\color{white}\circle*{15}}\put(-34,86.7){$S$}
\put(-30,190){\color{white}\circle*{15}}\put(-34,186.7){$S$}

\put(330,100){\color{white}\circle*{15}}\put(326,96.7){$S$}
\put(330,200){\color{white}\circle*{15}}\put(324,196.7){$2S$}


\color{yellow}
\linethickness{3mm}
\put(30,240){\line(1,0){20}}
\put(50,40){\line(1,0){20}}

\color{gray!50}
\linethickness{0.6mm}

\put(10,260){\line(1,0){40}}
\put(10,240){\line(1,0){40}}
\put(30,220){\line(1,0){40}}
\put(30,200){\line(1,0){20}}
\put(30,180){\line(1,0){20}}
\put(50,80){\line(1,0){20}}
\put(50,60){\line(1,0){20}}
\put(50,40){\line(1,0){40}}
\put(50,20){\line(1,0){40}}

\put(10,240){\line(0,1){35}}
\put(30,180){\line(0,1){95}}
\put(50,5){\line(0,1){270}}
\put(70,5){\line(0,1){95}}
\put(90,5){\line(0,1){35}}

\color{black}
\linethickness{0.7mm}

\put(50,240){\line(1,0){20}}
\put(50,220){\line(1,0){20}}
\put(50,200){\line(1,0){20}}
\put(30,180){\line(1,0){40}}
\put(30,160){\line(1,0){40}}
\put(30,140){\line(1,0){40}}
\put(30,120){\line(1,0){40}}
\put(30,100){\line(1,0){40}}
\put(30,80){\line(1,0){20}}
\put(30,60){\line(1,0){20}}
\put(30,40){\line(1,0){20}}

\put(30,40){\line(0,1){140}}
\put(50,40){\line(0,1){200}}
\put(70,100){\line(0,1){140}}


\color{white}\linethickness{0.5mm}
\put(0,213){\line(1,0){80}}
\put(0,210){\line(1,0){80}}
\put(0,207){\line(1,0){80}}

\put(0,73){\line(1,0){80}}
\put(0,70){\line(1,0){80}}
\put(0,67){\line(1,0){80}}

\color{blue}
\small

\put(30,260){\color{white}\circle*{10}}\put(27.5,256.7){1}

\put(50,220){\color{white}\circle*{10}}\put(47.5,216.7){1}

\put(50,200){\color{white}\circle*{10}}\put(47,200){\color{white}\circle*{10}}\put(53,200){\color{white}\circle*{10}}\put(42,196.7){$S\mi3$}

\put(50,180){\color{white}\circle*{10}}\put(47,180){\color{white}\circle*{10}}\put(53,180){\color{white}\circle*{10}}\put(42,176.7){$S\mi2$}

\put(50,160){\color{white}\circle*{10}}\put(47,160){\color{white}\circle*{10}}\put(53,160){\color{white}\circle*{10}}\put(43,156.7){$S\mi1$}

\put(50,140){\color{white}\circle*{10}}\put(46.6,136.7){$S$}

\put(50,120){\color{white}\circle*{10}}\put(47,120){\color{white}\circle*{10}}\put(53,120){\color{white}\circle*{10}}\put(43,116.7){$S\mi1$}

\put(50,100){\color{white}\circle*{10}}\put(47,100){\color{white}\circle*{10}}\put(53,100){\color{white}\circle*{10}}\put(42,96.7){$S\mi2$}

\put(50,80){\color{white}\circle*{10}}\put(47,80){\color{white}\circle*{10}}\put(53,80){\color{white}\circle*{10}}\put(42,76.7){$S\mi3$}

\put(50,60){\color{white}\circle*{10}}\put(47.5,56.7){1}

\put(70,20){\color{white}\circle*{10}}\put(67.5,16.7){1}

\color{gray}

\put(10,260){\color{white}\circle*{10}}\put(7.5,257){0}
\put(50,260){\color{white}\circle*{10}}\put(47.5,257){0}

\put(10,240){\color{white}\circle*{10}}\put(7.5,237){0}
\put(30,240){\color{white}\circle*{10}}\put(27.5,237){0}
\put(50,240){\color{white}\circle*{10}}\put(47.5,237){0}
\put(70,240){\color{white}\circle*{10}}\put(67.5,237){0}

\put(30,220){\color{white}\circle*{10}}\put(27.5,217){0}
\put(70,220){\color{white}\circle*{10}}\put(67.5,217){0}

\put(30,200){\color{white}\circle*{10}}\put(27.5,197){0}
\put(70,200){\color{white}\circle*{10}}\put(67.5,197){0}

\put(30,180){\color{white}\circle*{10}}\put(27.5,177){0}
\put(70,180){\color{white}\circle*{10}}\put(67.5,177){0}

\put(30,160){\color{white}\circle*{10}}\put(27.5,157){0}
\put(70,160){\color{white}\circle*{10}}\put(67.5,157){0}

\put(30,140){\color{white}\circle*{10}}\put(27.5,137){0}
\put(70,140){\color{white}\circle*{10}}\put(67.5,137){0}

\put(30,120){\color{white}\circle*{10}}\put(27.5,117){0}
\put(70,120){\color{white}\circle*{10}}\put(67.5,117){0}

\put(30,100){\color{white}\circle*{10}}\put(27.5,97){0}
\put(70,100){\color{white}\circle*{10}}\put(67.5,97){0}

\put(30,80){\color{white}\circle*{10}}\put(27.5,77){0}
\put(70,80){\color{white}\circle*{10}}\put(67.5,77){0}

\put(30,60){\color{white}\circle*{10}}\put(27.5,57){0}
\put(70,60){\color{white}\circle*{10}}\put(67.5,57){0}

\put(30,40){\color{white}\circle*{10}}\put(27.5,37){0}
\put(50,40){\color{white}\circle*{10}}\put(47.5,37){0}
\put(70,40){\color{white}\circle*{10}}\put(67.5,37){0}
\put(90,40){\color{white}\circle*{10}}\put(87.5,37){0}

\put(50,20){\color{white}\circle*{10}}\put(47.5,17){0}
\put(90,20){\color{white}\circle*{10}}\put(87.5,17){0}

\color{OliveGreen}
\thicklines
\put(65,45){\vector(-1,1){10}}
\put(65,65){\vector(-1,1){10}}
\put(65,85){\vector(-1,1){10}}
\put(65,105){\vector(-1,1){10}}
\put(65,125){\vector(-1,1){10}}


\color{yellow}
\linethickness{3mm}
\put(230,60){\line(1,0){20}}

\color{gray!50}
\linethickness{0.6mm}

\put(170,260){\line(1,0){60}}
\put(190,240){\line(1,0){40}}
\put(190,220){\line(1,0){40}}
\put(190,200){\line(1,0){40}}
\put(190,180){\line(1,0){40}}
\put(230,80){\line(1,0){20}}
\put(230,60){\line(1,0){60}}
\put(230,40){\line(1,0){60}}
\put(230,20){\line(1,0){60}}

\put(170,260){\line(0,1){15}}
\put(190,180){\line(0,1){95}}
\put(210,180){\line(0,1){95}}
\put(230,5){\line(0,1){270}}
\put(250,5){\line(0,1){95}}
\put(270,5){\line(0,1){55}}
\put(290,5){\line(0,1){55}}

\color{black}
\linethickness{0.6mm}

\put(230,260){\line(1,0){20}}
\put(230,240){\line(1,0){20}}
\put(230,220){\line(1,0){20}}
\put(230,200){\line(1,0){20}}
\put(190,180){\line(1,0){60}}
\put(190,160){\line(1,0){60}}
\put(190,140){\line(1,0){60}}
\put(190,120){\line(1,0){60}}
\put(190,100){\line(1,0){60}}
\put(190,80){\line(1,0){40}}
\put(190,60){\line(1,0){40}}

\put(190,60){\line(0,1){120}}
\put(210,60){\line(0,1){120}}
\put(230,60){\line(0,1){200}}
\put(250,100){\line(0,1){160}}

\color{white}
\put(180,213){\line(1,0){80}}
\put(180,210){\line(1,0){80}}
\put(180,207){\line(1,0){80}}

\put(180,93){\line(1,0){80}}
\put(180,90){\line(1,0){80}}
\put(180,87){\line(1,0){80}}

\color{blue}
\small

\put(210,260){\color{white}\circle*{10}}\put(207,260){\color{white}\circle*{10}}\put(213,260){\color{white}\circle*{10}}\put(204,256.7){\footnotesize$3S$}

\put(210,240){\color{white}\circle*{10}}\put(206,240){\color{white}\circle*{10}}\put(214,240){\color{white}\circle*{10}}\put(201,236.7){\footnotesize$3S\mi1$}
\put(230,240){\color{white}\circle*{10}}\put(227.5,236.7){\footnotesize 1}

\put(210,220){\color{white}\circle*{10}}\put(206,220){\color{white}\circle*{10}}\put(214,220){\color{white}\circle*{10}}\put(200.5,216.7){\footnotesize$3S\mi2$}
\put(230,220){\color{white}\circle*{10}}\put(227.5,216.7){\footnotesize 2}

\put(210,200){\color{white}\circle*{10}}\put(206,200){\color{white}\circle*{10}}\put(213,200){\color{white}\circle*{10}}\put(201,196.7){\footnotesize$S\!\!+\!\!3$}
\put(230,200){\color{white}\circle*{10}}\put(227,200){\color{white}\circle*{10}}\put(236,200){\color{white}\circle*{10}}\put(222,196.7){\footnotesize$2S\mi3$}

\put(210,180){\color{white}\circle*{10}}\put(206,180){\color{white}\circle*{10}}\put(213,180){\color{white}\circle*{10}}\put(201,176.7){\footnotesize$S\!\!+\!\!2$}
\put(230,180){\color{white}\circle*{10}}\put(227,180){\color{white}\circle*{10}}\put(236,180){\color{white}\circle*{10}}\put(222,176.7){\footnotesize$2S\mi2$}

\put(210,160){\color{white}\circle*{10}}\put(206,160){\color{white}\circle*{10}}\put(213,160){\color{white}\circle*{10}}\put(201.5,156.7){\footnotesize$S\!\!+\!\!1$}
\put(230,160){\color{white}\circle*{10}}\put(227,160){\color{white}\circle*{10}}\put(236,160){\color{white}\circle*{10}}\put(222,156.7){\footnotesize$2S\mi1$}

\put(210,140){\color{white}\circle*{10}}\put(207,136.7){\footnotesize$S$}
\put(230,140){\color{white}\circle*{12}}\put(225,136.7){\footnotesize$2S$}

\put(210,120){\color{white}\circle*{10}}\put(208,120){\color{white}\circle*{10}}\put(212,120){\color{white}\circle*{10}}\put(203,116.7){\footnotesize$S\mi1$}
\put(230,120){\color{white}\circle*{10}}\put(227,120){\color{white}\circle*{10}}\put(236,120){\color{white}\circle*{10}}\put(222,116.7){\footnotesize$2S\mi2$}

\put(210,100){\color{white}\circle*{10}}\put(208,100){\color{white}\circle*{10}}\put(212,100){\color{white}\circle*{10}}\put(202.5,96.7){\footnotesize$S\mi2$}
\put(230,100){\color{white}\circle*{10}}\put(227,100){\color{white}\circle*{10}}\put(236,100){\color{white}\circle*{10}}\put(222,96.7){\footnotesize$2S\mi4$}

\put(210,80){\color{white}\circle*{10}}\put(207.5,76.7){1}
\put(230,80){\color{white}\circle*{10}}\put(227.5,76.7){2}

\put(250,40){\color{white}\circle*{10}}\put(247.5,36.7){2}
\put(270,40){\color{white}\circle*{10}}\put(267.5,36.7){1}

\put(250,20){\color{white}\circle*{10}}\put(247.5,16.7){4}
\put(270,20){\color{white}\circle*{10}}\put(267.5,16.7){2}

\color{gray}

\put(170,260){\color{white}\circle*{10}}\put(167.5,257){0}
\put(190,260){\color{white}\circle*{10}}\put(187.5,257){0}
\put(230,260){\color{white}\circle*{10}}\put(227.5,257){0}
\put(250,260){\color{white}\circle*{10}}\put(247.5,257){0}

\put(190,240){\color{white}\circle*{10}}\put(187.5,237){0}
\put(250,240){\color{white}\circle*{10}}\put(247.5,237){0}

\put(190,220){\color{white}\circle*{10}}\put(187.5,217){0}
\put(250,220){\color{white}\circle*{10}}\put(247.5,217){0}

\put(190,200){\color{white}\circle*{10}}\put(187.5,197){0}
\put(250,200){\color{white}\circle*{10}}\put(247.5,197){0}

\put(190,180){\color{white}\circle*{10}}\put(187.5,177){0}
\put(250,180){\color{white}\circle*{10}}\put(247.5,177){0}

\put(190,160){\color{white}\circle*{10}}\put(187.5,157){0}
\put(250,160){\color{white}\circle*{10}}\put(247.5,157){0}

\put(190,140){\color{white}\circle*{10}}\put(187.5,137){0}
\put(250,140){\color{white}\circle*{10}}\put(247.5,137){0}

\put(190,120){\color{white}\circle*{10}}\put(187.5,117){0}
\put(250,120){\color{white}\circle*{10}}\put(247.5,117){0}

\put(190,100){\color{white}\circle*{10}}\put(187.5,97){0}
\put(250,100){\color{white}\circle*{10}}\put(247.5,97){0}

\put(190,80){\color{white}\circle*{10}}\put(187.5,77){0}
\put(250,80){\color{white}\circle*{10}}\put(247.5,77){0}

\put(190,60){\color{white}\circle*{10}}\put(187.5,57){0}
\put(210,60){\color{white}\circle*{10}}\put(207.5,57){0}
\put(230,60){\color{white}\circle*{10}}\put(227.5,57){0}
\put(250,60){\color{white}\circle*{10}}\put(247.5,57){0}
\put(270,60){\color{white}\circle*{10}}\put(267.5,57){0}
\put(290,60){\color{white}\circle*{10}}\put(287.5,57){0}

\put(230,40){\color{white}\circle*{10}}\put(227.5,37){0}
\put(290,40){\color{white}\circle*{10}}\put(287.5,37){0}

\put(230,20){\color{white}\circle*{10}}\put(227.5,17){0}
\put(290,20){\color{white}\circle*{10}}\put(287.5,17){0}

\color{OliveGreen}
\thicklines
\put(245,65){\vector(-1,1){10}}
\put(245,85){\vector(-1,1){10}}
\put(245,105){\vector(-1,1){10}}
\put(245,125){\vector(-1,1){10}}

\end{picture}
\vspace{-2mm}
\caption{Two examples of Young diagrams for which there exists a path between the left and right boundary of the diagram where no Bethe roots are encountered. Consequently, the Q-system is unique and can be generated trivially from QQ-relations. The left diagram corresponds to \eqref{tw2n} and the right to \eqref{L3ser}. 
The number of roots in the distinguished Q-functions are shown, and the central nodes are {\color{magenta!70}encircled}. Paths with no Bethe roots are marked in yellow. The {\color{OliveGreen}green} arrows show how to generate the central Q-functions via QQ-relations.}
\label{fig:t2YD}
\end{figure}

The first example is all multiplets containing operators of length two, corresponding to the oscillator numbers
\be\label{tw2n}
n^{2222}=[0,S\!-\!2|1,1,1,1|S\!-\!2,0]\,.
\ee
Note that the HWS in the grading $1133$ has the form $\mathcal{D}_{12}^S\ZZ^2$, i.e. it is the $\sl(2)$ twist-two operator with spin $S$. As seen in figure \ref{fig:t2YD}, there exist paths between the left and right boundaries of the diagram without any roots, and the full Q-system can thus be generated without solving algebraic equations from the trivial ansatz on such a path. The central Q-function can be written as 
\be
\dQ_{2,2}(S)=\nabla^S \prod_{k=1}^S \left(u+\frac i2-ik \right)^2 \,,
\ee
where we define the difference operator, $\nabla$, by $\nabla f(u) \equiv f\left(u\right)-f\left(u+i\right)$. 
The obtained result coincides with the known answer in terms of Hahn polynomials \cite{Faddeev:1994zg,Korchemsky:1994um,Eden:2006rx}, and it solves the Baxter equation
\be
\left(u+\frac{i}{2}\right)^2 \dQ^{[2]}+\left(u-\frac{i}{2}\right)^2 \dQ^{[-2]} + \left( -2u^2 +S(S+1) +\frac{1}{2} \right) \dQ = 0\,.
\ee

A second example in the $\mathcal{N}=4$ SYM spectrum are multiplets containing length three states with oscillator numbers
\be \label{L3ser}
n^{2222}=[S\!-\!2,S\!-\!2|1,1,1,1|2S\!-\!2,0]\,.
\ee
The corresponding Young diagram is shown in figure \ref{fig:t2YD}. Note that there is an equivalent series of multiplets with $n^{2222}=[0,2S-2|2,2,2,2|S-2,S-2]$ corresponding to a rotation of the diagram by $\pi$. 
The multiplets \eqref{L3ser} contain operators from the $su(2,1|2)$ sector, and the HWS in the grading 0033  has the field content $\DD_{11}^{S}\DD_{12}^{S} \ZZ^3$. 
As depicted in figure \ref{fig:t2YD}, the central Q-function can be generated immediately and has the form
\be
\dQ_{2,2}(S) \propto \nabla^{S} \prod_{k=1}^{S} \left(u+\frac{i}{2} -i k \right)^3 \,.
\ee


Note that the often studied twist-three operators, corresponding to oscillator numbers $n^{2222}=[0,S-2|2,2,1,1|S-2,0]$, for which the central Q-function can be written as a Wilson polynomial, do not have this property. Indeed, multiple solutions exist for these quantum numbers.

\subsection{Transfer from Young diagram Q-system to $\psu(2,2|4)$ Q-system}\label{sec:Qshort}
The quantum spectral curve at finite coupling is formulated in terms of the $\psu(2,2|4)$ Q-system, whereas the Q-system on Young diagrams is not  accessible given that the weights become non-integers\footnote{One can define Young diagrams with non-integer weights, and even use Schwinger oscillators to construct unitary representations \cite{Gunaydin:2017lhg}. But there are several things that complicate matters: Fock space itself depends on the anomalous dimension (there is no simple tensor product structure anymore), QQ-relations on the diagrams require integer spacings, and the analytic structure of the Q-functions is no longer simple for the distinguished $\dQ$ that we would expect to use. Finally recall that the symmetry reduces to $\psu(2,2|4)$ whereas Young diagrams reflect $\pu(2,2|4)\oplus \algu(1)$ symmetry. Therefore, at least at our current level of understanding, we cannot define and use Q-systems on Young diagrams at finite coupling.}.

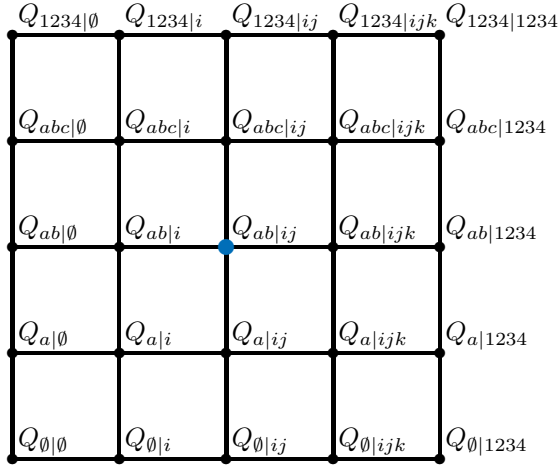
\begin{figure}[t]
\centering
\begin{picture}(200,180)

\linethickness{0.4mm}

\put(10,10){\line(1,0){160}}
\put(10,50){\line(1,0){160}}
\put(10,90){\line(1,0){160}}
\put(10,130){\line(1,0){160}}
\put(10,170){\line(1,0){160}}

\put(10,10){\line(0,1){160}}
\put(50,10){\line(0,1){160}}
\put(90,10){\line(0,1){160}}
\put(130,10){\line(0,1){160}}
\put(170,10){\line(0,1){160}}

\put(10,10){\circle*{4}}
\put(10,50){\circle*{4}}
\put(10,90){\circle*{4}}
\put(10,130){\circle*{4}}
\put(10,170){\circle*{4}}
\put(50,10){\circle*{4}}
\put(50,50){\circle*{4}}
\put(50,90){\circle*{4}}
\put(50,130){\circle*{4}}
\put(50,170){\circle*{4}}
\put(90,10){\circle*{4}}
\put(90,50){\circle*{4}}
\put(90,90){\color{NavyBlue}\circle*{6}}
\put(90,130){\circle*{4}}
\put(90,170){\circle*{4}}
\put(130,10){\circle*{4}}
\put(130,50){\circle*{4}}
\put(130,90){\circle*{4}}
\put(130,130){\circle*{4}}
\put(130,170){\circle*{4}}
\put(170,10){\circle*{4}}
\put(170,50){\circle*{4}}
\put(170,90){\circle*{4}}
\put(170,130){\circle*{4}}
\put(170,170){\circle*{4}}

\small
\put(12,15){$Q_{\emptyset|\emp}$}
\put(52,15){$Q_{\emptyset|i}$}
\put(92,15){$Q_{\emptyset|ij}$}
\put(132,15){$Q_{\emptyset|ijk}$}
\put(172,15){$Q_{\emptyset|1234}$}

\put(12,55){$Q_{a|\emptyset}$}
\put(52,55){$Q_{a|i}$}
\put(92,55){$Q_{a|ij}$}
\put(132,55){$Q_{a|ijk}$}
\put(172,55){$Q_{a|1234}$}

\put(12,95){$Q_{ab|\emptyset}$}
\put(52,95){$Q_{ab|i}$}
\put(92,95){$Q_{ab|ij}$}
\put(132,95){$Q_{ab|ijk}$}
\put(172,95){$Q_{ab|1234}$}

\put(12,135){$Q_{abc|\emptyset}$}
\put(52,135){$Q_{abc|i}$}
\put(92,135){$Q_{abc|ij}$}
\put(132,135){$Q_{abc|ijk}$}
\put(172,135){$Q_{abc|1234}$}

\put(12,175){$Q_{1234|\emptyset}$}
\put(52,175){$Q_{1234|i}$}
\put(92,175){$Q_{1234|ij}$}
\put(132,175){$Q_{1234|ijk}$}
\put(172,175){$Q_{1234|1234}$}

\end{picture}
\caption{The $\psu(2,2|4)$ Q-system. The momentum-carrying node is marked in {\color{NavyBlue}blue}.}\label{fig:psu224Q}
\end{figure}

The $\psu(2,2|4)$ Q-system consists of 256 Q-functions and can be written on a $4\times4$ square, see figure \ref{fig:psu224Q}. Importantly, the following constraint should be satisfied:
\be\label{qdet}
Q_{\emptyset|\emptyset}=Q_{1234|1234}=1\,.
\ee
Our goal is to generate all the 256 Q-functions at zero coupling from distinguished Q-functions on the Young diagram. In this section we explain how to get the distinguished Q-functions on the $4\times 4$ square, whereas the next section explains how to find all other ones.

\begin{figure}[t]
\centering
\begin{picture}(100,80)

\color{NavyBlue!20}
\linethickness{0.4mm}
\multiput(20,20)(0,1){40}{\line(1,0){40}}

\color{gray}
\linethickness{.1mm}

\multiput(0,20)(0,10){5}{\line(1,0){80}}
\put(20,10){\line(1,0){20}}
\put(40,70){\line(1,0){20}}
\put(10,20){\line(0,1){40}}
\put(20,0){\line(0,1){60}}
\put(30,0){\line(0,1){60}}
\put(40,0){\line(0,1){80}}
\put(50,20){\line(0,1){60}}
\put(60,20){\line(0,1){60}}
\put(70,20){\line(0,1){40}}

\color{black}

\linethickness{0.7mm}

\put(20,20){\line(1,0){50}}
\put(10,30){\line(1,0){60}}
\put(10,40){\line(1,0){50}}
\put(10,50){\line(1,0){50}}
\put(10,60){\line(1,0){50}}
\put(40,70){\line(1,0){10}}

\put(10,30){\line(0,1){30}}
\put(20,20){\line(0,1){40}}
\put(30,20){\line(0,1){40}}
\put(40,20){\line(0,1){50}}
\put(50,20){\line(0,1){50}}
\put(60,20){\line(0,1){40}}
\put(70,20){\line(0,1){10}}

\end{picture}
\caption{The Young diagram for the long multiplet $n^{2222}=[0,0|3,2,2,2|1,0]$. The diagram covers the whole $4\times4$ square and thus the $\psu(2,2|4)$ $\dQ$'s are identical to those on the Young diagram.}
\label{fig:longYD}
\end{figure}
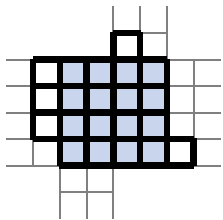

For multiplets that are long at $g=0$, there is no issue of making the transition. Young diagrams of such multiplets cover the $4\times4$ square, and the distinguished $\psu(2,2|4)$ Q-functions coincide exactly with the Young diagram $\dQ$-functions on the square, see an example in figure \ref{fig:longYD}.

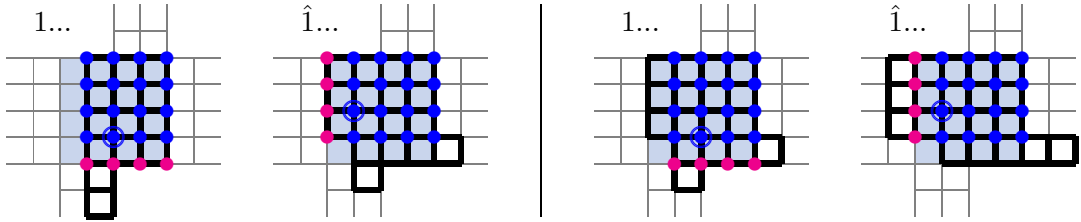
\begin{figure}[t]
\centering
\begin{picture}(300,80)

\color{NavyBlue!20}
\linethickness{0.4mm}
\multiput(-30,20)(0,1){40}{\line(1,0){40}}
\multiput(70,20)(0,1){40}{\line(1,0){40}}
\multiput(190,20)(0,1){40}{\line(1,0){40}}
\multiput(290,20)(0,1){40}{\line(1,0){40}}

\color{gray}
\linethickness{.1mm}

\put(0,0){
\multiput(-50,20)(0,10){5}{\line(1,0){80}}
\put(-30,10){\line(1,0){20}}
\put(-10,70){\line(1,0){20}}
\put(-40,20){\line(0,1){40}}
\put(-30,0){\line(0,1){60}}
\put(-20,0){\line(0,1){60}}
\put(-10,0){\line(0,1){80}}
\put(0,20){\line(0,1){60}}
\put(10,20){\line(0,1){60}}
\put(20,20){\line(0,1){40}}
}

\put(100,0){
\multiput(-50,20)(0,10){5}{\line(1,0){80}}
\put(-30,10){\line(1,0){20}}
\put(-10,70){\line(1,0){20}}
\put(-40,20){\line(0,1){40}}
\put(-30,0){\line(0,1){60}}
\put(-20,0){\line(0,1){60}}
\put(-10,0){\line(0,1){80}}
\put(0,20){\line(0,1){60}}
\put(10,20){\line(0,1){60}}
\put(20,20){\line(0,1){40}}
}

\put(220,0){
\multiput(-50,20)(0,10){5}{\line(1,0){80}}
\put(-30,10){\line(1,0){20}}
\put(-10,70){\line(1,0){20}}
\put(-40,20){\line(0,1){40}}
\put(-30,0){\line(0,1){60}}
\put(-20,0){\line(0,1){60}}
\put(-10,0){\line(0,1){80}}
\put(0,20){\line(0,1){60}}
\put(10,20){\line(0,1){60}}
\put(20,20){\line(0,1){40}}
}

\put(320,0){
\multiput(-50,20)(0,10){5}{\line(1,0){80}}
\put(-30,10){\line(1,0){20}}
\put(-10,70){\line(1,0){20}}
\put(-40,20){\line(0,1){40}}
\put(-30,0){\line(0,1){60}}
\put(-20,0){\line(0,1){60}}
\put(-10,0){\line(0,1){80}}
\put(0,20){\line(0,1){60}}
\put(10,20){\line(0,1){60}}
\put(20,20){\line(0,1){40}}
}

\color{black}

\put(150,00){\line(0,1){80}}

\linethickness{0.7mm}

\put(-20,0){\line(1,0){10}}
\put(-20,10){\line(1,0){10}}
\put(-20,20){\line(1,0){30}}
\put(-20,30){\line(1,0){30}}
\put(-20,40){\line(1,0){30}}
\put(-20,50){\line(1,0){30}}
\put(-20,60){\line(1,0){30}}

\put(-20,0){\line(0,1){60}}
\put(-10,0){\line(0,1){60}}
\put(0,20){\line(0,1){40}}
\put(10,20){\line(0,1){40}}

\put(0,0){
\put(80,10){\line(1,0){10}}
\put(80,20){\line(1,0){40}}
\put(70,30){\line(1,0){50}}
\put(70,40){\line(1,0){40}}
\put(70,50){\line(1,0){40}}
\put(70,60){\line(1,0){40}}

\put(70,30){\line(0,1){30}}
\put(80,10){\line(0,1){50}}
\put(90,10){\line(0,1){50}}
\put(100,20){\line(0,1){40}}
\put(110,20){\line(0,1){40}}
\put(120,20){\line(0,1){10}}
}

\put(120,0){
\put(80,10){\line(1,0){10}}
\put(80,20){\line(1,0){40}}
\put(70,30){\line(1,0){50}}
\put(70,40){\line(1,0){40}}
\put(70,50){\line(1,0){40}}
\put(70,60){\line(1,0){40}}

\put(70,30){\line(0,1){30}}
\put(80,10){\line(0,1){50}}
\put(90,10){\line(0,1){50}}
\put(100,20){\line(0,1){40}}
\put(110,20){\line(0,1){40}}
\put(120,20){\line(0,1){10}}
}

\put(220,0){
\put(80,20){\line(1,0){50}}
\put(60,30){\line(1,0){70}}
\put(60,40){\line(1,0){50}}
\put(60,50){\line(1,0){50}}
\put(60,60){\line(1,0){50}}

\put(60,30){\line(0,1){30}}
\put(70,30){\line(0,1){30}}
\put(80,20){\line(0,1){40}}
\put(90,20){\line(0,1){40}}
\put(100,20){\line(0,1){40}}
\put(110,20){\line(0,1){40}}
\put(120,20){\line(0,1){10}}
\put(130,20){\line(0,1){10}}
}

\put(-40,70){$1...$}
\put(60,70){$\hat{1}...$}
\put(180,70){$1...$}
\put(280,70){$\hat{1}...$}

\color{magenta}
\multiput(-20,20)(10,0){4}{\circle*{5}}
\multiput(200,20)(10,0){4}{\circle*{5}}

\multiput(70,30)(0,10){4}{\circle*{5}}
\multiput(290,30)(0,10){4}{\circle*{5}}

\color{blue}

\put(-10,30){\circle{8}}
\put(-10,30){\circle{7}}
\put(80,40){\circle{8}}
\put(80,40){\circle{7}}

\put(220,0){
\put(-10,30){\circle{8}}
\put(-10,30){\circle{7}}
\put(80,40){\circle{8}}
\put(80,40){\circle{7}}
}
 
\color{blue}
\matrixput(-20,30)(0,10){4}(10,0){4}{\circle*{5}}
\matrixput(80,30)(0,10){4}(10,0){4}{\circle*{5}}
\matrixput(200,30)(0,10){4}(10,0){4}{\circle*{5}}
\matrixput(300,30)(0,10){4}(10,0){4}{\circle*{5}}

\end{picture}
\caption{Defining the $\psu(2,2|4)$ Q-system for the case of a short representation. The two diagrams on the left are a pair of multiplets that joins at finite coupling. The two diagrams on the right are another such pair. On each diagram, the marked nodes are the 20 distinguished Q-functions which are transferred from the Young diagram to the $\psu(2,2|4)$ Q-system. The $\dQ$'s that must coincide (up to symmetry rescalings) on both Young diagrams  are marked in {\color{blue}blue}. The $\dQ$'s that can be transferred to the $\psu(2,2|4)$ Q-system from only one of the diagrams are marked in {\color{magenta}red}. The node that has a trivial Bethe root is encircled; the position of this node determines the grading in which the multiplet is going to join (note that the second and the third Young diagrams are identical but the gradings are different).}
\label{fig:shortYD}
\end{figure}

However, things are not that obvious for  short multiplets. For one thing, their Young diagrams do not cover the 4$\times$4 square, cf.\ figure \ref{fig:shortYD}. For another thing, there are several (two or four) different short multiplets that join into a long one at finite coupling, so their Young diagram Q-systems should somehow lead to the same $\psu(2,2|4)$ Q-system at finite coupling. Equivalent issues were explored and understood in detail on the level of asymptotic Bethe Ansatz equations \cite{Beisert:2005fw}, and here we give an equivalent analysis on the level of Q-systems.

\subsubsection*{Quantum numbers from analytic structure of Q-functions}
Let us do some reverse engineering first: suppose  all the 25 distinguished functions $\dQ_{a,s}$, $0\leq a,s\leq 4$, are known and we ask what Young diagram Q-system this data corresponds to. To identify the shape of the  diagram, it suffices to identify eight charges, and this should be done  {\it from the analytic properties} of $\dQ_{a,s}$. 

The fundamental weights can be read off from the asymptotic behaviour \eqref{powercount}. However, one should be careful. The QQ-relations are invariant under certain symmetries \cite{Gromov:2014caa,Kazakov:2015efa} which are summarised in section~\ref{sec:sym}. One of them implies the rescaling
\be\label{gaugedQ}
\dQ_{a,s}\to  (u^{[s-a]_{\rm D}})^{\Lambda}\, \dQ_{a,s}\,.
\ee
It hence shifts  the fundamental weights: $\{\lambda,\nu\}\to \{\lambda+\Lambda,\nu-\Lambda\}$.

Similarly, the factorisation \eqref{Qstruc} should define the charge $L$, but we could  rescale $q_{a,s}$ by $\Phi_{a,s}$ thus changing the length.

The two mentioned ambiguities parallel precisely the transformations \eqref{bothL}. The transformation \eqref{L1} corresponds to
\begin{subequations}
\label{qsymboth}
\be\label{qsym1}
\dQ_{a,s}\to\dQ_{a,s}\,,\quad q_{a,s}\to q_{a,s}\,\Phi_{a,s}\,,\quad L\to L-1\,,
\ee
while \eqref{L2} corresponds to
\be\label{qsym2}
\dQ_{a,s}\to\frac{\dQ_{a,s}}{u^{[s-a]_{\rm D}}}\,,\quad q_{a,s}\to q_{a,s}\frac{\Phi_{a,s}}{u^{[s-a]_{\rm D}}}\,,\quad L\to L-1\,.
\ee
\end{subequations}

The symmetries \eqref{qsymboth} of the Q-system are fixed  by the demand that all $q_{a,s}$  are polynomials. This is a very strong constraint. For instance  \eqref{qsym1} would not violate the polynomiality of $q_{a,s}$ only for a very specific placement of multiple Bethe roots: simultaneously it should be that $q_{1,2}=q_{0,1}=q_{0,2}^+=q_{0,2}^-=0$ at $u=0$. Even in the unlikely event that some transformations \eqref{qsymboth} do not violate polynomiality, we understand this as an accidental degeneracy -- two different multiplets in representations related by corresponding weight shifts \eqref{bothL} happen to have the same spectrum.

The transformations \eqref{qsymboth} are not used to relate short multiplets subject to joining. We note that transformations \eqref{bothL} did not serve this goal either. Instead, we considered in section~\ref{sec:short} an analog of \eqref{bothL} applied to a Young diagram of a smaller rank algebra. Equivalently, as we shall see, joining of short multiplets is related to transformations \eqref{qsymboth} applied to a subset of $q_{a,s}$ only.

\subsubsection*{Trivial roots from compatibility of short multiplets}

To not clutter the formulas, we consider only the case $\lambda_1+\nu_1=0$ in detail. This shortening implies  $\dQ_{1,1}=1$, as one can observe by a simple power counting \eqref{powercount}. Moreover, QSC requires $\dQ_{0,0}=1$ as well, cf.\ \eqref{qdet}. The QQ-relations \eqref{QQ3} then tell us that
\be
\dQ_{1,0}\dQ_{0,1}=0\,,
\ee
i.e.\ either $\dQ_{1,0}$ or $\dQ_{0,1}$ should be zero. Algebra-based Q-systems for short multiplets can have vanishing Q-functions, as discussed in \cite{Kazakov:2015efa}. It was also pointed out in \cite{Kazakov:2015efa} that either possibility can be realised depending on the choice of grading: the Q-functions that belong to the Dynkin path should be non-zero whereas Q-functions outside the Dynkin path may be zero if it is so dictated by the QQ-relations.

Applying this argument to our context: working in a grading of the type $1...$ (in long-hand notation) corresponds to choosing $\dQ_{1,0}=0$. Then, in consequence of the QQ-relations, we must set $\dQ_{2,0}=\dQ_{3,0}=\dQ_{4,0}=0$. In contrast, working in a grading of the type $\hat 1...$ corresponds to $\dQ_{0,s}=0$ for $s=1,2,3,4$.

In the perturbative solution of the QSC, setting $\dQ_{a,0}$ or $\dQ_{0,s}$ to zero is simply a question of a subset of the $\dQ$'s being suppressed by a factor of $g^2$. This factor can be shuffled around by the symmetries of the Q-system, and as a result we get two distinct but still well-defined Q-systems at zero coupling. These Q-systems are naturally associated with the grading choice, and thus they should describe two different emerging short multiplets.

The Young diagram Q-system does not contain zero Q-functions, but also  $\dQ_{0,0}$ does not belong to the diagram at all for short multiplets. What happens is that we {\it change boundary conditions} to swap between the Young diagram and algebra-based Q-system. Assuming there is only one shortening, a block of 20 non-zero $\dQ_{a,s}$ coincide on the Young diagram and the QSC Q-system. In the QSC, we supplement this set by setting $\dQ_{0,0}=1$ and the remaining four functions to zero. On the Young diagram, we can find the other $\dQ_{a,s}$ from the QQ-relations and the ansatz \eqref{Qstruc}, with $q_{a,s}=1\neq 0$ on the left and right boundary of the Young diagram.

In either  the $1...$ or $\hat 1...$ choice of grading, $\dQ_{a,s}\neq 0$ when both $a$ and $s$ are non-zero. These $16$ Q-functions must be the same in both Q-systems, otherwise getting the unique $\psu(2,2|4)$ Q-system at finite coupling would be impossible. A paradox is that while these functions coincide, they still  generate two non-equal Young diagrams. The answer to the paradox is that the value of $L$ is different. Indeed, as it follows from \eqref{sm12}, if a multiplet in grading $1...$ has length $L$, a compatible multiplet for joining should be of length $L+1$ in grading $\hat 1...$. Hence one has the following constraint
\begin{subequations}
\be\label{qqsa}
q_{a,s}^{1...}=\Phi_{a,s}\,q_{a,s}^{\hat 1...}\,,\ \ \  a\,,s> 0\,,
\ee 
which is a restricted version of \eqref{qsym1}. It imposes restrictions on the structure of Bethe roots. The benchmark relation is $q_{1,2}^{1...}=u\,q_{1,2}^{\hat 1...}$, which implies that $q_{1,2}^{1...}$ has a trivial zero (i.e.\ a zero at the origin). Also, from $q_{2,1}^{1...}=u^{-1}\,q_{2,1}^{\hat 1...}$, we see that $q_{2,1}^{\hat 1...}$ has a trivial zero.

The analysis of the $\lambda_4+\nu_4=0$ shortening goes in full analogy. Now we compare $\dQ_{a,s}$ with  $a,s< 4$, and they should be equal up to the rescaling \eqref{gaugedQ} which results in the following restriction on $q_{a,s}$:
\be
q_{a,s}^{...4}=\frac{\Phi_{a,s}}{u^{[s-a]_{\rm D}}}\,q_{a,s}^{...\hat 4}\,,\ \ \  a\,,s< 4\,.
\ee
\end{subequations}

To summarise, the procedure to process the Q-system for short multiplets is the following:
\begin{itemize}
\item Given a Q-system on a Young diagram, decide the grading according to:
\be
\begin{tabular}{rccl}
$\lambda_1+\nu_1=0\,,$ & $q_{1,2}$ \text{ has root at  } $u=0$ & $\Rightarrow$ & $1...$\,,\\
$\lambda_1+\nu_1=0\,,$ & $q_{2,1}$ \text{ has root at  } $u=0$ & $\Rightarrow$ & $\hat 1...$\,,\\
$\lambda_4+\nu_4=0\,,$ & $q_{3,2}$ \text{ has root at  } $u=0$ & $\Rightarrow$ & $...4$\,,\\
$\lambda_4+\nu_4=0\,,$ & $q_{2,3}$ \text{ has root at  } $u=0$ & $\Rightarrow$ & $...\hat 4$\,.
\end{tabular}
\ee
\item Given the grading, choose which $\dQ_{a,s}$ should vanish in the $\psu(2,2|4)$ Q-system:
\be
\begin{tabular}{rccl}
$1...$ & $\Rightarrow$ & $\dQ_{a>0,0}=0$\,,\\
$\hat 1...$ & $\Rightarrow$ & $\dQ_{0,s>0}=0$\,,\\
$...4$ & $\Rightarrow$ & $\dQ_{a<4,4}=0$\,,\\
$...\hat 4$ & $\Rightarrow$ & $\dQ_{4,s<4}=0$\,.
\end{tabular}
\ee
\item Set $\dQ_{0,0}=\dQ_{4,4}=1$. 
\item All other $\dQ_{a,s}$ should be the same as $\dQ_{a,s}$ on the Young diagram.
\end{itemize}

\subsubsection*{Trivial roots and the zero-momentum condition}
To only consider states that do not vanish when cyclicity of the trace is imposed, one can impose the {\it zero-momentum condition} (ZMC):
\be
\lim_{u\to0} \frac{\dQ_{2,2}\left(u+\frac{i}{2}\right)}{\dQ_{2,2}\left(u-\frac{i}{2}\right)}=1\,.
\ee
When $\lambda_1+\nu_1=0$, this condition guarantees the existence of a root at $u=0$ in either $q_{1,2}$ or $q_{2,1}$. This is a consequence of the fermionic QQ-relation that relates $\dQ_{2,2}$ and $\dQ_{1,1}=1$:
\be
\dQ_{2,2}^+ - \dQ_{2,2}^-=\dQ_{1,2}\,\dQ_{2,1}=q_{1,2}\,q_{2,1}\,.
\ee
At $u=0$ the left-hand side vanishes due to the ZMC, and thus the right-hand side must contain a factor of $u$.
Likewise, the ZMC guarantees a root at $u=0$ in either $q_{2,3}$ or $q_{3,2}$ when $\lambda_4+\nu_4=0$.

We therefore observe that the ZMC, besides being hard-coded into the analytic structure of QSC \cite{Gromov:2013pga}, is also the condition which guarantees the presence of trivial zeros --  a necessary property for short multiplets to be capable of joining.

\subsection{Full $\psu(2,2|4)$ Q-system}\label{sec:psuQsys}


To obtain the full Q-system from the distinguished Q-functions one needs to solve 12 first-order difference equations. We will here focus on determining $Q_{a|\emptyset}$ and $Q_{\emptyset|j}$ as the remaining Q-functions are easily reconstructed from these.

\subsubsection*{$Q_{a|\emptyset}$ from $\dQ$}
All four $Q_{a|\emp}$ belong to the left compact Young diagram and should thus be rational. Their structure is
\be
Q_{a|\emptyset}=\frac{q_{a|\emp}}{u^L}\,,\quad q_{a|\emp} = \sum_{k=0}^{p_{a|\emp}} c_k u^k\,.
\ee
The asymptotic powers $p_{a|\emp}$ can be found from power counting in the QQ-relations and are
\be
p_{2|\emp}=L-n^{2222}_{\fff_2}-1\,,\quad p_{3|\emp}=L-n^{2222}_{\fff_3}\,,\quad p_{4|\emp}=L-n^{2222}_{\fff_4}+1\,.
\ee
An easy way to find $Q_{a|\emp}$ is to simply fit polynomials to
\be
Q_{A|\emptyset}=\det_{1\le a,j\le |A|}\left(Q_{a|\emp}^{[|A|+1-2j]}\right)\,.
\ee
In practice, one can first fix $Q_{2|\emp}$ from $Q_{12|\emp}=\dQ_{2,0}$, then fix $Q_{3|\emp}$ from $Q_{123|\emp}=\dQ_{3,0}$, and finally $Q_{4|\emp}$ from $Q_{1234|\emp}=\dQ_{4,0}$.

\subsubsection*{$Q_{\emptyset|j}$ from $\dQ$}
$Q_{\emp|2}$ belongs to the left compact diagram and should be rational. Its structure is
\be
Q_{\emp|2}=u^L q_{\emp|2}\,\quad q_{\emp|2} = \sum_{k=0}^{p_{\emp|2}} c_k u^k\,,\quad p_{\emp|2}=n_{\bbb_2}^{2222}+1\,,
\ee
and it can be found by fitting a polynomial to
\be
Q_{\emp|12}=Q_{\emp|1}^+Q_{\emp|2}^- - Q_{\emp|1}^-Q_{\emp|2}^+\,.
\ee

$Q_{\emp|3}$ and $Q_{\emp|4}$ do not belong to the left or right compact diagram and are thus not expected to be rational. The basic QQ-relation $Q_{\emp|3}Q_{\emp|312}=...$ can be rewritten as\footnote{Define the difference operator, $\nabla$, by $\nabla(f) \equiv f-f^{[2]}$.
The operation $\Psi$ is defined as the inverse of this function,
$\Psi (\nabla (f)) = f + \mathcal{P}$,
where $\mathcal{P}$ is an arbitrary $i$-periodic function. The $\Psi$-operation can be represented as
$\Psi(f) = \sum_{n=0}^\infty f^{[2n]}$ when the sum converges. See also \cite{Marboe:2014gma}.
}
\be
Q_{\emp|3}=Q_{\emp|2}\Psi\left(\frac{Q_{\emp|1}Q_{\emp|123}}{Q_{\emp|12}^-Q_{\emp|12}^+}\right)-Q_{\emp|1}\Psi\left(\frac{Q_{\emp|2}Q_{\emp|123}}{Q_{\emp|12}^-Q_{\emp|12}^+}\right)\,. \label{Q3gen}
\ee
As we explain in \ref{ap:diffeq}, evaluating these $\Psi$-operations leads to poles only at $i \mathbb{Z}$. Also, the way we choose to define $\Psi$ leads to poles only in the lower half-plane, $\Im(u)<0$.

Similarly the QQ-relation $Q_{\emp|4}Q_{\emp|412}=...$ can be rewritten as
\be
Q_{\emp|4}=Q_{\emp|2}\Psi\left(\frac{Q_{\emp|1}Q_{\emp|124}}{Q_{\emp|12}^-Q_{\emp|12}^+}\right)-Q_{\emp|1}\Psi\left(\frac{Q_{\emp|2}Q_{\emp|124}}{Q_{\emp|12}^-Q_{\emp|12}^+}\right)\,. \label{Q4gen}
\ee
$Q_{\emp|124}$ belongs to the right compact Young diagram, i.e.\ it is rational, and it can be found by fitting a polynomial to
\be
Q_{\emp|12}Q_{\emp|1234}=Q_{\emp|123}^+Q_{\emp|124}^- - Q_{\emp|123}^-Q_{\emp|124}^+\,.
\ee

\subsubsection*{$Q_{a|j}$ from $Q_{a|\emp}$ and $Q_{\emp|j}$}
To generate $Q_{a|j}$ from $Q_{a|\emptyset}$ and $Q_{\emp|j}$ one simply has to solve the first order difference equation \eqref{QQ3},
\be\label{Qaidiff}
Q_{a|j}=-\Psi\left(Q_{a|\emptyset}^+Q_{\emptyset|j}^+\right)\,.
\ee

\subsubsection*{The rest}\label{sec:therest}
All other $\psu(2,2|4)$ Q-functions can be generated by taking determinants of $Q_{a|\emp}$, $Q_{\emp|j}$ and $Q_{a|j}$ \cite{Gromov:2014caa}:
\be\label{detf}
Q_{a_1,...,a_m|j_1,...,j_n} = \left\{
\begin{matrix}
\epsilon^{k_1,...,k_n}
\prod_{r=1}^{m} Q_{a_r|j_{k_r}}^{[\pm n\mp m]} \, \prod_{s=1}^{n-m} Q_{\emp|j_{k_{m+s}}}^{[n-m+1-2s]}  & m<n \\[10pt]


\epsilon^{k_1,...,k_m} \prod_{r=1}^m Q_{a_{k_r}|j_r} & m=n \\[10pt]	

\epsilon^{k_1,...,k_m}
\prod_{r=1}^{n} Q_{a_{k_r}|j_r}^{[\pm n\mp m]} \, \prod_{s=1}^{m-n} Q_{a_{k_{n+s}}|\emp}^{[m-n+1-2s]}  & m>n


\end{matrix} \right. \,.
\ee
Note that only poles in the lower half-plane, at $u=-i\mathbb{N}$ for an odd number of indices and at $u=-i(\frac{1}{2}+\mathbb{N})$ for an even number of indices, develop.
 
At first sight, solving the difference equations \eqref{Qaidiff} introduces constant ambiguities, but these can be fixed by recovering the distinguished Q-functions via \eqref{detf} and comparing to the original values found via the algorithm of \secref{sec:alg}, which were used to generate $Q_{a|\emp}$ and $Q_{\emp|i}$.


\subsection{Symmetries} \label{sec:sym}
Certain transformations of the Q-functions leave the QQ-relations invariant. We mention only those that do not spoil the analytic structure of the Q-functions, and we refer to \cite{Gromov:2014caa,Kazakov:2015efa} for a more complete treatment. 

\subsubsection*{$x$-rescalings}
As a specific case of the gauge transformations described in \cite{Gromov:2014caa,Kazakov:2015efa}, the QQ-relations and monodromy properties at finite coupling are invariant under the rescalings
\be
Q_{A|J}\to x^{ [\, |J|-|A|\, ]_{\rm D}}\, Q_{A|J}\,,
\ee
where $x$ is the Zhukovsky variable defined by $x+\frac{1}{x}=\frac{u}{g}$. At weak coupling the leading contribution to $x=\frac{u}{g}+\CO(g)$ is simply proportional to $u$. So, as the overall normalisation of the Q-functions is irrelevant in our discussion, the transformation of the one-loop Q-system  simply amounts to
\be
Q_{A|J}\to u^{ [\, |J|-|A|\, ]_{\rm D}} \, Q_{A|J}\,. \label{xresc}
\ee
When it is specified to $\dQ_{a,s}$, it becomes \eqref{gaugedQ}.

\subsubsection*{$H$-symmetry}
The generation of the full Q-system from $\dQ$ required the solution of a number of difference equations. The solution of such equations always introduce an arbitrary $i$-periodic function. The Q-functions all have integer powers in their large $u$ asymptotics. If we furthermore require that the functions are analytic in the upper half-plane, $\Im(u)>0$, then the only allowed $i$-periodic function is a constant.  The corresponding symmetry is dubbed $H$-symmetry and corresponds to the transformations
\be\label{Hsym}
Q_{a_1,...,a_m|j_1,...,j_n}\to H_{a_1}^{b_1}\cdots H_{a_m}^{b_m} \hat{H}_{j_1}^{k_1}\cdots \hat{H}_{j_n}^{k_n}  Q_{b_1,...,b_m|k_1,...,k_n}\,.
\ee
There are certain restrictions on these transformations. The one relevant in this paper is that  the distinguished Q-functions should not be affected by \eqref{Hsym}. In fact, the distinguished Q-functions are most properly defined as such $Q_{A|J}$ that $Q_{A|J}/\Phi_{a,s}$ is a polynomial and it is the polynomial of the lowest possible degree among all\footnote{To be precise, it is the one of lowest degree in the Young diagram Q-system.} polynomials  $Q_{A|J}/\Phi_{a,s}$ with $|A|=a,|J|=s$. Then $H$-rotations are used to reconcile notation \eqref{dqdef} with this property.

There are other restrictions on $H$-rotations which are needed  to properly reflect the match  of large-$u$ asymptotic of Q-system with representation theory, but the symmetry is never fully constrained, and we need to make some agreement to fix it when solving equations in section~\ref{sec:therest}.

\subsubsection*{Relations between solutions}
As the QSC equations are invariant under the parity transformation $u\leftrightarrow -u$, then if $\{Q_{A|I}(u)\}$ is a solution to the Q-system, $\{Q_{A|I}(-u)\}$ will be as well. In some cases, the two Q-systems will be identical, and we call such solutions {\it parity invariant}.

Similarly, the Hodge dual of $\{Q_{A|I}\}$, which we denote by $\{Q^H_{A|I}\}\equiv \{Q^{A|I}\}$ and define through
\be
Q^{A|J}\equiv (-1)^{|A||J|}\epsilon^{\bar{A}A}\epsilon^{\bar{J}J}Q_{\bar{A}|\bar{J}}\,,
\ee
will also be a solution, though in general to a Q-system with different boundary conditions. However, for some states the two sets of Q-functions are identical, and we call these {\it Hodge invariant}.

\begin{figure}[t]\centering
\begin{picture}(100,60)
\thicklines

\put(0,40){$\{Q_{A|I}(u)\}$}
\put(0,0){$\{Q_{A|I}^H(u)\}$}
\put(70,40){$\{Q_{A|I}(-u)\}$}
\put(70,0){$\{Q_{A|I}^H(-u)\}$}

\put(53,40){\large$\leftrightarrow$}
\put(53,0){\large$\leftrightarrow$}
\put(23,20){\large$\updownarrow$}
\put(93,20){\large$\updownarrow$}

\end{picture}
\caption{All solutions come in sets related by parity and Hodge transformation. In some cases, one or all of these transformations map a solution to itself.}
\label{fig:syms}
\end{figure}

Hodge invariance can occur when the left and right compact Young diagrams are identical, i.e.\ in the $2222$ grading
\be
n_{\fff_1}&=&L-n_{\fff_4}\no\\
n_{\fff_2}&=&L-n_{\fff_3}\no\\
n_{\bbb_1}&=&n_{\aaa_2}\no\\
n_{\bbb_2}&=&n_{\aaa_1}\,. \label{LRsym}
\ee
This is also referred to as {\it left/right symmetry}. When a state is left/right symmetric \eqref{LRsym}, it implies that either $\{Q_{A|I}(u)\}=\{Q_{A|I}^H(u)\}$ or $\{Q_{A|I}(u)\}=\{Q_{A|I}^H(-u)\}$ (or both). An overview of the related solutions is given in figure \ref{fig:syms}.


\section{Conclusions}
We have provided a classification of the symmetry multiplets that appear in the spectrum of the AdS$_5$/CFT$_4$ correspondence. We focused on the interpretation of these states as single-trace operators in SYM at $g=0$, and we found it convenient to use oscillator numbers, or equivalently Young diagrams, to label the multiplets.

The notion of  the  extension of a non-compact Young diagram  naturally emerged in our studies from the basic power-counting of the distinguished Q-functions. Given its natural appearance, we believe that it should be a useful combinatorial object deserving further study, e.g.\ to study characters and combinatorics of non-compact representations. In fact, we have already benefited from the Young diagram extensions: using that such an extension encodes equivalently a representation of a compact $\su(N)$ algebra, we  generated the explicit multiplet content of SYM using solely $\su(N)$ characters.

The explicit computation of the spectrum has previously been limited by the inefficiency of solving nested Bethe equations. By a more careful treatment of the underlying Q-system, we have managed to circumvent the direct usage of these equations and, along with significantly improved computation speed, we also cured some of the diseases of the Bethe equations related to under- or over-counting. The proposed method straightforwardly produces the number of solutions  predicted by combinatorics. The solutions that can be found within fifteen minutes on a standard laptop are marked in \ref{ap:spec}. We refer the reader to our supplemental \texttt{Mathematica} notebook for the explicit results.

The findings of this paper will serve as the seed for our upcoming work about the planar AdS$_5$/CFT$_4$ spectrum. This will contain an implemented algorithm to calculate perturbative corrections for any state to, in principle, any order \cite{spectrumII}. We also expect that  our classification and perturbative results can serve as the starting point for a concrete algorithm to determine the numerical spectrum of any operator at arbitrary coupling, by a generalisation of \cite{Gromov:2015wca} or \cite{Hegedus:2016eop}.

\vspace{0.8cm}
\subsection*{Acknowledgements}
We thank S\'ebastien Leurent, David Meidinger, Matthias Staudacher and Stijn van Tongeren for discussions and comments.
C.M. would like to thank IPhT, C.E.A.-Saclay, where part of this work was done, for hospitality.  The work of C.M. was partially supported by the People Programme (Marie Curie Actions)
of the European Union's Seventh Framework Programme
FP7/2007-2013/ under REA Grant Agreement
No 317089 (GATIS).

\newpage

\appendix
\addtocontents{toc}{\protect\setcounter{tocdepth}{1}}


\section{Representation theory details}


\subsection{Quantum numbers in the literature} \label{ap:qn}
In this appendix we provide a dictionary between the oscillator numbers $n$ and often encountered parametrisations of the quantum numbers in the literature.
\subsubsection*{$S^5$ and AdS$_5$ spins: $J$, $S$ and $\Delta_0$}
A typical notation for the quantum numbers is the $SO(6)\times SO(2,4)$ Cartan
charges $\{J_1,J_2,J_3|\Delta,S_1,S_2\}$. These are related to the oscillator numbers through
\be\label{Cartandef}
J_1 &=& \frac{n_{\fff_1}+n_{\fff_2}-n_{\fff_3}-n_{\fff_4}}{2} \no\\
J_2 &=& \frac{n_{\fff_1}-n_{\fff_2}+n_{\fff_3}-n_{\fff_4}}{2} \no\\
J_3 &=& \frac{-n_{\fff_1}+n_{\fff_2}+n_{\fff_3}-n_{\fff_4}}{2} \no\\
\Delta_0 &=& L+\frac{n_\bbb+n_\aaa}{2} = \frac{n_{\fff}}{2} +n_\aaa \no\\
S_1 &=& \frac{-n_{\bbb_1}+n_{\bbb_2}+n_{\aaa_1}-n_{\aaa_2}}{2} \no\\
S_2 &=& \frac{-n_{\bbb_1}+n_{\bbb_2}-n_{\aaa_1}+n_{\aaa_2}}{2}\,.
\ee
They alone are  enough to specify the multiplet at finite coupling, but not at zero coupling, except in the case when both shortening conditions \eqref{bothU} are satisfied.

\subsubsection*{Dynkin labels}
Another convention encountered in the literature, e.g.\ \cite{Beisert:2003te}, is to use the $\mathfrak{so}(4)$ Dynkin labels $[s_1,s_2]$ and $\mathfrak{so}(6)$ labels $[q_1,p,q_2]$, which are related to the oscillator numbers by
\begin{align}
s_1=n_{\bbb_2}-n_{\bbb_1},\quad s_2=n_{\aaa_1}-n_{\aaa_2};\quad q_1=n_{\fff_1}-n_{\fff_2},\quad p=n_{\fff_2}-n_{\fff_3},\quad q_2=n_{\fff_3}-n_{\fff_4}\,.\no\\ \label{dynklab}
\end{align}
Additionally  $\Delta_0$, $L$, and the hypercharge
\be\label{Hyperchargeref}
B=\frac{n_\bbb-n_\aaa}{2}\,
\ee
are specified. Then the multiplet is described by the data\footnote{Sometimes, a parity label is also given to a state related to its eigenvalue under the parity transformation $\Pi | \Phi_1 \Phi_2 \cdots \Phi_N \rangle=(-1)^N | \Phi_N \Phi_{N-1} \cdots \Phi_1 \rangle$. This does not influence the weights, but it is related to parity properties of the Q-system related to the multiplet.} $[\Delta_0;s_1,s_2;q_1,p,q_2]_{L}^B$.


\subsection{Efficient computation of the sum of states} \label{ap:char}
In this appendix we explain how the relevant terms in the sum of states 
\be
\Delta_{\text{V}} \, Z = \sum_\lambda c_\lambda W_\lambda \label{apsum}
\ee
can be generated for high rank of the considered group.

\begin{itemize}
\item[\bf Step 1] List all possible dominant terms. These correspond to all strictly decreasing partitions of $\Delta_{\text{max}}\cdot L + \Delta_{\text{max}}(2\Delta_{\text{max}}-1)$ into $2\Delta_{\text{max}}$ numbers, with the restriction that the first number does not exceed $L+2\Delta_{\text{max}}-1$ (for $\Delta_{\text{max}}>2$ the last $\Delta_{\text{max}}-2$ numbers satisfy a stronger upper bound).

Example: For $\Delta_{\text{max}}=2$ and $L=2$, the possible dominant terms (all allowed length-4 partitions of 10) are $x_1^4\,x_2^3\,x_3^2\,x_4$, $x_1^5\,x_2^4\,x_3 $, and $x_1^5\,x_2^3\,x_3^2$.

\item[\bf Step 2] For each dominant term, list all possible contributions from $\Delta_V$ and the corresponding contribution from $Z$. No row in the contribution from $Z$ can exceed $L$.

Example: The term $x_1^5\,x_2^3\,x_3^2$ can arise from $x_1^3\,x_2^2\,x_3\cdot x_1^2\,x_2\,x_3$ and from $x_1^3\,x_2\,x_3^2\cdot x_1^2\,x_2^2$.

\item[\bf Step 3] The terms from $\Delta_V$ come with a factor of $\pm1$. Each term in $Z$ comes with a coefficient that can be found by counting the number of ways that the term can be constructed from building blocks of the kind $x_{i_1}^d x_{i_2}^d\cdots x_{i_{\Delta_{\text{max}}}}^d$.

Example: $x_1^2x_2x_3$ must stem from $\chi_1(x^1)^2$ and can come from $x_1x_2\cdot x_1 x_3$ and $x_1x_3\cdot x_1 x_2$. Thus its coefficient in $Z$ is 2. $x_1^2x_2^2$ can stem from $\chi_1(x^1)^2$, and can only arise as $x_1x_2\cdot x_1x_2$, but it can also stem from $\chi_1(x^2)^1$, again with coefficient 1. So its coefficient in $Z$ is also 2.

\item[\bf Step 4] For each dominant term add up all contributions to $\Delta_V Z$.

Example: The term $x_1^5\,x_2^3\,x_3^2$ comes with a coefficient $2\cdot (-1) + 2\cdot 1=0$ in $\Delta_V Z$.

\end{itemize}
An implementation of this algorithm can be found in the ancillary \texttt{Mathematica} notebook.

\newpage

\subsection{Multiplets with $\Delta_0^{2222} \le 8$} \label{ap:spec}
Multiplets for which the corresponding $\dQ$-system is not found by our \texttt{Mathematica}-implementation of the solution algorithm described in section \ref{sec:alg} within 15 minutes on a standard laptop are marked in {\color{gray}grey}.

\begin{table}[h!]
\centering
\footnotesize
\def\arraystretch{1.1}
\begin{tabular}{|c||l|} \hline
$\Delta_0^{2222}$&Multiplets \\\hline\hline

2 & 
\begin{tabular}{l l l l} 
$1\cdot [0,0|1,1,1,1|0,0]$
\end{tabular}
\\\hline

3 & 
\begin{tabular}{l l l l}
$1\cdot [0,0|2,2,1,1|0,0]$
\end{tabular}
\\\hline

4 & 
\begin{tabular}{l l l l} 
$1\cdot[0,0|1,1,1,1|2,0]$ & $1\cdot[0,0|3,2,2,1|0,0]$ & 
$1\cdot[0,2|1,1,1,1|2,0]$ & $1\cdot[0,2|2,2,2,2|0,0]$ \\ 
$2\cdot[0,0|2,2,2,2|0,0]$ & $2\cdot[0,0|3,3,1,1|0,0]$ & 
$2\cdot[0,1|2,2,1,1|1,0]$
\end{tabular}
\\\hline

5 &  
\begin{tabular}{l l l l} 
$1\cdot[0,0|3,1,1,1|2,0]$ & $1\cdot[0,2|2,2,1,1|2,0]$ & 
$1\cdot[0,2|3,3,3,1|0,0]$ & $2\cdot[0,0|2,2,1,1|2,0]$ \\
$2\cdot[0,0|3,3,3,1|0,0]$ & $2\cdot[0,0|4,2,2,2|0,0]$ & 
$2\cdot[0,0|4,3,2,1|0,0]$ & $2\cdot[0,0|4,4,1,1|0,0]$ \\
$2\cdot[0,1|2,2,2,2|1,0]$ & $2\cdot[0,1|3,3,1,1|1,0]$ & 
$2\cdot[0,2|3,3,2,2|0,0]$ & $4\cdot[0,0|3,3,2,2|0,0]$ \\ 
$4\cdot[0,1|3,2,2,1|1,0]$
\end{tabular}
\\\hline

$\frac{11}{2}$ &  
\begin{tabular}{rrrr}
$2\cdot[0,0|4,3,1,1|1,0]$ & $2\cdot[0,1|3,2,1,1|2,0]$ &
$2\cdot[0,1|4,4,2,1|0,0]$ & $2\cdot[0,2|2,1,1,1|3,0]$ \\
$2\cdot[0,2|3,3,2,1|1,0]$ & $2\cdot[0,3|2,2,2,1|2,0]$ &
$4\cdot[0,0|3,2,2,2|1,0]$ & $4\cdot[0,0|3,3,2,1|1,0]$ \\
$4\cdot[0,1|2,2,2,1|2,0]$ & $4\cdot[0,1|3,3,3,2|0,0]$ &
$4\cdot[0,1|4,3,2,2|0,0]$ & $4\cdot[0,2|3,2,2,2|1,0]$
\end{tabular}
\\\hline

6 &
\begin{tabular}{rrrr}
$1\cdot[0,0|1,1,1,1|4,0]$ & $1\cdot[0,0|4,2,1,1|2,0]$ &
$1\cdot[0,2|1,1,1,1|4,0]$ & $1\cdot[0,2|4,4,3,1|0,0]$ \\ 
$1\cdot[0,4|1,1,1,1|4,0]$ & $1\cdot[0,4|2,2,2,2|1,1]$ & 
$1\cdot[0,4|2,2,2,2|2,0]$ & $1\cdot[0,4|3,3,3,3|0,0]$ \\ 
$1\cdot[1,1|1,1,1,1|4,0]$ & $2\cdot[0,0|2,2,2,2|1,1]$ & 
$2\cdot[0,2|2,2,2,2|1,1]$ & $2\cdot[0,2|3,2,2,1|2,0]$ \\
$2\cdot[0,3|2,2,1,1|3,0]$ & $2\cdot[1,1|2,2,2,2|1,1]$ & 
$2\cdot[1,1|2,2,2,2|2,0]$ & $2\cdot[1,1|3,3,3,3|0,0]$ \\
$3\cdot[0,0|2,2,2,2|2,0]$ & $3\cdot[0,0|4,4,3,1|0,0]$ & 
$3\cdot[0,0|5,3,2,2|0,0]$ & $3\cdot[0,0|5,3,3,1|0,0]$ \\ 
$3\cdot[0,0|5,4,2,1|0,0]$ & $3\cdot[0,0|5,5,1,1|0,0]$ & 
$3\cdot[0,2|3,3,3,3|0,0]$ & $4\cdot[0,0|3,3,1,1|2,0]$ \\
$4\cdot[0,1|2,2,1,1|3,0]$ & $4\cdot[0,1|3,3,3,1|1,0]$ & 
$4\cdot[0,1|4,2,2,2|1,0]$ & $4\cdot[0,1|4,4,1,1|1,0]$ \\ 
$4\cdot[0,2|4,4,2,2|0,0]$ & $4\cdot[0,3|3,3,2,2|1,0]$ & 
$5\cdot[0,0|3,3,3,3|0,0]$ & $5\cdot[0,2|3,3,1,1|2,0]$ \\ 
$6\cdot[0,0|3,2,2,1|2,0]$ & $6\cdot[0,2|4,3,3,2|0,0]$ & 
$7\cdot[0,2|2,2,2,2|2,0]$ & $8\cdot[0,1|4,3,2,1|1,0]$ \\
$9\cdot[0,0|4,3,3,2|0,0]$ & $10\cdot[0,0|4,4,2,2|0,0]$ & 
$16\cdot[0,1|3,3,2,2|1,0]$
\end{tabular}
\\\hline

$\frac{13}{2}$ &  
\begin{tabular}{rrrr}
$2\cdot[0,0|2,2,2,1|3,0]$ & $2\cdot[0,0|5,2,2,2|1,0]$ & 
$2\cdot[0,0|5,4,1,1|1,0]$ & $2\cdot[0,1|2,1,1,1|4,0]$ \\
$2\cdot[0,1|4,4,4,1|0,0]$ & $2\cdot[0,1|5,5,2,1|0,0]$ & 
$2\cdot[0,3|3,2,2,2|1,1]$ & $2\cdot[0,3|4,3,3,3|0,0]$ \\
$2\cdot[0,4|3,3,3,2|1,0]$ & $2\cdot[1,1|2,2,2,1|3,0]$ & 
$4\cdot[0,0|3,2,1,1|3,0]$ & $4\cdot[0,0|5,3,2,1|1,0]$ \\
$4\cdot[0,1|4,3,1,1|2,0]$ & $4\cdot[0,1|5,4,3,1|0,0]$ & 
$4\cdot[0,2|3,2,1,1|3,0]$ & $4\cdot[0,2|4,4,2,1|1,0]$ \\
$4\cdot[0,3|3,3,2,1|2,0]$ & $4\cdot[0,3|4,4,3,2|0,0]$ & 
$6\cdot[0,1|4,2,2,1|2,0]$ & $6\cdot[0,2|2,2,2,1|3,0]$ \\
$6\cdot[0,2|4,3,3,1|1,0]$ & $6\cdot[0,3|3,2,2,2|2,0]$ & 
$8\cdot[0,1|3,2,2,2|1,1]$ & $8\cdot[1,1|3,3,3,2|1,0]$ \\
$10\cdot[0,0|4,3,3,1|1,0]$ & $10\cdot[0,0|4,4,2,1|1,0]$ & 
$10\cdot[0,1|5,3,3,2|0,0]$ & $10\cdot[0,1|5,4,2,2|0,0]$ \\
$12\cdot[0,0|3,3,3,2|1,0]$ & $12\cdot[0,1|4,3,3,3|0,0]$ & 
$16\cdot[0,1|3,3,2,1|2,0]$ & $16\cdot[0,2|4,3,2,2|1,0]$ \\
$20\cdot[0,1|3,2,2,2|2,0]$ & $20\cdot[0,2|3,3,3,2|1,0]$ & 
$24\cdot[0,0|4,3,2,2|1,0]$ & $24\cdot[0,1|4,4,3,2|0,0]$
\end{tabular}
\\\hline

\end{tabular}
\caption{Spectrum of unprotected multiplets $[n_\bbb|n_\fff|n_\aaa]^{2222}$ with $\Delta_0^{2222}\le \frac{13}{2}$. We find complete agreement with the results in \cite{Beisert:2003te}. Our algorithm finds the $\dQ$-system for all these multiplets in a matter of minutes.}
\label{table:spec8}
\end{table}

\newpage

\begin{table}[h!]
\centering
\footnotesize
\def\arraystretch{1.1}
\begin{tabular}{|c||l|} \hline
$\Delta_0^{2222}$&Multiplets \\\hline\hline
7 &  
\begin{tabular}{rrrr} 
$2\cdot[0,0|2,2,1,1|4,0]$ & $2\cdot[0,0|3,1,1,1|4,0]$ & 
$2\cdot[0,0|5,2,2,1|2,0]$ & $2\cdot[0,2|3,1,1,1|4,0]$ \\ 
$2\cdot[0,2|5,4,4,1|0,0]$ & $2\cdot[0,3|1,1,1,1|5,0]$ & 
$2\cdot[0,4|3,3,2,2|1,1]$ & $2\cdot[0,4|3,3,3,1|2,0]$ \\ 
$2\cdot[0,4|4,4,3,3|0,0]$ & $2\cdot[0,4|4,4,4,2|0,0]$ & 
$2\cdot[0,5|2,2,2,2|3,0]$ & $2\cdot[1,1|2,2,1,1|4,0]$ \\ 
$3\cdot[0,0|5,3,1,1|2,0]$ & $3\cdot[0,0|6,6,1,1|0,0]$ & 
$3\cdot[0,2|5,5,3,1|0,0]$ & $3\cdot[0,4|2,2,1,1|4,0]$ \\ 
$3\cdot[1,2|2,2,2,2|2,1]$ & $4\cdot[0,0|6,4,3,1|0,0]$ & 
$4\cdot[0,1|4,2,1,1|3,0]$ & $4\cdot[0,3|3,3,1,1|3,0]$ \\ 
$4\cdot[0,3|4,4,3,1|1,0]$ & $5\cdot[0,0|5,4,4,1|0,0]$ & 
$5\cdot[0,0|6,3,3,2|0,0]$ & $6\cdot[0,0|6,5,2,1|0,0]$ \\ 
$6\cdot[0,1|5,5,1,1|1,0]$ & $6\cdot[0,2|4,2,2,2|1,1]$ & 
$6\cdot[0,3|2,2,2,2|2,1]$ & $6\cdot[1,1|3,3,3,1|2,0]$ \\ 
$6\cdot[1,2|2,2,2,2|3,0]$ & $7\cdot[0,1|2,2,2,2|2,1]$ & 
$7\cdot[0,2|2,2,1,1|4,0]$ & $7\cdot[0,2|4,4,1,1|2,0]$ \\ 
$7\cdot[0,4|3,3,2,2|2,0]$ & $7\cdot[1,2|3,3,3,3|1,0]$ & 
$8\cdot[0,0|5,5,3,1|0,0]$ & $8\cdot[0,0|6,4,2,2|0,0]$ \\ 
$8\cdot[0,1|5,3,3,1|1,0]$ & $8\cdot[1,1|3,3,2,2|1,1]$ & 
$9\cdot[0,0|4,2,2,2|1,1]$ & $9\cdot[1,1|4,4,4,2|0,0]$ \\ 
$10\cdot[0,0|3,3,2,2|1,1]$ & $10\cdot[0,0|3,3,3,1|2,0]$ & 
$10\cdot[0,0|4,4,1,1|2,0]$ & $10\cdot[0,2|5,3,3,3|0,0]$ \\ 
$10\cdot[0,2|5,5,2,2|0,0]$ & $10\cdot[0,3|2,2,2,2|3,0]$ & 
$10\cdot[0,3|3,2,2,1|3,0]$ & $10\cdot[1,1|4,4,3,3|0,0]$ \\ 
$12\cdot[0,1|2,2,2,2|3,0]$ & $12\cdot[0,1|3,3,1,1|3,0]$ & 
$12\cdot[0,3|3,3,3,3|1,0]$ & $12\cdot[0,3|4,4,2,2|1,0]$ \\ 
$14\cdot[0,2|3,3,3,1|2,0]$ & $14\cdot[0,2|4,2,2,2|2,0]$ & 
$15\cdot[0,0|4,4,4,2|0,0]$ & $15\cdot[0,0|5,3,3,3|0,0]$ \\ 
$16\cdot[0,0|4,2,2,2|2,0]$ & $16\cdot[0,2|4,4,4,2|0,0]$ & 
\color{gray}$18\cdot[0,0|4,4,3,3|0,0]$ & $18\cdot[0,0|5,5,2,2|0,0]$ \\ 
$18\cdot[0,1|3,2,2,1|3,0]$ & $18\cdot[0,1|5,4,2,1|1,0]$ & 
$18\cdot[0,2|3,3,2,2|1,1]$ & $18\cdot[0,2|4,3,2,1|2,0]$ \\ 
$18\cdot[0,3|4,3,3,2|1,0]$ & \color{gray}$18\cdot[1,1|3,3,2,2|2,0]$ & 
$22\cdot[0,0|4,3,2,1|2,0]$ & $22\cdot[0,1|4,4,3,1|1,0]$ \\ 
$22\cdot[0,1|5,3,2,2|1,0]$ & $22\cdot[0,2|5,4,3,2|0,0]$ & 
\color{gray}$23\cdot[0,1|3,3,3,3|1,0]$ & \color{gray}$24\cdot[0,0|3,3,2,2|2,0]$ \\ 
\color{gray}$24\cdot[0,2|4,4,3,3|0,0]$ & \color{gray}$42\cdot[0,0|5,4,3,2|0,0]$ & 
\color{gray}$43\cdot[0,2|3,3,2,2|2,0]$ & \color{gray} $52\cdot[0,1|4,4,2,2|1,0]$ \\ 
\color{gray}$74\cdot[0,1|4,3,3,2|1,0]$
\end{tabular}
\\\hline

$\frac{15}{2}$ &  
\begin{tabular}{rrrr}
$2\cdot[0,4|2,1,1,1|5,0]$ & $2\cdot[0,5|2,2,2,1|4,0]$ & \
$2\cdot[0,5|3,3,3,2|1,1]$ & $2\cdot[1,1|2,1,1,1|5,0]$ \\
$4\cdot[0,0|6,5,1,1|1,0]$ & $4\cdot[0,1|5,2,2,2|1,1]$ & \
$4\cdot[0,1|6,6,2,1|0,0]$ & $4\cdot[0,2|2,1,1,1|5,0]$ \\
$4\cdot[0,5|3,3,3,2|2,0]$ & $4\cdot[1,1|4,4,4,1|1,0]$ & \
$6\cdot[0,0|6,3,3,1|1,0]$ & $6\cdot[0,1|6,4,4,1|0,0]$ \\
$6\cdot[0,3|3,2,1,1|4,0]$ & $6\cdot[0,4|3,3,2,1|3,0]$ & \
$7\cdot[0,4|3,2,2,2|2,1]$ & $7\cdot[1,2|2,2,2,1|4,0]$ \\
$8\cdot[0,0|6,3,2,2|1,0]$ & $8\cdot[0,0|6,4,2,1|1,0]$ & \
$8\cdot[0,1|5,5,4,1|0,0]$ & $8\cdot[0,1|6,5,3,1|0,0]$ \\
$10\cdot[0,0|4,4,4,1|1,0]$ & $10\cdot[0,1|5,4,1,1|2,0]$ & \
$10\cdot[0,1|6,3,3,3|0,0]$ & $10\cdot[0,2|4,2,2,1|3,0]$ \\
$10\cdot[0,2|5,5,2,1|1,0]$ & $10\cdot[0,3|2,2,2,1|4,0]$ & \
$10\cdot[0,3|4,3,3,1|2,0]$ & $10\cdot[0,4|3,2,2,2|3,0]$ \\
$11\cdot[0,0|3,2,2,2|2,1]$ & $11\cdot[0,1|2,2,2,1|4,0]$ & \
$11\cdot[0,4|4,3,3,3|1,0]$ & $11\cdot[1,2|4,4,4,3|0,0]$ \\
\color{gray}$12\cdot[0,0|4,3,1,1|3,0]$ & $12\cdot[0,1|3,2,1,1|4,0]$ & \
$12\cdot[0,1|5,2,2,2|2,0]$ & $12\cdot[0,2|4,3,1,1|3,0]$ \\
$12\cdot[0,2|4,4,4,1|1,0]$ & $12\cdot[0,3|4,4,2,1|2,0]$ & \
\color{gray}$12\cdot[0,3|5,5,3,2|0,0]$ & $12\cdot[0,4|4,4,3,2|1,0]$ \\
$14\cdot[0,0|4,2,2,1|3,0]$ & $14\cdot[0,3|5,4,4,2|0,0]$ & \
\color{gray}$14\cdot[1,1|3,2,2,2|2,1]$ & \color{gray}$14\cdot[1,2|3,3,3,2|1,1]$ \\
$16\cdot[0,0|3,3,2,1|3,0]$ & $16\cdot[0,3|4,3,2,2|1,1]$ & \
$16\cdot[0,3|5,4,3,3|0,0]$ & $16\cdot[1,1|3,3,2,1|3,0]$ \\
$18\cdot[0,0|3,2,2,2|3,0]$ & $18\cdot[0,1|5,3,2,1|2,0]$ & \
\color{gray}$18\cdot[0,2|5,4,3,1|1,0]$ & $18\cdot[0,3|4,4,4,3|0,0]$ \\
$20\cdot[0,0|5,5,2,1|1,0]$ & $20\cdot[0,1|6,5,2,2|0,0]$ & \
\color{gray}$20\cdot[0,3|3,3,3,2|1,1]$ & \color{gray}$20\cdot[1,1|3,2,2,2|3,0]$ \\
\color{gray}$31\cdot[0,2|3,2,2,2|2,1]$ & \color{gray}$31\cdot[1,2|3,3,3,2|2,0]$ & \
$38\cdot[0,0|5,4,3,1|1,0]$ & \color{gray}$38\cdot[0,1|3,3,3,2|1,1]$ \\
$38\cdot[0,1|6,4,3,2|0,0]$ & $38\cdot[0,2|3,3,2,1|3,0]$ & \
\color{gray}$38\cdot[0,3|4,3,2,2|2,0]$ & \color{gray}$38\cdot[1,1|4,3,3,3|1,0]$ \\
\color{gray}$43\cdot[0,0|4,3,3,3|1,0]$ & \color{gray}$43\cdot[0,1|4,4,4,3|0,0]$ & \
\color{gray}$48\cdot[0,1|4,3,3,1|2,0]$ & $48\cdot[0,1|4,4,2,1|2,0]$ \\
\color{gray}$48\cdot[0,2|3,2,2,2|3,0]$ & \color{gray}$48\cdot[0,2|5,3,3,2|1,0]$ & \
$48\cdot[0,2|5,4,2,2|1,0]$ & \color{gray}$48\cdot[0,3|3,3,3,2|2,0]$ \\
\color{gray}$52\cdot[0,1|4,3,2,2|1,1]$ & \color{gray}$52\cdot[1,1|4,4,3,2|1,0]$ & \
\color{gray}$72\cdot[0,0|5,3,3,2|1,0]$ & \color{gray}$72\cdot[0,0|5,4,2,2|1,0]$ \\
\color{gray}$72\cdot[0,1|5,4,4,2|0,0]$ & \color{gray}$72\cdot[0,1|5,5,3,2|0,0]$ & \
\color{gray}$77\cdot[0,1|3,3,3,2|2,0]$ & \color{gray}$77\cdot[0,2|4,3,3,3|1,0]$ \\
\color{gray}$80\cdot[0,0|4,4,3,2|1,0]$ & \color{gray}$80\cdot[0,1|5,4,3,3|0,0]$ & \
\color{gray}$120\cdot[0,1|4,3,2,2|2,0]$ & \color{gray}$120\cdot[0,2|4,4,3,2|1,0]$
\end{tabular}

\\\hline

\end{tabular}
\caption{Spectrum of unprotected multiplets $[n_\bbb|n_\fff|n_\aaa]^{2222}$ with $\Delta_0^{2222} = 7$ and $\Delta_0^{2222} = \frac{15}{2}$.}
\label{table:spec8}
\end{table}

\newpage

\begin{table}[h!]
\centering
\footnotesize
\def\arraystretch{1.1}
\begin{tabular}{|c||l|} \hline
$\Delta_0^{2222}$&Multiplets \\\hline\hline

8 &  
\begin{tabular}{rrrr}
$1\cdot[0,0|1,1,1,1|6,0]$ & $1\cdot[0,0|5,1,1,1|4,0]$ & 
$1\cdot[0,4|1,1,1,1|6,0]$ & $1\cdot[0,4|5,5,5,1|0,0]$ \\ 
$1\cdot[0,6|1,1,1,1|6,0]$ & $1\cdot[0,6|2,2,2,2|2,2]$ & 
$1\cdot[0,6|2,2,2,2|3,1]$ & $1\cdot[0,6|2,2,2,2|4,0]$ \\
$1\cdot[0,6|3,3,3,3|1,1]$ & $1\cdot[0,6|4,4,4,4|0,0]$ & 
$1\cdot[1,1|1,1,1,1|6,0]$ & $1\cdot[1,3|1,1,1,1|6,0]$ \\ 
$1\cdot[2,2|1,1,1,1|6,0]$ & $2\cdot[0,2|1,1,1,1|6,0]$ & 
$2\cdot[0,3|3,1,1,1|5,0]$ & $2\cdot[0,5|2,2,1,1|5,0]$ \\
$2\cdot[0,5|3,3,3,1|3,0]$ & $2\cdot[0,6|3,3,3,3|2,0]$ & 
$2\cdot[1,3|2,2,2,2|2,2]$ & $2\cdot[2,2|2,2,2,2|3,1]$ \\ 
$3\cdot[0,0|6,2,2,2|1,1]$ & $3\cdot[0,0|6,2,2,2|2,0]$ & 
$3\cdot[0,2|5,5,5,1|0,0]$ & $3\cdot[1,1|5,5,5,1|0,0]$ \\ 
$3\cdot[2,2|2,2,2,2|2,2]$ & $4\cdot[0,0|7,4,4,1|0,0]$ & 
$4\cdot[0,0|7,7,1,1|0,0]$ & $4\cdot[0,1|3,1,1,1|5,0]$ \\ 
$4\cdot[0,5|4,4,4,2|1,0]$ & $5\cdot[0,0|2,2,2,2|2,2]$ & 
$5\cdot[0,0|2,2,2,2|3,1]$ & $5\cdot[0,0|6,4,1,1|2,0]$ \\ 
$5\cdot[0,2|6,6,3,1|0,0]$ & $5\cdot[0,4|2,2,2,2|2,2]$ & 
$5\cdot[1,1|2,2,2,2|2,2]$ & $5\cdot[1,3|4,4,4,4|0,0]$ \\ 
$5\cdot[2,2|2,2,2,2|4,0]$ & \color{gray}$5\cdot[2,2|3,3,3,3|1,1]$ & 
$5\cdot[2,2|4,4,4,4|0,0]$ & $6\cdot[0,0|5,5,5,1|0,0]$ \\ 
$6\cdot[0,0|7,3,3,3|0,0]$ & $6\cdot[0,5|3,3,2,2|2,1]$ & 
$6\cdot[1,2|2,2,1,1|5,0]$ & $7\cdot[0,0|2,2,2,2|4,0]$ \\ 
$7\cdot[0,0|4,2,1,1|4,0]$ & $7\cdot[0,0|7,6,2,1|0,0]$ & 
$7\cdot[0,4|4,4,4,4|0,0]$ & $7\cdot[0,4|5,5,4,2|0,0]$ \\ 
$8\cdot[0,0|6,3,2,1|2,0]$ & $8\cdot[0,1|2,2,1,1|5,0]$ & 
$8\cdot[0,1|6,6,1,1|1,0]$ & $8\cdot[0,2|4,2,1,1|4,0]$ \\ 
$8\cdot[0,2|6,5,4,1|0,0]$ & $8\cdot[0,4|3,3,1,1|4,0]$ & 
$8\cdot[0,4|4,4,3,1|2,0]$ & $8\cdot[0,5|4,4,3,3|1,0]$ \\ 
$9\cdot[0,4|3,2,2,1|4,0]$ & $10\cdot[0,0|3,3,1,1|4,0]$ & 
$10\cdot[0,0|7,5,3,1|0,0]$ & $10\cdot[0,1|5,2,2,1|3,0]$ \\
$10\cdot[0,1|5,3,1,1|3,0]$ & $10\cdot[0,2|2,2,2,2|2,2]$ & 
$10\cdot[0,3|2,2,1,1|5,0]$ & $10\cdot[0,3|5,4,4,1|1,0]$ \\ 
$10\cdot[0,3|5,5,3,1|1,0]$ & $10\cdot[0,4|2,2,2,2|3,1]$ & 
$10\cdot[0,4|4,4,2,2|1,1]$ & \color{gray}$10\cdot[0,4|5,5,3,3|0,0]$ \\
$10\cdot[0,5|3,3,2,2|3,0]$ & $10\cdot[1,1|3,3,1,1|4,0]$ & 
$10\cdot[1,3|2,2,2,2|3,1]$ & $10\cdot[1,3|2,2,2,2|4,0]$ \\
$10\cdot[2,2|3,3,3,3|2,0]$ & $12\cdot[0,2|5,5,1,1|2,0]$ & 
$12\cdot[0,3|4,4,1,1|3,0]$ & $13\cdot[0,0|3,2,2,1|4,0]$ \\
$13\cdot[0,4|5,4,4,3|0,0]$ & $14\cdot[0,0|6,6,3,1|0,0]$ & 
$14\cdot[0,0|7,5,2,2|0,0]$ & \color{gray}$14\cdot[1,1|2,2,2,2|3,1]$ \\
\color{gray}$14\cdot[1,3|3,3,3,3|1,1]$ & $15\cdot[0,0|5,5,1,1|2,0]$ & 
$15\cdot[0,2|6,6,2,2|0,0]$ & \color{gray}$16\cdot[0,4|3,3,3,3|1,1]$ \\
\color{gray}$16\cdot[1,1|2,2,2,2|4,0]$ & $18\cdot[0,0|6,5,4,1|0,0]$ & 
$18\cdot[0,0|7,4,3,2|0,0]$ & $18\cdot[0,3|4,2,2,2|2,1]$ \\
$18\cdot[0,4|2,2,2,2|4,0]$ & \color{gray}$18\cdot[0,4|4,3,3,2|1,1]$ & 
\color{gray}$18\cdot[1,1|3,2,2,1|4,0]$ & $18\cdot[1,2|3,3,3,1|3,0]$ \\
\color{gray}$20\cdot[0,0|3,3,3,3|1,1]$ & \color{gray}$20\cdot[0,0|4,4,4,4|0,0]$ &
\color{gray}$20\cdot[1,1|4,4,4,4|0,0]$ & $24\cdot[0,2|2,2,2,2|3,1]$ \\ 
$24\cdot[1,3|3,3,3,3|2,0]$ & $26\cdot[0,2|3,3,1,1|4,0]$ & 
$26\cdot[0,4|4,4,2,2|2,0]$ & $28\cdot[0,1|4,4,1,1|3,0]$ \\ 
$28\cdot[0,3|5,5,2,2|1,0]$ & $30\cdot[0,1|6,5,2,1|1,0]$ & 
$30\cdot[0,2|2,2,2,2|4,0]$ & $30\cdot[0,3|3,3,3,1|3,0]$ \\ 
$30\cdot[0,3|4,2,2,2|3,0]$ & $30\cdot[0,4|3,3,3,3|2,0]$ & 
\color{gray}$33\cdot[0,0|3,3,3,3|2,0]$ & \color{gray}$33\cdot[0,2|4,4,4,4|0,0]$ \\
\color{gray}$34\cdot[0,0|6,6,2,2|0,0]$ & $34\cdot[0,2|5,3,2,2|1,1]$ & 
$34\cdot[0,3|4,3,2,1|3,0]$ & $34\cdot[1,1|4,4,3,1|2,0]$ \\
\color{gray}$35\cdot[1,1|3,3,3,3|1,1]$ & $36\cdot[0,1|6,4,3,1|1,0]$ &
$38\cdot[0,2|5,3,3,1|2,0]$ & $39\cdot[1,1|4,4,2,2|1,1]$ \\ 
\color{gray}$40\cdot[1,2|3,3,2,2|2,1]$ & $41\cdot[0,2|3,2,2,1|4,0]$ & 
$41\cdot[0,4|4,3,3,2|2,0]$ & \color{gray}$43\cdot[0,0|4,4,2,2|1,1]$ \\ 
\color{gray}$43\cdot[1,1|5,5,3,3|0,0]$ & \color{gray}$44\cdot[0,1|3,3,3,1|3,0]$ & 
\color{gray}$44\cdot[0,3|5,3,3,3|1,0]$ & \color{gray}$45\cdot[0,0|5,3,2,2|1,1]$ \\
$45\cdot[0,2|5,4,2,1|2,0]$ & \color{gray}$45\cdot[1,1|5,5,4,2|0,0]$ & 
\color{gray}$46\cdot[0,0|5,3,3,1|2,0]$ & \color{gray}$46\cdot[0,2|6,4,4,2|0,0]$ \\
\color{gray}$49\cdot[0,0|4,4,3,1|2,0]$ & \color{gray}$49\cdot[0,2|6,4,3,3|0,0]$ & 
$52\cdot[0,1|5,4,4,1|1,0]$ & $52\cdot[0,1|6,3,3,2|1,0]$ \\ 
\color{gray}$54\cdot[0,1|4,2,2,2|2,1]$ & \color{gray}$54\cdot[0,2|3,3,3,3|1,1]$ & 
\color{gray}$54\cdot[1,1|3,3,3,3|2,0]$ & \color{gray}$54\cdot[1,1|4,3,3,2|1,1]$ \\
\color{gray}$54\cdot[1,2|4,4,4,2|1,0]$ & \color{gray}$60\cdot[0,0|4,3,3,2|1,1]$ & 
$60\cdot[0,1|5,5,3,1|1,0]$ & \color{gray}$60\cdot[0,1|6,4,2,2|1,0]$ \\
\color{gray}$60\cdot[0,3|3,3,2,2|2,1]$ & \color{gray}$60\cdot[1,1|5,4,4,3|0,0]$ & 
\color{gray}$60\cdot[1,2|3,3,2,2|3,0]$ & \color{gray}$63\cdot[0,0|5,4,2,1|2,0]$ \\ 
\color{gray}$63\cdot[0,2|6,5,3,2|0,0]$ & \color{gray}$71\cdot[0,0|5,5,3,3|0,0]$ & 
\color{gray}$72\cdot[0,2|4,4,2,2|1,1]$ & \color{gray}$72\cdot[1,1|4,4,2,2|2,0]$ \\
\color{gray}$73\cdot[0,2|4,4,3,1|2,0]$ & \color{gray}$73\cdot[0,2|5,3,2,2|2,0]$ & 
\color{gray}$74\cdot[0,1|4,2,2,2|3,0]$ & \color{gray}$74\cdot[0,3|4,4,4,2|1,0]$ \\ 
\color{gray}$77\cdot[0,0|5,4,4,3|0,0]$ & \color{gray}$77\cdot[0,0|5,5,4,2|0,0]$ & 
\color{gray}$77\cdot[0,0|6,4,3,3|0,0]$ & \color{gray}$80\cdot[0,1|3,3,2,2|2,1]$ \\
\color{gray}$80\cdot[1,2|4,4,3,3|1,0]$ & \color{gray}$84\cdot[0,0|6,4,4,2|0,0]$ & 
\color{gray}$90\cdot[0,0|4,4,2,2|2,0]$ & \color{gray}$90\cdot[0,2|5,5,3,3|0,0]$ \\ 
\color{gray}$92\cdot[0,3|3,3,2,2|3,0]$ &\color{gray} $96\cdot[0,0|5,3,2,2|2,0]$ & 
\color{gray}$96\cdot[0,1|4,3,2,1|3,0]$ & \color{gray}$96\cdot[0,2|5,5,4,2|0,0]$ \\ 
\color{gray}$96\cdot[0,3|5,4,3,2|1,0]$ & \color{gray}$101\cdot[0,2|3,3,3,3|2,0]$ & 
\color{gray}$110\cdot[0,1|3,3,2,2|3,0]$ & \color{gray}$110\cdot[0,3|4,4,3,3|1,0]$ \\
\color{gray}$112\cdot[0,0|6,5,3,2|0,0]$ & \color{gray}$120\cdot[0,0|4,3,3,2|2,0]$ & 
\color{gray}$120\cdot[0,2|5,4,4,3|0,0]$ & \color{gray}$122\cdot[0,2|4,3,3,2|1,1]$ \\
\color{gray}$122\cdot[1,1|4,3,3,2|2,0]$ & \color{gray}$130\cdot[0,1|5,5,2,2|1,0]$ & 
\color{gray}$160\cdot[0,1|4,4,4,2|1,0]$ & \color{gray}$160\cdot[0,1|5,3,3,3|1,0]$ \\
\color{gray}$176\cdot[0,2|4,4,2,2|2,0]$ &\color{gray} $210\cdot[0,1|4,4,3,3|1,0]$ & 
\color{gray}$256\cdot[0,2|4,3,3,2|2,0]$ & \color{gray}$378\cdot[0,1|5,4,3,2|1,0]$
\end{tabular}

\\\hline

\end{tabular}
\caption{Spectrum of unprotected multiplets $[n_\bbb | n_\fff | n_\aaa]^{2222}$ with $\Delta_0^{2222}=8$.}
\label{table:spec8}
\end{table}


\section{Difference equations} 

\subsection{Cancellations of apparent poles at Bethe roots}\label{ap:diffeq}
Consider the standard QQ-relation
\be
Q_{ab}Q_\emptyset = Q_a^+Q_b^- - Q_a^- Q_b^+ \quad \Rightarrow \quad Q_b = Q_a \Psi \left( \frac{Q_{ab}^+Q_{\emptyset}^+}{Q_a Q_a^{[2]}} \right)\label{basicQQ}
\ee
and use that all Q-functions are allowed to have poles only at $i\mathbb{Z}$ (or $i(\frac{1}{2}+\mathbb{Z})$ for an even number of indices). Then it should be that apparent poles coming from non-trivial zeros of $Q_a$ in the argument of $\Psi$ cancel out. We are going to derive an expression which explicitly accounts for this cancellation.

We consider only the case when all Q-functions on the right-hand side of \eqref{basicQQ} are rational. If $Q_{ab}$ involves $\Psi$-functions, an equivalent nested argument can be generated. Rewrite the Q-functions as
\be
Q_{A}=\frac{q_A p_A}{r_A}\,,
\ee
where $p_A$ and $r_A$ are fused factors of $u$ of the type $\prod (u+in)^k$ (or $\prod (u+in+\frac{i}{2})^k$ for $|A|$ even), where $n\in\mathbb{Z}$, and $q_A$ contains no such factors.  
Then it is possible to uniquely split the argument of the $\Psi$-summation in \eqref{basicQQ} as
\be
\frac{Q_{ab}^+Q_{\emptyset}^+}{Q_a Q_a^{[2]}} = \frac{q_{ab}^+q_\emptyset^+}{q_a q_a^{++}} \underbrace{\frac{p_{ab}^+p_\emptyset^+r_a r_a^{++}}{p_a p_a^{++} r_{ab}^+ r_\emptyset^+}}_{P/R} = \frac{A}{q_a}+ \frac{B}{q_a^{++}} + \frac{C}{R} + D\,,
\ee
where $R$ and $P$ are polynomials, and where $A$ and $B$ are polynomials of lower degree than $q_a$, $C$ is a polynomial of lower degree than $R$, and $D$ is a polynomial of at most the total asymptotic degree of the left-hand side. The polynomials $A$, $B$, $C$ and $D$ can be fixed by simply matching coefficients of individual powers in
\be
q_{ab}^+q_{\emptyset}^+ P = A q_a^{++} R + B q_a R + C q_a q_a^{++} + D q_a q_a^{++} R \,.
\ee
Now,
\be
Q_b &=& Q_a \Psi \left( \frac{Q_{ab}^+Q_{\emptyset}^+}{Q_a Q_a^{[2]}} \right) = Q_a \left( \frac{A}{q_a} + \Psi \left( \frac{A^{++}+B}{q_a^{++}} \right) +\Psi\left( \frac{C}{R} + D \right) \right) \no \\
&=& \frac{p_a A}{r_a} + Q_a \Psi\left( \frac{A^{++}+B}{q_a^{++}} \right) + Q_a \Psi\left( \frac{C}{R} + D \right)\,.
\ee
In this expression, the only potential poles away from $i\mathbb{Z}$ arise from the second term. Then it should be that $A^{++}+B=0$, since $A^{++}+B$ is of lower degree than $q_a$ and thus unable to cancel the poles otherwise. Conclusion:
\be
Q_b = \frac{p_a A}{r_a}+ Q_a \Psi\left( \frac{C}{R} + D \right) \,.
\ee

\bibliography{bibliography}
\bibliographystyle{elsarticle-num}
\biboptions{sort&compress}

\end{document}